# Society's Resilience to a Total Loss of Agriculture

Adin Richards

**Note to the reader**

This report was largely written in late 2023 and edited in early 2024. There may be data which is out of date at the time of this report's publication, but the author still endorses its overall conclusions.

## Abstract

While technology and trade have made modern food systems increasingly resilient to disruptions, it is unknown if human society could survive the most extreme threats to agriculture, such as from severe climate change or nuclear/biological warfare. One way that society could withstand such disruptions is to make food without agriculture, such as from edible microbes or chemically synthesized food. Here, I evaluate the feasibility of rapidly scaling up nonagricultural food production in response to a sudden disaster. I find that even in an idealized worst-case scenario where all current agricultural production ends instantaneously and cannot be restarted, it may be possible for at least some countries to begin producing enough food without agriculture to feed their populations before existing food supplies run out. This means that it may be possible to survive a disaster in which all plants suddenly die. As a proof of principle, for the US, producing edible bacteria grown on natural gas appears to be one of the best options for quickly making food without agriculture. Feeding the entire US population this way would require ~16% of US annual natural gas production, ~6% of US annual electricity consumption, and ~$530 billion (~1.9% of US GDP) worth of food production facilities. Given the US's natural gas reserves, this method of food production could potentially sustain the current US population for over five centuries. Several key findings from this report remain uncertain, and if a disaster hits at the worst possible time with respect to annual variation in stored food supplies, there may not be enough time to scale up other food production options. However, this report suggests that some countries are close to, or may already be at, the point where they could survive just about any agricultural disaster, as long as modern industry remains intact. I find that there are several interventions that could more thoroughly close this window of vulnerability: designing and piloting alternative food production facilities in advance of a disaster, improving the nutritional quality of alternative foods through research and development, and increasing crop stock levels.

# Summary of findings

- It is an open question whether society could weather a global agricultural disaster, such as those that could theoretically be posed by extreme climate change, nuclear winter, or biological warfare. To get more clarity on this question, this report examines whether society could survive the most extreme version of such a disaster: a sudden and total loss of agriculture.

- This report focuses on whether the United States (US), given current manufacturing capacity, infrastructure, and supply chains, could begin producing enough nonagricultural food to feed its population before existing food supplies ran out in a disaster affecting agriculture. The US was chosen for this analysis because it is likely among the best-positioned countries to weather an extreme disaster (and, among those, has the most available data), and therefore if it is not possible to scale up food production without agriculture in the US, it is likely not possible anywhere. As this report's findings suggest that alternative food production is feasible in the US, future work should explore optimized resilience strategies for the US and the rest of the world.

- The report discusses food production options in several loss of agriculture scenarios. The following sources of food are precluded in all cases: growing new plants outdoors, raising new livestock, hunting wild terrestrial animals, and harvesting non-woody plant biomass. I assess "total-loss" scenarios in which both harvesting of woody plant biomass and fishing are also impossible, and I look at options when these two constraints are relaxed. I also look at the value of advanced warning of a disaster and how that impacts survival odds.

- I find that in a loss of agriculture scenario, one of the most scalable ways to produce food in the US is through the fermentation of natural gas to grow edible methanotrophic bacteria, a technology currently used to make "single-cell protein" products, mostly for livestock feed. Even if there is no advanced warning and a disaster hits at the worst time with respect to the US crop calendar so that food stocks are at an annual minimum, the US likely has the manufacturing and construction capacity to make the facilities needed to meet its nutrition needs before exhausting food reserves. However, there is little room for error, and it may be critical to have engineering plans in place to start construction quickly enough. With some advanced warning, feeding the US with industrially produced foods seems feasible. Current single-cell protein technology does not provide complete nutrition, so some other foods or supplements would also have to be scaled up, which I evaluate to be possible but challenging. Alternatively, it seems possible to engineer strains of natural gas-fed microbes to provide better nutrition, although I have not evaluated feasibility in depth.

- I briefly extend my analysis to propose strategies for improving resilience to agricultural shocks. Options include designing plans for scaling up industrial foods, improving industrial food technology, promoting the use of industrial foods in appropriate market niches during non-disaster times, and increasing food reserves.

# Contents

# Scenarios considered

In this report, I look at options for society to cope with an unprecedented loss of agriculture scenario. The potential causes of a global agricultural disaster (e.g., nuclear winter, climate change, widespread biological warfare) would be unlikely to lead to such extreme outcomes. However, I study a "total loss" scenario because it in many ways provides an instructive "worst-case" starting point for analyzing society's resilience.

To study agriculture loss in isolation, I consider an abstracted disaster that only affects agriculture, rather than a more complex threat that also directly causes population loss, infrastructure damage, or other issues. A disaster affecting agriculture could take many forms, but a discussion of specific threats is outside the scope of this report. I focus on how societies could respond to such a disaster by scaling up the production of nonagricultural food sources. I am mostly concerned with food production options in a scenario where a disaster makes it impossible to grow any crops, harvest even non-crop plant biomass, or draw on marine food sources like fish. I also generally assume that a disaster hits instantaneously, meaning there is no warning or time to prepare. However, I include discussions of how access to fish (appendix 7.0 Fishing) and non-crop plant biomass impacts food production potential, and I also provide a brief note on the impact that an advanced warning of a disaster would have on the time available to begin nonagricultural food production (appendix 2.6 Note on advance warning potential).

To evaluate food production options available in the above scenarios, I focus specifically on the US's ability to begin producing food without outdoor grown plants before existing food supplies are drawn down. I assume that the US has to scale up alternative food production without access to international trade (autarky). I focus on the US as a test case because it is likely one of the countries most capable of quickly scaling up food production and weathering the disasters considered in this report, but I also consider other countries (see in appendix 20.0). This is because it has a much higher ratio of stored food to population size than most other countries (see appendix 3.0 Note on food stores in other large countries), and has a fairly diversified industrial base. The large food reserves give the US more time to scale up alternative foods before existing stockpiles run out, and the diversified industrial base means that the US may be able to maintain most of its industry without trade.[1] Exploring the ability of other countries to weather a disaster like this is an essential future research direction.

Summarizing, the core scenario makes the following assumptions:

- All crops die instantly, and no crops can be grown in the future (an idealized worst-case agricultural disaster)
- There's no fishing or use of non-crop plant biomass (although I analyze cases where these assumptions are relaxed)
- Infrastructure is otherwise undamaged (to evaluate the effects of an agricultural disaster in isolation)
- Human populations are unharmed by the disaster, but vulnerable to starvation (again to evaluate the effects of an agricultural disaster in isolation)
- The US has to operate without access to international trade (since I'm evaluating only one country's response capacity)

It's important to note that these assumptions were made for the purposes of a) evaluating the limits of society's capacity to weather a hypothetical threat to agriculture and b) to simplify my analysis. They do not reflect any specific threat model or what I think is most likely. For example, a truly "instantaneous" loss of agriculture is unrealistic, trade may decline but complete autarky seems unlikely, and a disaster resulting from an armed conflict would also cause population losses and infrastructure damage.

# US food supplies could last nearly two years if managed well

In a loss of agriculture scenario, a population would have to begin producing food industrially before food stores run out. The amount of stored food varies substantially within a year, as crop stocks are replenished after harvests and depleted throughout the rest of the year. US food stores are at their lowest in September and highest in December, since the country's two most important crops, corn and soy, are harvested in the fall. The amount of food stored in September could feed the whole US population for 21–22 months if waste was reduced, consumption rationed to an average of 2,100 kcal per person per day, and all edible crops only fed to people (rather than being used for livestock feed or biofuel production). Under these same assumptions, the amount of food in stock in December could feed the US for almost 70 months. For more detail on these estimates, see appendix 2.0, and see appendix 1.0 for a brief discussion of current US food production and allocation. In some of the disaster scenarios considered in this report, the US could supplement this stored food with short-term food sources like increased fishing or turning wood into sugar. These options would not sustainably replace agricultural food production, but could act as temporary stopgaps, potentially adding a few extra months of food to current reserves.

---

[1] Most of this report is concerned with what food production options are possible given modern industry and supply chains. However, in appendices 17.0 and 18.0, I explore nonagricultural food production options that may be possible in cases where society collapses in a disaster and survivors do not have access to this infrastructure. I evaluate several cases, ranging from what extremely isolated individuals or small groups of people may be able to accomplish up to scenarios where a population center maintains some local industry, but is cut off from outside trade and national infrastructure.

## US stored food

In the US, corn and soy are the two dominant crop species, accounting for almost 78% of the calories produced from crops. To look at a worst-case scenario, I calculated the number of calories available in stored crops in September, right before most corn and soy are harvested. Most wheat, which is the third most important crop and makes up ~6% of calories produced from crops, is harvested a bit earlier. This means that wheat stocks are actually at their highest in September. However, given the dominance of corn and soy in US agriculture, September would be the worst month for an instantaneous disaster to strike. The graph below[2] shows the number of months that the amount of stored crops at different times of year could feed the US, assuming that these were the only source of food, that the US cut consumer waste from its current ~33% to ~13%, and that everyone ate at an average of 2,100 kcal per day. See appendix 1.2 for a discussion of current food waste levels, appendices 2.3 Food waste and 2.4 Food spoilage for my assumptions for post-disaster food waste and spoilage rates, and appendix 4.0 Average adult calorie needs for a note on post-disaster food consumption.

---

[2] The data for this chart can be found on tab 1 of this Google Sheet.

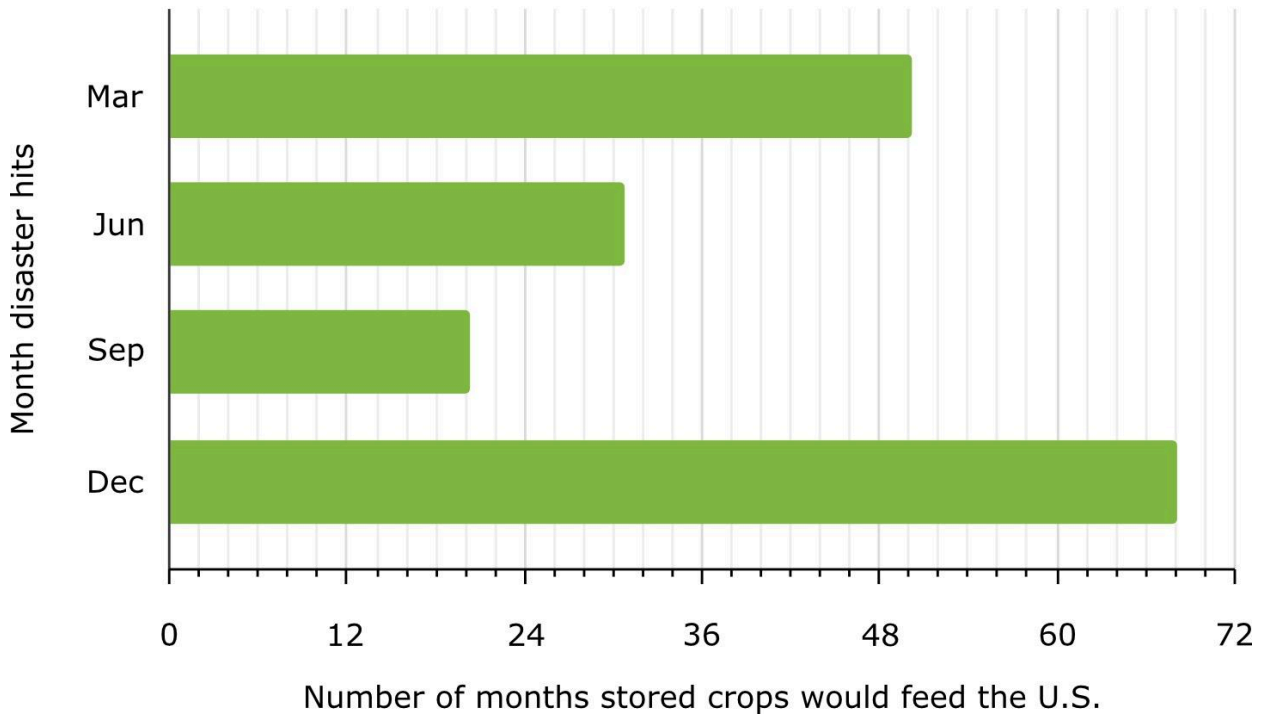

**Number of months stored crops would feed the U.S. depending on when disaster hits**

*Number of months stored crops would feed the U.S.*

There is considerably less food stored in the US in the form of livestock and animal products than there is from stored crops and crop products. At any given time, the amount of meat, dairy, and eggs in the US constitutes enough calories to feed all Americans for about one month, assuming no spoilage or waste, if all that the US ate for that time were these animal products. The US also has around 1.5 billion chickens, 74 million pigs, and 92 million cows. These three species represent around 95% of US livestock by headcount, so I focus on them. If all chickens, pigs, and cows could be slaughtered, processed, and stored, this represents about 165 days, or over 5 months, of food. In a scenario where all plants die, livestock should not be kept alive, because there are significant energy losses from feeding crops to animals and then eating them. If only crops died, but other plants like grass were still alive, then some livestock production may continue. The amount of time that the US would have to slaughter, process, and store livestock depends on the scenario in question, but I discuss in appendix 2.4 that a conservative estimate would only see the US storing about 30% of these calories, providing just ~1.5 months of food.

The chart below shows how long stored crops, animal products, and slaughtered livestock could feed the US population. I show how long existing food supplies would last if there was no food waste or spoilage, and how long I expect food to last given my assumptions about food waste and spoilage. The chart only shows how long supplies in September, when food supplies are at their lowest, would feed the US population.

| Source | Without waste or spoilage (days) | With waste and spoilage (days) |
|---|---|---|
| | | |
| Stored crops | 709 | 617 |
| Stored animal products | 32 | 24 |
| Live animals | 165 | 44 |
| Total | 906 | 685 |

## Stopgap food sources

In scenarios where aquatic ecosystems do not collapse, fish may offer a short-term food supply. Seafood currently provides only ~0.8% of US calories. This means that if fishing rates were not increased, a year's worth of fishing would only provide about three days of food. However, I discuss in appendix 7.0 how much more fish the US fishing fleet could plausibly catch, and find that over the course of a year, the US may be able to get ~1 month's worth of food from additional fish, but this estimate is uncertain.

If the disaster kills plants but dead plant biomass is still usable, then this could present another stopgap food source. In such a scenario, most of the available plant matter would be wood. This is because trees constitute about 70% of terrestrial plant mass, and wood also degrades slower than other plants. Two ways that wood could provide calories is by using it to grow mushrooms, or turning it into sugar through a process of removing lignin and saccharifying cellulose and hemicellulose into glucose and xylose — "lignocellulosic sugar". The appendix includes a discussion of both options. I find that even though mushrooms are much more nutritious than sugar, they would be slower to scale and require around 15–25x more wood than lignocellulosic sugar production to produce a given amount of calories (see appendix 18.5 for a discussion of mushroom calorie production efficiency when grown on wood, and appendix 6.0 for a discussion of lignocellulosic sugar production). Therefore, producing lignocellulosic sugar is probably the better use of wood supplies.

Without building new facilities — and only retrofitting existing sites that currently make ethanol, beer, or pulp and paper — the US could, over the course of a year, make enough lignocellulosic sugar to meet ~52% of its annual calories. The amount of wood needed to do this would only be about 40% of what the US harvests in a year, so the infrastructure for collecting enough biomass is already in place. The US may not be able to retrofit all of its existing facilities for sugar production, so sugar produced this way could probably contribute about three months of food in the short term (see appendix 8.1 for a discussion). In such a scenario, the sugar would be consumed along with other foods, and I discuss how much sugar can be included in different diets in appendix 6.1. Later in the report, I also describe scenarios where sugar from wood is used as a longer-term food source, either directly or as a feedstock for fungal or microbial foods. The table below summarizes my estimate for how many additional months of food the two stopgap food sources discussed here — fish and lignocellulosic sugar — could provide the US over the course of a year (e.g., fishing for a year, producing sugar for a year).

| Stopgap | Days of additional food |
|---|---:|
| Additional fishing | 30 |

| Lignocellulosic sugar | 91 |
| --- | --- |

## Overall food supplies

If a disaster strikes without warning at the worst time for US food stores (around September), then good coordination and management could extend US food supplies for roughly 21–26 months, depending on whether food sources like fishing and lignocellulosic sugar are available. This means that in the most dire cases where no new plants can be grown outdoors, the US may have under two years to begin producing food without agriculture at a level high enough to meet all US calorie requirements. This is shown in the figure below,[3] which includes different food sources and how long they would last.[4]

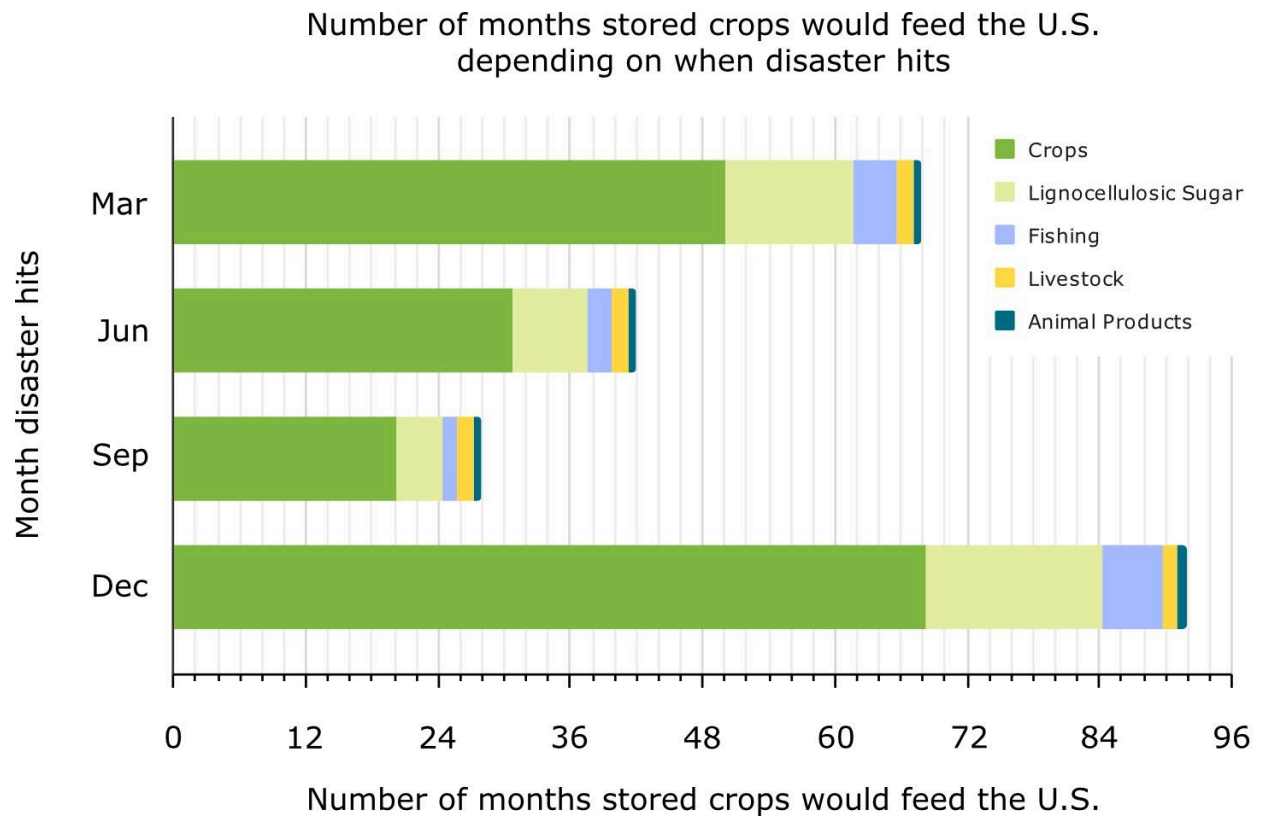

Number of months stored crops would feed the U.S. depending on when disaster hits

# Methane-fed microbes appear to be a particularly scalable resilient food in the scenarios considered

In the most severe scenarios that I am considering, any food production would have to be completely independent of plants grown outdoors. In some less severe cases, non-crop plant matter could be used, either because plants are killed but not destroyed, so their dead biomass can be used as an input to food production, or because the threat spares non-crop plants so they can be freely grown and harvested. While this report focuses on food production options that do not rely on any plants, since these are most robust across all scenarios, I also discuss options opened up by access to non-crop plant biomass.

For food that does not rely on plants grown outdoors, there are three major options:

1. Grow plants, or other photosynthetic organisms, in protected indoor environments
2. Abiotically synthesize nutrients from non-plant-derived feedstocks
3. Grow edible chemotrophs on non-plant-derived feedstocks

I review several options in each category and look at the major material or energy inputs required to produce each. To simplify comparisons of the scalability of different food options, I first examine what would be needed to produce enough of each food to meet all US calories with it, even though none of the foods that I evaluate could actually serve as a population's only food source. I also conservatively assume for now that if making enough of a given food to meet US calorie needs would require more of a given material or more energy than the US currently makes, that food is likely a poor choice as an option for quickly producing calories in an emergency. I discuss the limits of this approach in appendix 10.0. Several foods appear promising, but edible microbes grown on natural gas is the option that most clearly meets my stringent criteria. This technology is already used at an industrial scale to produce livestock feed, and there are methods for ensuring that it is safe for human consumption (appendix 11.0). For more detail on the food sources described below, see appendix 9.0.

After finding that methane-fed microbial food may be especially scalable given the current availability of the materials needed to produce it, I then evaluated its overall nutritional value (appendix 11.0 Methane-fed microbial food nutrition), and homed in on more specific questions of equipment and facility requirements and scale-up speeds.

## Growing crops or algae in biosecure environments

Massive energy requirements are the main reason that crops or other photosynthetic organisms cannot be produced at sufficient scale to feed the US in a setting that is robust against biological threats (see appendix 9.3 Indoor agriculture). If grown fully on artificial light, the electricity needed to grow enough indoor crops to feed the US for a year is almost 7x the US's annual electricity production, even without accounting for any inefficiencies, losses to heat, or the energy required to maintain a suitable indoor growing environment (e.g., temperature control). Even crops grown under natural sunlight in greenhouses require substantial amounts of energy. While the sun provides the energy for plant growth, additional energy has to be used to ventilate and maintain the greenhouses. Some lower-tech greenhouses do not require large energy inputs, but they require natural ventilation and so are not biosecure. High-tech greenhouses similar to those in use today would require ~125% of all US electricity to produce enough food to feed the US, making them unsuitable at a large scale.

Aquatic single-cell photosynthesizers, like *Arthrospira* and *Chlorella* species, grow more efficiently than crops, converting sunlight into biomass at about 3–4x the efficiency of most land plants. This suggests that producing all US calories from algae grown on artificial light would ~1.7x the US's current annual electricity production, again ignoring inefficiencies, losses to heat, nutrient circulation, harvesting, and other complications. Growing algae in open outdoor ponds requires very little electricity, but may not be possible in scenarios where the threat affects aquatic photosynthesizers. A compromise option — growing algae in closed transparent tubes outdoors so they can use sunlight but are protected from potential threats — may be possible, and would probably require less than 20% of current US annual electricity production to power the needed circulation systems. However, providing enough nutrients to algae would require significantly more fertilizer, especially potassium, than the US currently produces. As discussed in appendix 9.3.2, it may be possible to quickly increase US fertilizer production, but under my strict criteria for assessing food scale-up potential (appendix 10.0), the need to increase the output of multiple industries simultaneously probably makes algae a poor choice for rapid food production in a disaster.

| Indoor agriculture option | Likely bottleneck |
|---|---|
| Crops grown with artificial light/in greenhouses | To grow enough food with fully indoor ag, the US would have to use something like 7x its current total electricity production.<br><br>To grow enough food with greenhouses that have at least some filtration system to keep out biothreats, the US would have to use ~125% of its current electricity.<br><br>There would also be various material constraints that I didn't look into, like light bulbs or cover material for greenhouses. |
| Algae | If grown with artificial light, it would need ~1.7x US electricity to produce all US calories.<br><br>Would need less electricity to grow outdoors, but this may not be biosecure, depending on the threat.<br><br>Either way, severely nutrient-limited, such as requiring ~35x more potassium per year than is currently produced in the US. |

## Chemically synthesized food

All three macronutrients — fat, carbohydrate, and protein — have been synthesized from abiotic substrates, as have several vitamins. However, the energy or material requirements for scaling up all of these options go well beyond current US capabilities, making them poor choices for quickly meeting US calorie needs (see appendix 9.1 Nutrient synthesis). As noted above, to simplify the comparison of different synthetic food options, here I look at what would be needed to produce enough of one macronutrient to meet all US calorie needs, even though this of course would not be a viable diet.

Sugar has been produced from $CO_2$ through electrosynthesis, although only at an experimental scale. Assuming a maximum theoretical efficiency of the intermediate steps, the electrosynthesis of glucose from $CO_2$ would require around ⅓ of US annual electricity production to make a year's worth of calories for all Americans. More realistic estimates of energy efficiency would put the energy requirements at over 2x current US annual electricity production. Even if high efficiencies could be achieved, production would also require greater material input than could likely be supported by existing US infrastructure, like large amounts of $H_2$ gas production and $CO_2$ capture.

One essential amino acid, methionine, is currently produced abiotically, and the US makes almost 250,000 tonnes (kt) of synthetic methionine each year. However, achieving the ~260x increase in methionine production needed to make enough to meet all US calorie needs would require that the US also increase its annual supply of methionine's major precursors, methanethiol (a methanol derivative) and acrolein (a propene derivative), by two orders of magnitude. This probably puts the rapid scale-up of synthetic amino acids out of reach.

Fats can be produced from Fischer-Tropsch wax, which has historically been made from petroleum or liquified coal. In fact, during WWII, Germany produced "coal butter", which was reportedly eaten by some soldiers in amounts up to ~700 kcal per day. Recently proposed production methods have reportedly improved on the efficiency of the techniques used in WWII and can use other feedstocks like natural gas. With these modern production processes, all US calories could be made with just ~7% of US annual natural gas supplies or ~19% of US annual coal production. However, production requires glycerol. While other countries make glycerol from propylene, a petroleum derivative, all US glycerol comes from biodiesel production, and so would not be available in the disasters considered here since there would be no new crops. The US would therefore have to start making glycerol from propylene. Current US propylene production is sufficient to make enough glycerol, but this would require scaling up another chemical process on top of scaling up the production of Fischer-Tropsch wax and synthetic fat itself. This probably precludes significant synthetic fat production as a quick solution, but some amount of fat production seems plausible.

There are also several processes currently used to synthesize vitamins; these are discussed in appendix 12.0.

The table below summarizes the findings from this section, and more detail is given in appendix 9.1.

| Chemically synthesized food | Likely material or energy bottleneck |
|---|---|
| Synthetic carbohydrates | Assuming a maximum theoretical efficiency of the intermediate steps, the electrosynthesis of glucose from $CO_2$ would require ~a third of current US annual electricity production to make a year's worth of calories for all Americans. Accounting for likely inefficiencies, production would need ~1.7x the amount of electricity currently produced in the US. |
| Synthetic amino acids | Only one of the nine essential amino acids that humans need, methionine, has been synthesized abiotically. Achieving the ~260x increase in |

| | methionine production needed to meet all US calories would require a ~100x increase in methanethiol production and a ~125x increase in acrolein. This in turn could only be done by increasing the production of chemicals used to make these inputs, like increasing methanol production by almost 4x. |
|---|---|
| Synthetic fat | All proposed processes require glycerol, which in the US comes from biofuel production. Since this production requires crops, and since the disasters considered here would force humans to directly consume crops instead of making biofuels, biofuel-derived glycerol would be inaccessible. <br><br> The US could make enough synthetic glycerol from its current annual production of propylene to then make enough synthetic fat for its calorie needs, but it would have to increase its annual chlorine production by ~2x. |

## Chemotrophic microbial food

The final option is to grow edible microbes on substrates that don't come from plants (appendix 9.2). A common product made through this type of biomass fermentation is "single-cell protein" (SCP), which is an industry term of art describing what is simply dried microbial cell mass that contains other macro and micro nutrients besides protein. Since the product is generally used to supplement livestock feed, high-protein products are most economical under normal market conditions. However, biomass fermentation can instead be used to produce high-fat products, as is sometimes done for making biodiesel, and organisms could likely be selected for a more balanced nutritional profile. For now, I focus on current protein-heavy SCP products as a model alternative food, because there is more information about their production processes. In appendix 12.0 I discuss how different strains/species and growth conditions could be optimized to improve the nutritional value of the product.

Historically, the world's largest SCP plant used methanol as the feedstock.[5] However, making enough SCP to meet US calorie needs would require increasing annual methanol production by ~20x, which poses immediate scaling challenges. One company that has recently received approval to sell human-edible SCP in Singapore uses $CO_2$ and $H_2$ as the feedstocks for production, but making enough of this product to feed the US would require a 3x increase in US annual $H_2$ production and an almost 3x increase in annual $CO_2$ capture.[6] Methane, the main component of natural gas, has also been used as a carbon source for SCP, and the US would only have to use ~16% of its current annual production of natural gas to meet its calorie needs for a year, including the natural gas needed to run production facilities. There are several other non-plant-derived substrates that have been used for SCP production, such as crude oil and peat, but similar to methanol and $CO_2$, the US would have to significantly increase its production of these to make enough SCP to meet its calorie needs. A possible exception is diesel, but there is very little information on the use of this substrate as a feedstock for SCP, so for now I focus on natural gas-fed SCP. In appendix [18.4 Microbial food substrate options](#) I discuss other possible SCP substrates, but I have not exhaustively looked at all options.

---

[5] Reportedly, the USSR had several plants that were even larger than the largest known methanol-fed SCP facilities and these used paraffin wax (which comes from petroleum or coal), but much less information is available on the processes used at these plants. I discuss paraffin wax–fed SCP, and other less well–studied feedstock options, in appendix 18.4 Microbial food substrate options.

[6] This is not accounting for the fact that ~80% of captured CO2 is used for oil and gas recovery, which would have to continue during the disaster.

The table below summarizes the substrate availability in the US for the three main substrate options considered here.

| Substrate option | Current substrate availability in the US |
|---|---|
| Methanol | The US would need to ~20x current annual methanol production to make enough to feed all Americans. |
| $CO_2$ and $H_2$ | The US would have to ~2.7x the amount of $CO_2$ it captures to make enough SCP. <br><br> The US would also have to 3x $H_2$ production. |
| Natural gas (methane) | Using only ~16% of current US natural gas production, the US could feed itself on microbes grown on natural gas. |

# Methane-fed microbe production scale-up is probably not limited by energy, material, or equipment requirements

In the previous [section,](#) I found that microbes grown on natural gas may be a uniquely scalable way to produce food without agriculture in the US test case. This is because the US only needs around one-sixth of its current natural gas production to meet all of its calorie needs with methane-fed SCP. Of course, growing all of this microbial mass requires substantial amounts of other resources. Operating hundreds of large, world class–sized fermentation sites would require a sizable share of all US electricity, and the equipment, much of it bespoke for the particular fermentation process, would have to be manufactured quickly to start producing food before existing reserves run out. Making all of this equipment would require large amounts of raw materials, as would maintaining the nutrient-rich growth media needed for efficient microbial production. However, mass and energy balance calculations, market proxies for US equipment production potential, and historical precedent suggest that meeting all of these requirements may be within America's industrial capacity.

## Energy inputs

Given the caloric density of methane-fed SCP, feeding everyone in the US an average of 2,100 kcal per day would require making about 64 million tonnes (Mt) of SCP per year. Accounting for likely waste levels (appendix [2.3](#)), the necessary quantity is probably closer to ~73 Mt. The largest [proposed](#) facilities for making natural gas-fed SCP would produce about 100,000 tonnes per year (ktpa).[7] As a first pass, this report assumes that all facilities built in a disaster are also this size, although the optimal size depends on several factors, like US natural gas pipeline infrastructure, that are beyond the scope of this report. This means that the US would have to build over 700 production plants, at an estimated total cost of around $530 billion, or ~1.9% of US GDP (see appendix 13.1 Facility construction and operation costs). The electricity required to produce enough SCP would equal about 6% of all US electricity production (see appendix 13.2 Electricity). This is higher than the current electricity consumption of the US food and beverage industry (~4.2% of total consumption), but national projects have consumed a similarly large amount of the nation's power before. From 1943 to 1945, the Manhattan Project reportedly used around 14% of US electricity. At 6%, the energy demands of large-scale SCP production are unlikely to be prohibitive.

## Material requirements

The material demands of SCP production comprise both the needed nutrient media and equipment.

As discussed in appendix 13.3, a standard mix of nutrient and buffering mineral salts can likely be provided without exceeding US domestic production of any ingredient, even as the exact media composition can vary depending on the desired properties of the final product.

For equipment, the largest pieces are the stainless steel bioreactors. Making enough of these to grow ~73 Mt of SCP per year would require about 15% of US annual stainless steel production. This is quite a substantial demand, but represents less metal than is currently turned into heavy gauge tanks each year, and heavy gauge tanks are similar to the bioreactors needed for SCP production. Including other equipment requiring stainless steel puts the total demand at ~53% of total annual production. This is less than the ~60% of the US's stainless steel supply that is recycled each year, meaning that even in a scenario of US autarky, recycled steel alone would provide enough material to make all of the needed equipment. This is important because the US otherwise does not have enough domestic nickel and chromium primary production (i.e., from mines as opposed to from recycled material) to make the needed amount of stainless steel. For the other materials needed for equipment that I could identify, US domestic production is also sufficient, as detailed in appendix 13.3 Equipment and building materials.

## Equipment manufacturing

I find that the US is likely able to manufacture all of the equipment needed for methane-fed microbial food production within the roughly two years before existing food supplies would run out, if the disaster strikes at the worst time with respect to US food stores. This is based on the cost of individual types of equipment, such as fermenters, centrifuges, pipes and valves, etc. (see details on the equipment in appendix 13.5 Equipment). For most equipment categories, the cost of different types of equipment, scaled to the level needed to produce enough SCP to meet US calorie needs, is less than the annual sales of industries that currently produce these or similar pieces of equipment. For example, high-temperature processors are needed to treat SCP to make it safe for consumers, and the cost of all of the equipment needed to do this is less than the sales of US high-temperature food treatment equipment manufacturers. I take this as a rough indication that US domestic industries could manufacture the needed equipment.

---

[7] Currently, SCP is used as fish or livestock feed. To be suitable for human consumption, it has to go through additional processing, described in appendix 11.0. The models used in this report for estimating equipment requirements and facility costs account for this extra processing step. One hundred ktpa is around the same production capacity of some Soviet SCP plants that used petroleum waxes as a feedstock. In the UK, Imperial Chemical Industries used methanol to make SCP, with one plant having a capacity of up to ~75 ktpa.

The sales data of different industries may be an optimistic estimate of US production potential, since a large share of the equipment for microbial food production is specific to the needs of an SCP plant, and the equipment that an industry typically produces may not be suitable (e.g., wrong size or design). However, I believe that historical precedent for the ability of industry to shift to making particular products in a short period of time, including novel ones, actually makes my estimates somewhat conservative. During COVID-19, GM and Ford produced tens of thousands of ventilators, even though these are not machines that the automobile makers usually produce. One Ford car parts plant began production within one month of the pandemic's onset, and within five months had made 50,000 ventilators. This suggests that other industries may well be able to contribute to production capacity, so my method of looking only at industries currently making specific material needed may underestimate US manufacturing capabilities. In 1940, the US military placed a $9.2 billion order to US automakers to begin producing warplanes, some of which had completely novel designs. This order amounted to ~0.7% of US GDP at the time, and was fulfilled within a year. Once the US entered WWII, its increase in equipment production was even faster, again showing the ability of national production capabilities to quickly provide massive amounts of needed equipment in emergencies. Similarly, at its peak the Manhattan project cost ~0.4% of the US GDP, showing that substantial levels of research and development, as may be needed to refine and scale up nonagricultural food production in a disaster, can be quickly supported through a national effort. For reference, the cost of the equipment for making enough SCP to meet all US calorie needs is probably ~$117 billion, or ~0.4% of US GDP. I believe that the scale of manufacturing needed to make enough equipment for SCP production is within the reach of US producers, given my assessment of equipment requirements and relevant industry sizes.

## Overall production requirements

The table below summarizes the amount of different major inputs that are needed to produce enough methane-fed microbial food to meet all US calorie needs, as a percent of current US annual supply of each input. For electricity and natural gas, the figures in the table show ongoing annual requirements for microbial food production, while steel, construction capacity, and equipment manufacturing represent the percent of current US supply needed to make all of the SCP production facilities, and so would be a roughly one-time expense, modulo ongoing requirements for maintenance and repair. The percent of construction capacity is defined here as the construction cost of making all of the needed SCP facilities, as a percentage of the annual revenue of the US nonresidential building construction sector. Similarly, equipment costs are compared to the revenue of US fabricated metal product manufacturers to calculate the percent of US equipment manufacturing capacity needed for microbial food facilities. More specific comparisons are made in appendix 13.0. To provide some comparison, the table also shows the percentage of these inputs currently used by several other sectors in the US.[8]

---

[8] Data on inputs to different industries comes from the Bureau of Economic Analysis Input-Output data and the Energy Information Agency's Manufacturing Energy Consumption Survey data; the sales data comes from Census Bureau data on US business.

| | | Production of other industries, for comparison | | | |
|---|---|---|---|---|---|
| Input | Microbial Food | Food and agriculture | Total manufacturing | Chemical manufacturing* | Primary metal production* |
| % current US natural gas production | 15.8 | 3.8 | 29.8 | 13.2 | 6.4 |
| % current US electricity generation | 5.9 | 4.2 | 17.7 | 3.4 | 2.6 |
| % current US steel production | 37.2 | 0 | 58.9 | 0.2 | 13.3 |
| % US current construction capacity | 48.3 | 1.1 | 9.7 | 0.7 | 0.2 |
| % US current equipment manufacturing output | 34 | 5.6 | 48.1 | 3 | 2.1 |

\* These sectors are subsectors of the Total manufacturing sector shown to their left

Note that the table shows input requirements for scaling up microbial food as a fraction of annual US capacity or supplies (for steel, construction, and equipment manufacturing). In reality, the US would not actually have to, or likely be able to, make all SCP production plants in just a year (see next section), so in this sense the chart understates the inputs that would be available.

# Quickly constructing facilities to make methane-fed microbial food is difficult but may be possible, even in worst-case scenarios

Even if the US can produce the needed feedstock, material, energy, and equipment inputs to make enough natural gas-fed microbial food to meet its calorie needs, all of the facilities to make the food have to be constructed, and equipment installed and tested, before existing food reserves run out. If a disaster eliminates all current food sources without warning at roughly the worst time with respect to the US crop calendar, then with good management and rationing, food supplies should last about 22 months (as covered above, and discussed in more detail in appendix 2.0 US stored food). For now I'll take this figure as the time budget that the US has to get all needed food production facilities up and running. I also assume for now that the US tries to meet all of its calorie needs through gas fermentation, but other food sources would likely be used, depending on the specific scenario. Based on an approximate construction schedule for a large natural gas fermentation facility, it seems possible though not guaranteed that food production could be ramped up before pre-disaster food supplies run out even in a worst-case scenario.

## Construction time estimates from reference class buildings

As a quick first pass estimate for the construction time needed for a facility, several formulas relating facility cost to construction time can be used. Above, I discuss that building a facility making roughly 100,000 tonnes per year (ktpa) of microbial food should cost ~$753 million. While the actual desired facility size may differ from this, for now I'll assume that all of the facilities are built to this size. Given the ~$753 million price tag, one formula, which looks at data on general industrial building construction, suggests that construction would take ~16 months from start to finish. Another formula, which has been used by ALLFED[9] and refers to general chemical plants, gives a construction time of ~24 months. In a disaster, construction may be able to be sped up relative to normal operations, such as by working rotating shifts, waiving various permits, and other steps. However, in the scenario considered, the US would be building hundreds of new plants simultaneously, since, if all facilities make ~100 ktpa of microbial food, meeting all US calorie needs requires a little over 700 plants. In this case, additional bottlenecks may emerge that could slow down construction. Still, above and in appendix 13.0, I cover how the current annual output of relevant US industries in manufacturing, material production, equipment installation, and construction are likely able to manage the task.

## Custom construction time estimates

Importantly, the facilities referenced in the formulas above are less complicated than a gas fermentation plant. Therefore, while at Open Philanthropy, I worked with Synonym Bio, a financing and development platform for bio manufacturers, to create an approximate construction schedule for a 100 ktpa SCP plant. This schedule incorporated lead time quotes — the time from equipment order to delivery — from US manufacturers of the major pieces of equipment needed in a gas fermentation plant. These lead times assume current backlogs and delays for order fulfillment, since suppliers have to fulfill existing orders from other customers. In a disaster, the lead times could therefore probably be sped up, at least for some plants. But there would likely also be issues in having to make an unprecedented number of specific pieces of equipment all at once, so it is difficult to say what the net effect would be on delivery times (see appendix 14.0 Plant construction times for some discussion, but note that this is an area of ongoing research and current uncertainty for me). Based on typical experiences from site managers, the project schedule also includes conservative estimates for permitting requirements, site purchase contract negotiations, and other steps that may be able to be eliminated or significantly expedited in a disaster.

Synonym found that for a first-of-its-kind plant, a 100 ktpa gas fermentation facility would take almost 50 months to complete, and that this could be cut down to ~42 months if a similar plant had been made before so that engineers and managers could learn from past projects. There are several ways to further expedite the steps involved in bringing a plant to full operation that can very likely be counted on in an emergency, such as eliminating permitting or reducing the time taken for negotiations. However, these do not save significant amounts of time. The major factors driving total startup time are engineering, sourcing major pieces of equipment, construction, testing different equipment and systems, and serially ramping up parts of the plant. In appendix 14.0, I discuss how changes to the time typically required for these steps might impact the total time needed to begin producing food through gas fermentation. I estimate these changes could drive down total facility start-up time down to 28.8 months, and down to 22.4 months if engineers and managers can learn from the history of previous plant construction projects. As noted above, even the most optimistic timeline goes right up to the amount of time that US stored food might last if there are no other food sources and a disaster hits at the worst possible time. More pessimistic timelines probably exceed the US's time budget if a disaster hits in September, even if the US can quickly increase fishing and sugar production. The difference between the time requirements for a novel and a repeat facility shows that pre-disaster experience with resilient food production may substantially increase survival odds.

[9] The Alliance to Feed Everyone in Disasters, a nonprofit research organization dedicated to finding resilient food solutions. Members of ALLFED's team provided great feedback and input to this report (see acknowledgments).

These accelerated start-up time estimates hinge on several aggressive assumptions about how much different processes can be expedited in a disaster. I expect that some are unrealistic, at least if trying to generalize to all plants being built simultaneously. Some additional time savings could be achieved with advanced preparation on the engineering front, such as by having plans and site selections in place before a disaster hits. But I am not assuming this type of preparation in my mainline scenario. That even aggressive assumptions for time savings still put the total time at almost two years suggests that at best, there would be very little margin for error in a worst-case scenario. In appendix 13.6 Construction, I briefly discuss how evidence of accelerated factory construction rates in the US during WWII might inspire some optimism, but overall, in a worst-case scenario, constructing all of the needed facilities before reserves run out could be a tenuous prospect.

## Construction time estimates and food supplies

The time budget for constructing new food production facilities varies significantly depending on when a disaster hits. The chart below[10] compares food reserve levels, presented above (and with more detail in appendix 2.0), compared to four different estimates for the time needed to make methane-fed microbial food production plants. The first two use the aggressive assumptions about speed-up ability discussed above and in more detail in appendix 14.0, for both a first-of-its-kind plant and a plant that is built using lessons from past projects. I also show the times required for plants that follow an unadjusted construction schedule for a novel and a repeat plant. These unadjusted schedules assume typical permitting requirements, negotiation times, construction schedules, etc.

---

[10] The data for this chart can be found on tab 3 of this Google Sheet.

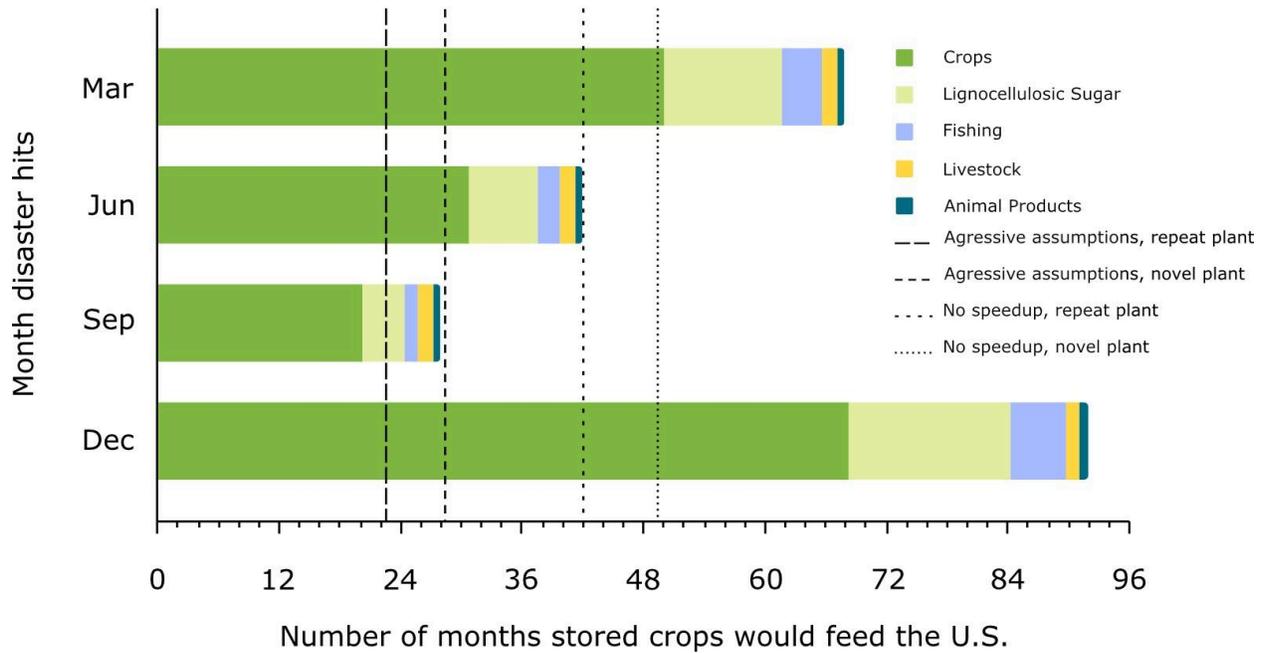

The chart above shows that even with no speed-up in construction times, there are parts of the year when existing US food supplies could last longer than the time needed to build new natural gas-fed microbial food production plants. This analysis ignores the increased food stock levels that might be available in scenarios where there is advanced warning of a disaster (appendix 2.6). Appendix 2.4 Food spoilage includes a discussion of why I expect most of these calories to store for extended periods of time, and in appendix 5.0 I analyze the nutritional value of major crops that would make up most of the diet in a disaster before nonagricultural foods like methane-fed SCP could be scaled up.

# Food from woody biomass extends the time available for scale-up of methane-fed microbial food, but cannot replace it in a long-term loss of agriculture scenario

As briefly mentioned above, there are two main ways that woody biomass can be useful as an additional resource for producing food in the scenarios considered here. The first is to turn it into sugar and consume it directly. The second is to use it as a substrate to grow other food, either by growing fungi on the wood, or by first turning the wood into sugar and then using it as a feedstock to produce fungal or bacterial food. Mushroom production on wood would provide a nutritious food source, but is difficult to scale given its inefficiency. Wood-derived sugar is relatively straightforward to produce, and cheap on a per-calorie basis compared to other food options, so supplementing a diet with sugar would reduce the overall difficulty of meeting US calorie needs. However, there's a limit to how large a part of the diet sugar can make up while still allowing people to meet their minimum nutrient requirements. Fermenting sugar (i.e., using it as a substrate for microbial growth) can provide many nutrients, and is a more well-known process than methane fermentation for food production. However, when added to the cost of producing the substrate sugar, this fermentation process is more expensive than producing food from natural gas, so would probably not be chosen as the main calorie source. Still, the diverse nutritional profile of products that can be made from sugar fermentation means that in scenarios where woody mass is available, these foods would probably make up some of the post-disaster diet. Overall, access to wood biomass would probably increase the odds of producing enough food in a disaster, but is not necessary and is less scalable than natural gas-fed microbial food.

## Growing mushrooms on wood

Modern mushroom production is not scalable to meet all US calories, given how much fertilizer and energy is currently used. Growing enough mushrooms to meet all US calorie needs would require increasing the production of Agaricus mushrooms (which represent about 97% of mushrooms grown in the US) by almost 4,500x. Doing this would require nearly double the US's annual electricity generation and several times more nitrogen fertilizer than it currently produces each year. As covered in appendix 8.0 Large-scale mushroom production, it would also require a substantial amount of biomass and space, either indoors or in caves. Some mushrooms, like oyster and shiitake mushrooms, can be simply grown on logs. While the process can be labor intensive and yields are less reliable than modern mushroom cultivation on compost, this can be done outdoors, reducing energy demands. This may be a helpful option in scenarios where society collapses and survivors have much less access to modern technology, as discussed in appendix 17.0. However, the volume of wood needed makes this a difficult food production option to scale. Growing enough oyster mushrooms to meet all US calories, for example, would require that the US increase its annual wood harvest by almost 10–20x. I estimate that scaling up mushroom production may also take a long time, and pests and pathogens could introduce significant yield uncertainties that make this a poor choice for an industrial society, especially compared to more efficient options for producing food from wood biomass. See the appendix for a discussion of wood-grown mushroom yields (appendix 18.5) and mushroom nutritional value (appendix 8.2).

## Making wood into sugar

Wood can be turned into sugar fairly efficiently. Even meeting all US calories from wood sugar for a year only requires about 70% of the wood that the country currently harvests each year.[11] As noted above, and discussed in more detail in appendix 6.0 Lignocellulosic sugar production, there is significant potential for using existing production capacity to turn wood into sugar. Without building new facilities, I estimate that the US could make about 32 Mt of wood sugar per year, totaling around 52% of US calories. In a 2,100 kcal per day diet otherwise consisting of methane-fed SCP, sugar could make up about a third of total calories without lowering fat below the minimum acceptable levels (see appendix 6.1 for discussion). As noted in appendix 12.0, it may be possible to cultivate strains with other nutrient profiles, potentially raising the amount that sugar could contribute to the diet. Sugar is relatively economical when using retrofitted facilities: repurposing facilities to make sugar costs about $710/tonne of annual sugar production capacity, while building new facilities is about 2.4x more expensive. This is still cheaper than the estimated ~$5,271 capital expenditure/tonne of annual food production capacity needed for methane-fed microbial food (see appendix 13.0), so adding in sugar to a post-disaster diet generally lowers the bar for meeting the US's calorie needs.

---

[11] In scenarios where all trees are dead and the US is harvesting dead biomass, this rate of wood consumption could last for about 77 years, ignoring wood rot (16.1). Much of the biomass would have decayed by this point, but this is also a higher rate of consumption than I anticipate since the US wouldn't be meeting all of its calorie needs from sugar.

## Fermenting lignocellulosic sugar

There are many edible organisms that can grow on simple carbohydrates. Quorn, a SCP product made from the fungi *Fusarium venenatum*, has been sold to consumers for decades and grows on sugar. Quorn is generally cheaper to produce than methane-fed microbial food, although it requires greater fermentation volume per unit product because it has a lower growth.[12] This means that the material demands for setting up large-scale production are higher. Producing enough sugar from wood to grow fungal foods would also require increasing annual wood harvests by about 34% to 116%, potentially posing an additional challenge to this production option. The total cost of making enough sugar production and fermentation capacity is also probably higher on a per calorie basis than methane-fed SCP, but possibly not by much. There are other possible options for turning wood into a useful substrate for fungal or bacterial biomass production, like turning wood into methanol, but these also do not compare favorably to using natural gas on an economic basis (see discussion in appendix 15.0 Sugar-fed fungal and bacterial food). The role for sugar fermentation–derived products in a post-disaster diet would probably be to supplement a predominantly methane-fed microbial foods diet, particularly for some vitamins and essential fatty acids (see appendix 15.1 Fungal SCP nutrition).

# Overall contribution of woody biomass to food production

I expect disasters where wood biomass is available to present a higher likelihood that large populations can produce food without agriculture. The main mechanisms by which access to wood increases these odds are a) providing more options, so making enough food doesn't hinge on one technology working, b) reducing the amount of calories that have to be produced from non-plant-derived feedstocks, and c) granting access to more diverse edible organisms, probably improving the dietary quality that could be achieved post-disaster.

# Summary: Current technology likely makes it possible to feed large populations without agriculture, but it is difficult to scale up alternative food production quickly

My work so far shows that it may be technically possible to make enough food to feed everyone in a country without any access to crops, even if society had to scale up alternative food production quickly in a disaster context. In the US, domestic energy and material supplies, and manufacturing and construction capacity, are likely sufficient to make enough alternative food to fully replace agriculture in a scenario where all plants die without warning. However, in my review of food production options, I only found a few that could likely scale quickly to produce enough food for a large population. Food supplies would hinge on just one or a few technologies, making for a fairly precarious situation. Also, time required to scale up nonagricultural foods poses a major challenge to meeting calorie needs during an agricultural disaster. A large-scale deployment of industrially produced foods probably takes at least 1.5–2 years if all goes well, and may take considerably longer. If a disaster hits without warning at about the worst time with respect to the US crop calendar, then even well-managed and rationed food supplies may only feed the whole US population for 1.5–2 years. Thus there is probably little room for error in trying to keep large populations alive and preserving modern industrial capacity in a worst-case scenario. However, a better-timed disaster would give society longer to respond, making the scale-up of other food production options significantly more feasible.

---

[12] However, there may be other options for foods produced from sugars that grow faster and so require less fermenter volume.

In appendix  I explore prospects for resilience in dire scenarios in which only small, isolated groups survive a loss of agriculture disaster. I find that, with a loss of critical sectors like transportation, mining, manufacturing, and utilities, survivors would be left with far fewer and worse options for producing food. As discussed below, more research may help improve these options. However, my findings underscore that adopting scalable foods that can feed everyone is important, both to minimize the loss of life and to maintain the large-scale industries that make technologically advanced foods possible.

# There are several ways to increase the likelihood of successfully scaling up food production in loss of agriculture scenarios

Put together, my assessment of food production options suggests that societies have several ways to respond to a loss of agriculture disaster, but are still vulnerable in worst-case scenarios. I expect that the odds of scaling up food production could be improved through several types of interventions. Broadly, these fall into the categories of a) continued research, b) advancing resilient food production for non-disaster applications, c) disaster response plans, and d) increasing food reserves. There may also be actions that could reduce the likelihood of a high-consequence attack on agriculture, but that is outside of the scope of this report.

For research, I believe that the main areas that deserve the most focus are:

1. Investigating the optimal response to different disasters
2. Improving the nutritional value of different foods
3. Determining the minimum equipment needed for small-scale resilient food production

A more detailed understanding of optimal responses — meaning responses that have a portfolio of food production options that minimize scale-up time, the need for manufacturing novel equipment, and nutritional challenges — could help with both directly improving disaster response and guiding which technologies should be researched now. Different scenarios, like ones with and without advanced warning or with or without access to dead plant biomass, could suggest fairly different choices in which food production options should be used. The approach that I took in looking for scalable food options could miss cases where it makes sense to increase the production of some input to one or more food production options above current levels (see appendix 10.0 for a discussion). Research into an optimal response would try to fix this shortcoming.

The most scalable option for food production at a society-wide level that I identified, methane-fed microbes, has a few nutritional shortcomings if relied on as a sole food source (appendix 11.0 Methane-fed microbial food nutrition). As I discuss in appendix 12.0, there are several options for addressing these shortcomings, both through strain optimization and by supplementing with other foods.[13] There are also some uncertainties about the effects of long-term consumption of foods discussed here. Resolving these uncertainties and developing more nutritious foods that could be made at scale in a loss of agriculture scenario could meaningfully cut down on the risk of ubiquitous and catastrophic malnutrition.

There has been very little research into small-scale production of food without access to agricultural substrates. The large food production facilities discussed here are massive industrial efforts that rely on extensive supply chains and national infrastructure. However, distributed and technologically simpler food production options would make societies more robust to disasters that affect both outdoor food production and social cohesion or national infrastructure. Research into these options could provide valuable insights into overall societal resilience to agricultural disasters.

I have not explored options for promoting resilient food adoption and developing disaster response plans as much as I have looked into the research gaps highlighted above. However, the more widely used a resilient food production technology is before a disaster, the more likely it can be quickly scaled up (see discussion on facility construction times). This could mean that promoting relevant technologies (such as microbial fermentation of natural gas) could be extremely valuable. This promotion could come in the form of various policies, such as subsidies, tax credits, or public R&D, or finding market opportunities for the deployment of these technologies. Separately, disaster response plans could be extremely valuable for improving the efficiency and timeliness of a response, highlighting weaknesses in a plan ahead of time, and providing assurance to people in a chaotic situation. I therefore would be interested in seeing some work in this direction, but haven't yet looked into all that this would entail.

I find that in the worst-case scenario, replacing agriculture with industrial foods before existing food supplies run out could be possible, but is by no means guaranteed. Since this is true in the US, where food supplies are much larger than they are in most other major countries (appendix 3.0 Note on food stores in other large countries), it is likely that other countries would also struggle to scale up alternative food production before stored supplies run out. (See appendix 20.0 for a preliminary discussion of other countries' odds of succeeding in this.) Larger food stores would give society more run time to scale up nonagricultural foods, so exploring policies to encourage greater buffer stocks could be one of the most important ways of increasing robustness to agricultural disasters. There is a long history of food stocks being higher than they currently are in the US and elsewhere. Increased international trade and the perfection of just-in-time delivery practices have reduced the importance of buffer stocks in non-emergency times, but returning to some of the policies that once encouraged greater food stores would be a useful hedge against agricultural disasters.

---

[13] The same problem holds for foods that could be made in low-tech settings if society collapses (appendix 18.0 Low-technology food production options without industry or crops), although I think that there is probably less room for research to improve options here.

# Acknowledgments

In writing this report, I received extremely valuable input from many people, especially the teams at the Alliance to Feed Everyone In Disasters (ALLFED) and Synonym Bio. Members of ALLFED's team, especially Professor David Denkenberger, Juan García Martínez, and Michael Hinge, reviewed earlier iterations of my work and provided helpful comments, feedback, and suggestions, and also provided independent insights into some of the research questions addressed in this report. Several members of Synonym Bio's engineering team, especially Crystal Bleecher, P.E., Phillip Crawford, P.E., and Nick Salvadore, developed a techno-economic assessment (TEA) of both natural gas and $CO_2$-fed single-cell protein (SCP) production plants. The team also developed a detailed schedule for engineering and building an SCP facility. I also had useful discussions with and received important feedback from Dr. David Humbird, P.E., Dr. Kathleen Alexander and Dr. Ian McKay (Savor-it), Dr. Shannon Nangle (Circe Biotech), Dr. Adam Leman (Good Food Institute), Dr. Ezhil Subbian (String Bio), Dr. Adam Marblestone (Convergent Research), and Conrad Kunadu (Oxford). I also want to thank my colleagues at Open Philanthropy who reviewed previous drafts and discussed and provided great feedback on my research priorities while I prepared this report. Any mistakes in the report are my own and are not the fault of my reviewers.

# Appendix

## 1.0 Current US food production

To provide context on current US agricultural production and food consumption, I walk through sources of US calories, current production of crops and livestock, food loss and waste, imports and exports, and non-food uses of crops. The data on crop production, trade, and use in this section comes from the UN's Food and Agriculture Organization (FAO), unless stated otherwise. Similar data is available from the USDA, but it does not cover as many products. The FAO also provides data on total calories delivered to consumers (i.e., calories that are not lost due to waste, use as livestock feed, etc), but for total calorie production I multiplied production quantities from the FAO by the calorie content of different foods taken from the USDA's nutrition database. For food waste figures, I used EPA and UN Environmental Programme (UNEP) data. Due to inconsistent data availability from 2022 and 2023, most of the data presented here is from 2021.

### 1.1 Current US food consumption

About 71% of US calorie consumption comes from crop products, 28% from terrestrial livestock and terrestrial livestock products, and ~1% from marine animals and their products. The table below breaks down calorie supply[14] into additional categories.

| Food type | % of US calories |
|---|---:|
| **Crops** | **70.8** |
| Grains | 20.9 |
| Crop oils | 19.3 |
| Refined sugars | 15.4 |
| Fruits and vegetables | 6.6 |
| Alcohol | 3.7 |
| Tubers | 2.4 |
| Nuts and seeds | 1.8 |
| Legumes | 0.7 |
| **Livestock** | **27.2** |
| Terrestrial livestock meat | 15.5 |
| Eggs and milk | 11.7 |

---

[14] I use the term calorie supply as opposed to consumption here because the data that I work with, from FAO, reports calories "delivered" to consumers — e.g., purchased at a store or restaurant — rather than calories consumed. This is because the data does not take into account personal waste. This means that the distribution of calories consumed might be slightly different than the distribution of calories supplied.

| | |
|---|---:|
| **Seafood** | **0.8** |
| Fish and crustaceans | 0.8 |

The data in the table above include calories supplied for direct human consumption, but not human-edible calories used to feed livestock.

US calories are concentrated among just a handful of major crop and livestock sources. The top three crop products that contribute to direct US calorie consumption are wheat and wheat products, soybean oil, and sugar (~[55%](#) from sugar beets and ~[45%](#) from sugar cane). These products alone provide ~32.7% of US calories. The top three animal products contributing to US calories, providing a combined ~19.9% of US calories, are milk, chicken, and pig. Corn makes up ~84.2% of the mass of human-edible products fed to livestock, and soy makes up ~3.7%, with a dozen or so other products contributing fairly small shares.

## 1.2 US food product destination

The previous section describes food that is produced and used directly for domestic human consumption. However, most food does not fall into this category. Of the food produced in the US, some is exported; fed to livestock; used as seed for future planting; lost in the production, processing, or harvest; used for non-food applications like making biofuels; or wasted by retailers or consumers. The table below shows the percentage of calories produced in the US from both crops and livestock that end up at each of these destinations. See tab 1 of this [Google Sheet](#) for relevant calculations and data.

| Destination | Percent of production |
|---|---:|
| Exported | 27.0 |
| Livestock feed | 21.3 |
| Non-food use (e.g., biofuels) | 17.5 |
| Consumed | 16.8 |
| Waste* | 10.6 |
| Losses | 3.3 |
| Seed | 1.0 |

*For waste, I could only find data on bulk mass amount, not on the percent of specific foods. This figure may be off, since it shows that 10.6% of the mass of food produced in the US is wasted, not necessarily that 10.6% of the calories produced in the US are wasted.

The table above shows that the US exports ~27% of the calories that it produces. It imports about a third as many calories, so that its net exports equal about 20.3% of its calorie production. Also note that the values do not sum to 100 (they add to 97.5) because of slight methodological differences between sources.

# 2.0 Stored food

## 2.1 US crop stocks

Here I calculate stored crop levels, and composition, in the US, with relevant data and calculations on tab 3 of this Google Sheet. The FAO provides country-level data on food storage levels. As in the previous appendix (1.0), similar data is available from USDA, but it is less comprehensive, so for most crops I use FAO data. However, the FAO only reports food stocks at their end-of-year levels, defined for crops as the time right before new crops are harvested. This means that for any given year, FAO stock levels represent the minimum stores for the products covered, since crop stocks are replenished after harvest. Because different crops are harvested at different times, adding together these minimum estimates actually provides an unrealistically low estimate for even the minimum level of stored crops at any given time during the year. Data on US crop stock variation throughout the year is available from USDA, but it only covers six major crops: corn, soy, wheat, barley, sorghum, and oats. From the data covered above, these crops make up a little over 28% of US calories for direct consumption, ~81% of the total crop calories produced in the US,[15] and also take up ~87% of US planted cropland. Therefore, the intra-annual variation in stock levels for these crops captures a substantial share of the total variation for all crops, but not all of it. For now, I conservatively hold the stock levels for all other crops constant at their end-of-year levels throughout the year, and look only at intra-year variation for these major crops.

As discussed in the section of the main text on US food supplies, total US crop stock levels are at their lowest around September, right before corn and soy harvests. Crop stock levels are highest right after these harvests finish, around December. To estimate the total number of crop calories in store at a given time, I take FAO and USDA data on crop stock levels and multiply these by data from the USDA on the calorie content of different foods. Then, to estimate how long these crop stocks could feed the US if no new food was being produced, I divide this number by the number of calories needed by all ~332 million Americans for a day. Not accounting for waste or spoilage, all stored crops added together should feed the US for a little over 23 months in September, and over 78 months in December (with data from 2021). The table below shows a rough breakdown of how many calories come from different types of crops in September.

| Crop type | % of stored calories |
|---|---|
| Grains | 65.1 |
| Oilseeds | 9.1 |
| Refined oils | 8.0 |
| Fruits and vegetables | 6.9 |
| Refined sugars | 6.9 |
| Tubers | 2.0 |

---

[15] The difference in calories consumed versus produced is due to the differential use of some crops, most notably corn, for purposes other than direct domestic human consumption, like livestock feed or biofuel production. For corn, it only makes up ~3.6% of US calories for direct consumption, but equals ~49.0% of calories produced from crops and livestock in the US.

| Pulses | 1.1 |
| Nuts and seeds | 0.9 |

As noted, the calculations above on how long food would last assume that no stored calories spoil or are wasted. The sections on waste and spoilage address this assumption, and in the main text I presented my overall estimate for how long crop stocks would actually last.

## 2.2 Animal products and livestock numbers

About 28% of US calories currently come from animals or animal products. In a disaster, calories could come from both stored animal products and living livestock that could be slaughtered. If a disaster kills all plants, then livestock should generally be slaughtered as quickly as possible, since all human-edible crops are more efficiently used as direct human food, and without food the livestock will die of starvation.[16] Some livestock can be fed non-human-edible plant matter, such as crop residues or grass, so they wouldn't have to go without food immediately. I address this briefly in the next section on food waste and spoilage, but for now focus on how many calories are theoretically recoverable from all livestock currently in the US.

For stored animal products, I again used FAO data on quantities stored and USDA data on the nutritional content of different products. This shows that the US has about 31 days worth of animal products stored, if these were all that Americans ate for that time.

There are almost 1.8 billion livestock animals in the US at any given time, and most (1.5 billion) are chickens. Based on the amount of meat that can be harvested from a typical animal and the USDA's data on the caloric content of different meats, I calculate that there is enough meat on livestock to meet the calorie needs of everyone in the US for about 165 days, if meat is all that they ate for that time.

See relevant data and calculations on tabs 4 and 5 of this Google Sheet.

## 2.3 Food waste

While the US currently wastes about a third of all food delivered to consumers, there is reason to believe that this amount could be significantly reduced in an emergency where food supplies are more scarce. Even though only about a third of the calories eaten by US consumers come from restaurants, restaurants account for roughly two thirds of this waste, so just a shift to at-home consumption would reduce waste from roughly a third to about a sixth of food delivered to consumers. There is also precedent for consumer waste falling significantly in response to disasters. For example, WWII food waste levels in the UK were only about 3%. However, this does not account for waste that occurs before food reaches consumers. Other work that has tried to estimate expected food waste in disasters assumes that total waste can reasonably fall to about 13%. For now, I use this figure in estimating both how long existing food would last and how much nonagricultural food has to be produced to meet minimum calorie requirements.

[16] The ratio of calories eaten by livestock to harvestable calories from animal products (meat, egg, or milk) is called the feed conversion ratio. For example, a feed conversion ratio or 10% means that for every 10 calories an animal consumes throughout its life, it produces 1 calorie of human-edible products. For chicken, pig, and cow meat, feed conversion ratios are on average 12%, 8%, and 3%, respectively. For milk and eggs, the ratios are about 40% and 22%. Not all calories eaten by livestock come from human-edible sources, since some animals can digest a larger array of plant material than humans can. But these conversion efficiencies mean that all stored human-edible foods are better used by being consumed by people directly rather than as livestock feed.

## 2.4 Food spoilage

Since stored food would have to last over a year in the scenarios considered, spoilage could be a significant issue. However, I find that if food is well managed, spoilage shouldn't reduce food supplies too much. Here I describe the assumptions I used when accounting for spoilage in my calculations of food availability.

If frozen, food is generally safe indefinitely, but may decline in quality. To avoid spoilage, it will likely be important to avoid overfilling US freezer capacity. Distributing food to consumers will also have to be well organized, since households likely do not have much freezer capacity compared to industrial facilities. Stored crops and crop products can generally last over a year without freezing. Animal products require more effort to preserve, but below I estimate that US freezer capacity should be sufficient to hold most or even all stored animal and crop products. I provide shelf life estimates for major components of US stored calories, but I expect most of these to be fairly irrelevant if reserves are well managed and largely frozen.

### 2.4.1 Stored crops and crop products

I believe that, if well managed, most of the calories in stored crops would last over a year, and possibly several. This is because standard storage conditions can preserve crop products for many months to a year, and even without freezing, temperature controlled silos can hold crops for much longer. Because of this, I expect that few crops and crop products will spoil in the first few months of a disaster, and subsequently the US would be able to move most products to long term storage facilities.

The largest source of calories from stored crops comes from stored grains. These have a very long shelf life. In standard storage conditions, corn has a shelf life of 9–10 months, but in very good conditions it can last over a decade. Wheat can similarly last over a decade under ideal conditions, but is usually stored for 6–12 months. The next largest source of stored crop calories are oilseed, the largest of which is soy. Soybeans typically store for about 6 months without reduced quality, and when frozen can last over a decade. Oils are the third largest source of stored crop calories, and again these last months to over a year in normal storage conditions. Refined sugars are resistant to microbial growth and have an indefinite shelf life in room temperature conditions. Fresh potatoes store for about a month without reduced quality, and can last for one to two years if dried. Nuts last for about a year at room temperature. The foods most vulnerable to spoilage are fruits and vegetables. Generally, I expect that in cases where food supplies are well managed, foods that spoil first would be eaten sooner, so that the calories are not wasted. However, since fruits and vegetables make up about 7.6% of US stored crop calories, their calories could feed the US for about 42 days if they were all that was eaten. Without additional measures taken (e.g., freezing), most fruits and vegetables would spoil before 42 days have passed. The overall picture is that the US should be able to avoid losing most of the calories available in stored crops and crop products to spoilage.

### 2.4.2 Stored animal products

The picture is likely less optimistic for livestock and livestock products. I expect that most of the calories stored in meat and animal products that are currently available in the US would last long enough to be eaten before spoilage, again because the freezer capacity in the US is large enough to hold these products. However, under conservative assumptions about slaughter and processing speed, perhaps just about a third of the calories potentially available in living US livestock could be harvested in a disaster where all plants die and stored crops are used only to feed humans directly and not to keep livestock alive.

For stored animal products, 60% of the stored calories come from tallow. This has a shelf life of at least [nine months](#) even stored at room temperature, and other sources claim a longer shelf life. The next largest stored animal product, by calories, is cow's milk cheese, making up about 11% of the calories from stored animal products. The shelf life of cheese varies by type, but USDA [guidelines](#) recommend six months in cool storage conditions. Whole milk has a shelf life of only about [one week](#), but since it makes up about 3.6% of the calories in stored animal products, its spoilage would only marginally reduce the number of calories that the US could get from stored animal products. Similarly, eggs are also quick to spoil,[17] but make up less than 1% of the calories in stored animal products. Overall, if well managed, I estimate that at least 90% of the calories from stored animal products could be eaten before they spoil, and I reduce this by another 13% to account for general waste.

Even if all livestock could be slaughtered, there should be ample room to store all of the meat that could be harvested. Cow, chicken, and pig meat makes up almost 97% of the potential calories stored on US livestock, and represent about 95% of the mass of meat, so I'll focus on these species. Given the density of [cow](#), [pig](#), and [chicken](#) meat, storing all of the meat that could potentially be harvested from these animals would occupy about 47.5 million m$^3$. This is about half of the estimated [~106 million m$^2$](#) of freezer space that the US has. A concerted effort to prioritize storing meat could likely get most of the meat into suitable storage where it can last for a long time. Meats can be [preserved indefinitely](#) if frozen, although their quality is reduced. Before experiencing significant losses in quality, frozen chicken can [last a year,](#) frozen pork lasts [4–6 months,](#) and beef can last up to a [year](#) (less time if it is processed). Even without freezing, meat could also be salted and dried for preservation.

For livestock, slaughter and processing capacity is likely a bigger bottleneck than the ability to store meat and prevent spoilage. In the wake of a disaster affecting crops, all human-edible food should be eaten directly to avoid the [losses](#) associated with feeding livestock. Therefore, there would be a race against the clock to slaughter and process livestock before they starve, which would be only a matter of days. We can look to current practices to provide a conservative estimate of how long that would take, focusing on cattle, pigs, and chickens. There are ~1.5 billion chickens alive in the US at a given time, which is about 16% of the [9.25 billion](#) slaughtered each year. This suggests that, at current slaughter rates, all currently living chickens could be killed in about 60 days. For pigs, the number living at a given time is ~59% of the [number slaughtered](#) each year, meaning that at current rates it would take about 214 days to kill all of the pigs. There are almost 2.8x the number of cattle alive than are killed each year, so without a substantial increase in the rate of slaughter, few could be killed quickly. Cattle represent about ⅔ of the total calories currently "stored" in live animals in the US. This means that if they could not be quickly slaughtered and processed, the total amount of meat that living livestock could contribute to US food stores in a disaster would be much lower.

At the slaughter rates suggested by the above figures, if we assume that all animals took 10 days to starve to death and their corpses could not be salvaged, only about 2.8% of the calories available in livestock could be harvested, adding just ~3 days to US food reserves. However, this is a conservative assessment. Given the excess of freezer space, animals could be killed and stored with no or limited processing, accelerating timelines. In addition, some livestock feed (which wouldn't be counted among the stored crops described above) is already on site at farms, which could delay starvation and enable more slaughtering and processing. Naively, I'll assume that even in a dire scenario ~25% of cattle, ~50% of pigs, and ~90% of chickens can be killed and processed before they starve and their carcasses spoil.

---

[17] Eggs spoil [especially fast](#) in the US because US sanitation practices remove a natural protective cuticle from eggs that other countries do not remove.

Note that these are really just placeholder values. I also use them to represent other processes or simplifications that cause waste or reduce my original estimates of the amount of calories available in livestock.[18] Going with these figures still means that only 30% of the calories available from livestock could actually be slaughtered, which would provide ~50 days of food.

## 2.5 Note on additional assumptions

For most of this report, I have assumed that the loss of agriculture disaster occurs roughly instantaneously. Under this idealized assumption, society goes from its normal use of agricultural products, briefly summarized for the US in appendix 1.0, to trying to conserve and ration all edible calories overnight. This type of rapid transformation — where the US goes from using just ~⅙ of the calories that it produces for direct domestic consumption to trying to bring this figure to as close to 100% as possible by minimizing waste, ending exports, halting the use of human-edible crops for livestock feed and biofuel production, etc. — would take some time, which I have not accounted for in this report. Conversely, I have conservatively assumed that calorie intake does not fall below an average of 2,100 kcal/day,[19] and I have ignored food stored in homes and other non-retail locations.[20] As a first pass for this report, I have assumed that these countervailing factors roughly cancel out in terms of how long food stores can last in the US.

## 2.6 Note on advanced warning potential

If a disaster affecting crop production did not hit instantaneously, then society could begin rationing and conserving existing food supplies while potentially also harvesting crops before they are affected by the threat. For illustrative purposes, consider a scenario in which a threat was detected in the US in September, but it did not kill all crops until after US soy and corn harvests had finished (by the end of October for soy and the end of November for corn). Because the US would benefit from these harvests and would have been rationing the reserves it held in September for three months, as opposed to drawing them down at the normal rate in a scenario where there was no advanced warning, the US would have about 100 months worth of crop calories stored in December, compared to the ~70 that would otherwise be stored in at that time.

In the above illustrative example, the US could also begin scaling up nonagricultural food production in September, although probably with some delay. This means that it would be further along by December than it would be in a scenario where a disaster hit instantaneously in December. A warning would also give the US more time to slaughter and process livestock, further adding to food reserves. Of course, the fact that a threat is detectable could mean that it is already affecting harvest levels. The specifics of this effect could vary substantially by the threat under consideration, and a case-by-case assessment of different threats is outside of the scope of this report.

---

[18] For example, in calculating calories, I assumed that all livestock currently alive are fully grown, but of course that's not accurate.

[19] In my calculations of average adult calorie requirements (appendix 4.0), I assume that US adults will fall to an average BMI of about 18.5, since this is the low end of a healthy weight. Averaged across men and women, this would bring mean adult body mass in the US down to ~53.5 kg. Since the average (again, across men and women) adult weight in the US is ~84 kg, this represents a mean weight loss of over 30 kg. Losing a kg of fat requires burning ~7700 kcals, or the "equivalent" of ~3.7 days at a 2,100 kcal/day level of consumption. If all of the 30 kg of weight that can be lost by an average US adult before they're at the low end of a healthy weight was fat, then they could hypothetically go without food for perhaps >100 days. Under medical supervision, there's at least one example of an individual fasting for >1 year. I do not expect deliberate prolonged fasts to be the optimal strategy, but I do expect that many members of the US population could sustain a daily calorie intake well below 2,100 kcals for a while, thereby stretching out the length of existing food reserves, perhaps substantially.

[20] The data that I've used for food stores only includes products that are held by producers and retailers. As far as the data is concerned, once food is sold, it is consumed or wasted by consumers, so food stored by consumers is ignored in my calculations.

It is important to note that the above example picks a nearly optimal time for the US to receive an advanced warning. Since corn and soy alone represent ~78%, and are generally harvested between September and December, advanced warning of a threat is most impactful if it allows the US to begin rationing and conserving its food supply and to harvest these crops. Warnings that do not allow this, such as a three-month warning in March, do not impact the total food supplies nearly as much, relative to what I report in my earlier discussion of US food reserves.

There are various ways that an analysis of scenarios where a threat does not strike instantaneously could significantly extend this brief discussion, but I leave that to further research. I'll note that my figure for the increased food storage made possible by an advance warning did not include a calculation of how advanced warning could increase the amount of calories from livestock that the US could harvest in a disaster,[21] but this would be a smaller contribution to overall food supplies than the change in the rate at which the US drew down its stored food levels.

## 3.0 Note on food stores in other large countries

The US has large, although not uniquely large, food stores compared to other countries (see tab 6 in this Google Sheet for relevant data and calculations). Looking at the nine other largest countries by population, China and Indonesia have larger food stocks than the US relative to their population, while the rest have smaller reserves. The table below compares how long stored food — both crop and livestock products — would feed the populations of each of the world's ten largest countries. I have not calculated the amount of food that could be harvested from existing livestock for countries other than the US, but going with the US example, this is unlikely a major source of calories compared to stored food, and the amount of calories that could be harvested quickly in a disaster varies by the specific scenario (see above for a discussion of the US case). As above, I used FAO data on food stores, which measure stocks of crops just before a new harvest. This means that all crop stocks are at their lowest levels for a given year (2021 in this case). Since crops have different harvest times, this presents an unrealistically low total estimate for the amount of stored food. Unfortunately, I have not yet found data on how much food is stored across the year for countries other than the US, so in the chart below I look only at how much food is stored based on FAO data. In the case of the US this brings the total amount of stored food down from an annual minimum of ~21 months to ~18 months, so the estimates from other countries can be expected to underestimate their total reserves even at their annual minimum, but I do not know by how much. For calculating the length of time that stored food would feed a country, I assumed that no food spoiled, that waste was reduced to ~13%, and that everyone ate at an average of 2,100 kcal/day.

---

[21] In a scenario where the US has some notice before plants die, it could collect a large amount of non-human-edible plant products (like crop residues or grass) as it tries to redirect all human-edible crop products away from livestock feed and toward direct human consumption. This may allow the US to keep livestock alive for far longer than I assumed in my discussion of an instantaneous disaster, which would mean that more livestock could be slaughtered and processed before being wasted. This would add to total calorie stores, but perhaps more importantly could help fill in some nutrient gaps present in a diet made mostly of major staple crops like corn and soy.

| Country | Months of stored food |
|---|---:|
| India | 4 |
| China | 20 |
| United States | 18 |
| Indonesia | 21 |
| Pakistan | 4 |
| Nigeria | 3 |
| Brazil | 6 |
| Bangladesh | 4 |
| Russia | 9 |
| Mexico | 7 |

The table above suggests that most of the countries included, which together represent ~38% of the world's population, would have significantly less time to begin producing food without agriculture in a disaster if they operate autarkically. If there was still international trade, then many countries would have longer, but some countries like the US would have less time.

In appendix 20.1.7 I discuss food stores in other countries, although note that the quality of data available for many smaller countries is substantially lower than what I could find for these larger countries.

# 4.0 Average adult calorie needs

Throughout this report, I have assumed that a 2,100 calorie per day diet is sufficient for the US population, on average. This is based on an FAO/WHO report on human energy needs by first finding an individual's basal metabolic rate (the amount of calories they require to maintain their bodyweight, if they were not moving at all) and then multiplying this by a factor called the Physical Activity Level (PAL). The CDC puts the lower end of the healthy BMI range for adults at 18.5, so given the US average adult height of 170 cm (averaging men and women), the typical adult could weigh 53.5 kg and still be healthy. For an 18–30 year old man, this gives a BMR of ~1,604 kcal/day, and ~1,279 kcal/day for women. Averaging the two gives a BMR of 1,442. For a sedentary person, this figure is multiplied by 1.4 to 1.6, giving a calorie requirement of 2,018 to 2,306. Older adults generally require less calories to maintain their weight, and children require more calories per kilogram of bodyweight to grow, but weigh less, so an age-weighted average would probably put the figure a bit lower, but for now I've rounded to ~2,100 calories per day. Note that this is an average, with many individuals requiring more or less than this.

# 5.0 Nutritional value of stored foods

Since corn, wheat, and soy make up about two thirds of the calories of stored crops in the US (going by stocks in September), I wanted to check what a diet consisting of these crops, eaten at a level proportional to their respective storage amount, would look like. If no other food was consumed, then this diet would be about 67% corn, 36% soy, and 7% wheat. The table below shows the nutritional breakdown of a 2,100 kcal/day diet consisting of these three crops, if eaten in this proportion. Nutrients are highlighted in green if they are present in an adequate amount, yellow if there is not enough, red if there is too much, and left clear if I do not have a good estimate for min or max daily intake values. The upper and lower intake values come from ALLFED.

| Nutrient | Unit | Amount in 2,100 kcal | Min daily intake | Max daily intake |
|---|---|---|---|---|
| Protein | g | 85 | 46 | 400 |
| Fat | g | 43.3 | 23 | 200 |
| Carbohydrate | g | 357.1 | 50 | 500 |
| Ca | mg | 378.1 | 50 | 16,500 |
| Fe | mg | 31.2 | 10 | 120 |
| Mg | mg | 882.8 | 400 | |
| P | mg | 1,810.7 | 550 | 4,000 |
| K | mg | 3,470.2 | 3500 | 7,000 |
| Na | mg | 138.9 | 200 | 2,300 |
| Zn | mg | 15.8 | 9.5 | 40 |
| Cu | mg | 3.4 | 0.09 | 10 |
| Mn | mg | 6.5 | 2.05 | 11 |
| Se | µg | 87.2 | 55 | 400 |
| Vitamin C | mg | 7.3 | 0.9 | 4,000 |
| Thiamin | mg | 2.7 | 0.88 | |
| Riboflavin | mg | 1.9 | 1.2 | |
| Niacin | mg | 18.4 | 12 | 500 |
| Pantothenic acid | mg | 3.1 | 5 | 20 |
| Vitamin B6 | mg | 2.9 | 1.3 | 500 |
| Folate | µg | 541.1 | 166 | 1,000 |
| Vitamin B12 | µg | 0 | 0.9 | 25 |
| Vitamin A | µg | 43.7 | 500 | 8,000 |
| Vitamin E | mg | 3.2 | 10 | 1,000 |
| Vitamin D | µg | 0 | 3.8 | 15,000 |
| Vitamin K | µg | 59.1 | 70 | |
| Omega-3 Fatty | g | 1.9 | 1.35 | |

| Acid | | | | |
|---|---|---|---|---|
| Omega-6 Fatty acid | g | 20.1 | 10 | |
| Folic acid | µg | 0 | | |
| Choline | mg | 154.8 | | |
| Retinol | µg | 0 | | |

This diet is fairly complete, and doesn't supply too much of any nutrient. However, it does come with a few important deficiencies. One of the deficiencies, potassium, is probably not an issue because the diet provides >99% of the recommended minimum daily intake, but I left it yellow out of strictness. Similarly, about 84% of one's recommended vitamin K is provided by this diet, so this may not be an issue. Sodium and vitamin D can probably be acquired or naturally synthesized in adequate amounts for most people. The remaining deficiencies, vitamins A, E, B-5 (pantothenic acid) and B12, can probably be addressed with the inclusion of a small amount of other food products.

It takes a very low dose of animal products to provide enough vitamin B12. Getting the minimum 0.9 µg/day only requires, for example, eating about 107 kcal/day of cow meat (~42 g). Just using this as an example, if cow meat was the only source of B12 in the US diet, then the amount of stored cow meat and the amount that I estimate could be collected from livestock in an emergency would provide enough B12 to 332 million Americans for almost 570 days. This alone should hold the population over until some alternative food production can begin, and I expect this to contain enough B12 (see appendix 11.0). Of course other animal products also provide B12, and I estimate that existing US B12 production through microbial biosynthesis could contribute significantly to an adequate supply. Providing 0.9 µg per person per day to 332 million Americans requires just 109 kg of B12 per year, and the US probably produces about 62.7 tonnes per year, based on market indicators.[22]

Vitamin A, while abundant in some plants and available in very high concentrations in animal organs, may be difficult to get in a restricted diet. However, I believe that industrially produced food can provide enough (see appendix 11.0), and that the US population could avoid acute deficiencies with only the sources available in stored food. Sweet potatoes are high in vitamin A, but the ~600 kt of sweet potatoes stored in the US, for example, could provide enough vitamin A for 332 million people for just about one month. Getting enough vitamin A each day only requires eating about 9 g of pig liver, but in the ~37 million pigs that I expect could be slaughtered and processed quickly, there is only enough pig liver to last the US 35 days. Chicken liver could provide enough for about 19 days, and cow liver enough for 75 days, given my estimates for slaughter and process capacity in an emergency.[23] Given that other stored foods have some vitamin A, I expect that most of the US population could avoid any acute effects of deficiency if stocks are well managed and industrial foods are scaled up quickly.

Without supplementation, the diet above provides about a third of the vitamin E needed. If everyone in the US ate corn oil as their only source of vitamin E, then the ~1.6 Mt stored in the US would provide enough vitamin E for ~109 days (eating ~44 g/day). The amount of soybean oil stored in the US would provide another ~25 days. Given that the vitamin is available from several other sources, I again believe that most of the US could avoid significant deficiencies even on a restricted diet of stored food.

---

[22] B12 sells for about $1.9 million/tonne, and the US production of B12 is worth about $119.3 million/year.

[23] Liver amount calculations: there are ~74 million pigs in the US at an average weight of ~129 kg; livers represent about 2.25% of this weight. I estimate that ~50% of pigs could be slaughtered and processed in a disaster. Chicken livers usually weigh ~27 g/kg chicken bodyweight, the average chicken in the US weighs ~2.6 kg, and there are ~1.5 billion chickens in the country at any given time. Cow livers are usually about 5.7 kg, and I expect that ~25% of the US's roughly 94 million cows could be slaughtered and processed.

For B5, the diet of just corn, wheat, and soy provides a little over 60% of the recommended amount, and the nutrient is available in many other foods. I expect that deficiencies are unlikely to occur.

## 6.0 Lignocellulosic sugar production

There is a large amount of potentially edible energy available in plant fibers, or lignocellulosic biomass, particularly in wood. Globally, there is about 315 Gt of tree mass. The US has ~7.7% of global forest biomass, or over 24 Gt. Other plants could provide substantial amounts of useful biomass, but for now I'll focus my analysis on trees since they represent 75% of terrestrial plant biomass and are already harvested in large amounts. Lignocellulose consists of lignin, hemicellulose, and cellulose. Breaking down this material can produce six and five carbon sugars that are human-edible, and some US facilities are already producing food grade sugars from lignocellulose. Estimates for the potential mass conversion of wood to sugar range from about 36% to 50%, on a dry mass basis. If all of the cellulosic sugar was six carbon sugars, e.g., glucose, then it would have a calorie density of ~4,000 kcal/kg. However, ALLFED finds that the inclusion of five carbon sugars like xylose reduce the calorie density to ~3,300 kcal/kg. Using this estimate and the range of conversion efficiencies given above, the amount of wood in the US could be made into enough calories to feed all Americans for 117 to 164 years if it was the only calorie source. The sugars produced are just a source of calories and provide no other nutritional value, but these figures show that there is significant potential for woody mass to supplement other food sources. In appendix 15.0 I discuss using these sugars as a carbon source to cultivate fungal and bacterial foods to provide more nutrients.

Some lignocellulosic sugars are available with little to no requirements for new construction, but not enough to meaningfully contribute to total calorie supply. The US currently produces 150 million gallons[24] of cellulosic ethanol per year. Glucose is an intermediate product in that process, produced when cellulose is broken down by the enzyme cellulase. Since ethanol weighs 2.96 kg/gallon, 150 million gallons equals 435,000 tonnes. The conversion rate of glucose to ethanol is typically ~46% (90% of theoretical yield of 51.4%) so 435,000 tonnes of ethanol requires ~946,000 tonnes of glucose. This would provide just about 1.5% of US caloric needs on a 2,100 calorie diet, not accounting for waste. This means that without any new construction, lignocellulosic sugar could only provide a few day's worth of calories for the US, if all currently produced lignocellulosic sugar is food grade.

While new construction would be required to provide more calories from woody mass, the amount of wood harvested would not have to increase, at least not substantially. To meet all US calorie requirements for a year, about 67% of the wood that the US currently harvests per year would have to be converted into sugars, going with the conversion rate of 36% from above. Before going into more detail on the possibility of constructing new cellulosic sugar production facilities or repurposing existing facilities, I looked at an evaluation of the power and materials needed for sugar production. The main inputs per tonne of cellulosic sugar production, the amount of these inputs that would be needed to ramp up production to meet US calorie requirements, and current US supply of these inputs is shown in the table below.

---

[24] The source reports 100 million gallons of cellulosic ethanol, but this figure is given on an energy equivalent basis (relative to gasoline, which is 50% more energy dense), so the actual production is around 150 million gallons of cellulosic ethanol.

| Input | Requirement/tonne cellulosic sugar production | Amount needed to meet all US calorie needs | Current US annual production or supply |
|---|---|---|---|
| Sulfuric Acid | 16.7 kg | 1.3 Mt | 22.8 Mt |
| Cellulase | 27.8 kg | 2.1 Mt | 0.4 Mt[25] |
| Ethanol | 27.8 g | 2,143 tonnes | 44 Gt |
| Electricity | 0.67 kWh | 0.05 tWh | 4297 tWh |

The data above show that, at current levels, the US only produces enough cellulase to cover ~20% of its total calorie needs (given the process described in the study above). The US could probably increase its production of cellulase, or it could use other processes that do not require cellulase. One such process is described in this article,[26] where acid hydrolysis, instead of cellulase, is used to break down cellulose. Based on this article, the table below shows different inputs needed, how much would be required to provide all US calories from wood sugar for a year, and how much of each input the US currently makes each year.

| Input | tonnes/tonne sugar | Mt/year to meet US calories for a year | Current US annual production or supply (Mt) |
|---|---|---|---|
| Sulfuric acid | 1 | 77 | 22.8 |
| Lime | 0.6 | 46.2 | 17 |
| Anthraquinone | 0.05 | 3.9 | 0.03 |
| Sodium hydroxide | 0.01 | 0.8 | 11.6 |
| Sodium sulfide | 0.03 | 2.3 | 0.3 |

[25] Note on my cellulase estimate: The US probably doesn't produce enough cellulase currently. I couldn't find data on US or global production, or even the US market. However, the global market is ~$1.6 billion. I found that global prices average ~$1,000/tonne, implying a global supply of ~1.6 million tonnes. The global enzyme market (all enzymes, not just cellulase) is $6.95b, so the cellulase market is worth ~23% of the enzyme market. The US enzyme market should be ~$1.82 billion (calculated from CAGR and 2016 estimate from USDA source). If cellulase is also ~23% of the US enzyme market, then the US cellulase market should be worth about $420 million. This would then suggest that production is ~420,000 tonnes per year, or 23.8% of what is needed for supplying all US calories with cellulosic sugar.

[26] The paper describes the production of ethanol from wood, but since the ethanol is made by first turning the cellulose into glucose (through acid hydrolysis) and then fermenting the glucose into ethanol, this process can be used to make sugar by just skipping the last step.

| | | | |
|---|---|---|---|
| Sodium carbonate | 0.1 | 7.7 | 12 |
| Sodium sulfate | 0.005 | 0.4 | 0.3 |

The data above suggest that the US may be better off increasing its cellulase production rather than increasing its annual production of several other inputs. When using glucose as a feedstock, cellulase can be produced with a yield of ~0.15 g cellulase/g glucose. Since less than 0.03 g of cellulase are needed per g of glucose produced from wood (above), this is a net gain, but shows that a considerable amount of sugar would be needed to make enough cellulase (about 14 Mt, or enough calories to otherwise make up ~22% of US calorie needs, without waste).[27] Production would also require a significant fermentation volume. Even if a reactor could run continuously for a year with 100% working volume, producing enough cellulase would need ~1.3 million m³ of fermenter space, given a productivity of 0.18 g / L h. With 10% downtime and 90% working volume, this rises to over 1.6 million m³ of fermenter space. For reference, the US has about 3.1 million m³ of fermentation capacity for brewing and perhaps about 560,000 m³ of fermentation space for ethanol production.[28] As discussed below, ethanol and brewing facilities could serve as sites for wood sugar production after some repurposing, so having to use these facilities to make cellulase could reduce US sugar production potential, at least in the short term.

The US could repurpose brewing, ethanol, and pulp and paper mill facilities to make sugar from lignocellulosic biomass, and a similar repurposing process has been proposed for turning a pulp and paper mill into a cellulosic ethanol plant. An ALLFED paper has estimated the lignocellulosic sugar production potential of different facilities based on their current size. Using these estimates, and the size of US corn ethanol, pulp and paper, and brewing industries, the table below shows how much sugar could be produced if all US plants were repurposed to make lignocellulosic sugar. See tab 7 of this Google Sheet for relevant calculations.

| Facility type | Conversion efficiency from retrofitting | Current annual capacity (Mt) | Annual sugar production potential (Mt) |
|---|---|---|---|
| Corn ethanol | 0.50 | 45.6 | 22.62 |
| Pulp and paper | 0.09 | 67.5 | 5.94 |
| Brewery | 0.18 | 20.5 | 3.73 |

This production, ~32.3 Mt per year in total, could provide up to ~52% of US calories (without waste), or closer to ~45% accounting for 13% waste. This much production would require ~900 kt of cellulase, or about 2.24x what I estimate is currently produced in the US each year. Producing the other ~500 kt per year would require ~390 million m³ of fermentation space. This is less than 13% of the volume of fermenter space used for brewing, so if the production of retrofitted brewers could be reduced by just 13%, then enough extra cellulase could be produced.

The cost of this repurposing depends on which facility is used. A paper modeling the costs of turning a pulp and paper mill into an ethanol plant found that doing so would take ~$79.3 million, if I subtract out the fermentation equipment that wouldn't be necessary for sugar production. Given the size of the facility examined in the paper, it could probably make ~41 kt of sugar per year. If this cost held for all repurposing projects, then making the 32.3 Mt of annual sugar production from above would cost almost $63 billion. This gives a per-tonne cost of about $1,950. ALLFED's paper found a much lower cost for repurposed facilities, with a capital expenditure of ~$290/tonne for repurposed pulp and paper mills. The article also found a capital expenditure of ~$690/tonne for repurposed corn ethanol plants and ~$1,475/tonne for repurposed breweries. Given the distribution of production capacity in the US from the table above, this gives an average ~$710/tonne, for a total capital expenditure of ~$22.9 billion. I use ALLFED's estimates in the analysis above, but note the potential for higher costs. I use ALLFED's estimates because they're generally consistent with the prices for wood sugar production estimated in this DOE review of sugar production from different substrates, which reviewed several studies estimating the cost of making wood sugar production facilities and found the highest cost to be ~$940/tonne, while other studies report much lower costs. While this is higher than ALLFED's estimate, ALLFED was looking at repurposing facilities rather than making new ones, so the cost should be lower (ALLFED was also accounting for higher costs incurred from attempting to minimize construction times during a disaster).

## 6.1 Note on the inclusion of sugar in a post-disaster diet

In scenarios where society has substantial access to non-crop plant biomass, the production of lignocellulosic sugar may be one of the cheapest and most quickly scalable sources of calories. However, it does not have any other nutritional value, so the percent of a diet that can be made of sugar is capped by the amount of other food sources needed to meet minimum daily nutrient requirements. This varies by diet, but as a quick note, I wanted to look at how much sugar could be included in a 2,100 kcal/day diet otherwise made up of a) major US staple crops (discussed in appendix 5.0), since these would make up the bulk of a post-disaster diet before other foods could be produced, and b) methane-fed single-celled protein (SCP, and see appendix 11.0 for a discussion of SCP nutrition), since this is one of the more scalable food options in the scenarios considered in this report, if society and industry are intact. Neither of these food options are a complete source of nutrients themselves. As I mentioned above for staple crops, the judicious use of other stored food sources could cover most nutritional needs, and I discuss in a separate appendix below various options for improving or supplementing an SCP diet. For a quick first pass, here I look at how much sugar could be included in a diet otherwise made up of staple crops or SCP without bringing the daily intake of fat or protein too low.

For major staple crops, given the composition of the diet discussed in appendix 5.0, sugar can make up to ~45% of the diet without lowering the fat and protein intake below the minimum recommended amount, although society may aim for more of a buffer than that. Also, in addition to the nutrients that are already too low in a diet made up of only corn, wheat, and soy (in the proportions in which these crops are stored in the US in September), several other nutrients would be too low in a diet that included that much sugar, but this gives a rough sense for how much sugar might be included in a diet before other alternative foods are produced.

For a diet otherwise consisting of SCP, sugar can make up ~34% of the diet before total fat levels are too low. As noted below, a diet made up of SCP has several serious challenges. The inclusion of sugar actually helps by reducing total protein and copper intake, but the diet would still be too low in essential fatty acids, fiber, and a few critical vitamins and minerals. I believe that there are options for addressing these deficiencies in the scenarios considered, but this is an area that I hope to research further.

In the current US diet, refined sugars make up a little over 15% of total calories consumed, and some human populations get up to 30% of their calories from sugar, as in the case of the Hadza of Southern Africa who eat large amounts of honey. Therefore it seems likely that sugar can be included in a diet up to at least 30% without serious ill effect. I have not looked into higher levels of consumption, but I suspect that they are also safe for most of the population, at least for a short period of time while the production of other foods is being scaled up.

# 7.0 Fishing

Here I discuss how much additional fishing might be able to add to a post-disaster diet, at least in the short term. Fish currently make up just about 0.7% of US calories, of which about 82.3% is produced domestically, with 93% being wild caught and the rest coming from farms. However, since aquatic ecosystems may not be affected by an attack on conventional agriculture, fish could be a valuable food source, at least to help provide more time to ramp up alternative food production technologies. For fish in the body size targeted for fishing (10g to 100kg), total global biomass (given in fig. 6) is ~4.7 Gt, or 34.8x the amount of fish caught each year (135 million tonnes).[29] Based on the number of calories currently produced per tonne of fish, this gives ~4 billion person-years worth of food (data on fish harvests and calories are from FAO).[30] Thus, there is a large reserve of calories in the form of fish biomass. The critical question is how quickly a large share of this biomass can be harvested with existing fishing infrastructure.

A standard way that fishing capacity is measured is to determine fish holding capacity in a fishing fleet, estimate how long it typically takes for a boat to fill its holding capacity with fish, determine how many trips per year this would allow, and multiply through. I attempted to apply this estimation method to the US.

America has 73,200 fishing vessels. The FAO estimates that these vessels have an average holding capacity of 101.3 tonnes, for a fleet total of 7.55 million tonnes. One source (NOAA) puts the length of the typical commercial fishing operation at about 25 days, suggesting that 14.6 trips could be taken each year (optimistically assuming that there are no delays from weather, the logistics of offloading a new catch, fixing equipment, or labor availability). Multiplying these figures together gives a rough estimate that the US fishing fleet has the capacity to catch 101.2 million tonnes, or ~21x the amount that the US currently catches. This implies that the US fishing fleet is only operating at a little under 5% of its hypothetical capacity. This estimate is actually very similar to findings from a study of Oregon's fishing fleet, which estimated that over a ten year period, fleet utilization averaged ~5.7%. There are a range of reasons why utilization rates could be so low; many of which have to do with limits on fishing (such as fishing only being legal during certain parts of the year, or limits on the number or type of fish that can be caught) imposed for conservation purposes.

Using a conservative estimate for the calories provided per kg of fish (see above comment), 100 million tonnes of fish would feed the US for about 102 days, providing ~28% of the US calorie annual requirements. Note that this is 100 days of food provided over the course of a year. My estimate for fishing capacity may be a bit optimistic by ignoring various possible logistical constraints or diminishing returns to additional fishing effort. I am using a conservative estimate for the calories per kg of caught fish, since my estimation method assumes that fish is wasted at the same level that it currently is. I also conservatively ignore the potential contribution of non commercial-fishing boats. While the US has a little over 72,000 fishing vessels, it has nearly 17 million registered recreational boats. Only about 850,000 of these are >8m in length, and so the per-vessel holding capacity is probably fairly low compared to commercial fishing boats. By sheer numbers, however, it seems possible that these vessels could still contribute to overall fishing effort. However, on net I imagine that ~100 days of food from fish is still fairly optimistic, and I estimate that the US may instead only be able to get ~25 days of food, given various constraints and logistical challenges. This is still a roughly 5x increase in US annual fish catching, so even this may be optimistic. I should also note that in some of the scenarios considered in this doc, the disaster may affect marine ecosystems and so preclude fishing, at least on a large scale.

---

[29] However, at least ~20% of this biomass is below the epipelagic zone, where most of commercial fishing takes place, so the practically available biomass is lower
[30] This likely underestimates the calories actually available in fish, since it assumes that fish is wasted at the same rate that it currently is.

## 7.1 Fish nutritional value

Fish would not likely represent a large portion of diets as society shifts to producing food that does not require any agricultural inputs. However, in some societal collapse scenarios, if the disaster in question does not disrupt marine ecosystems too much, fish may make up a large fraction of the diet of survivors.

It appears that while e.g., Inuit diets do contain a small amount of terrestrial plants, at least one study found no adverse health effects for a few subjects who ate no plants for a year. Getting adequate micronutrients seems to require eating organ meat and other animal parts that are not often consumed and so are not reflected in the nutritional content displayed below for a "mix of white fish species" from the USDA. Again, the minimum and maximum daily intakes come from this ALLFED paper.

| Nutrient | Unit | 2,100 kcal | Min daily intake | Max daily intake |
|---|---|---|---|---|
| Protein | g | 299.3 | 46 | 400 |
| Fat | g | 91.8 | 23 | 200 |
| Carbohydrate | g | 0 | 50 | 500 |
| Ca | mg | 407.5 | 50 | 16,500 |
| Fe | mg | 5.8 | 10 | 120 |
| Mg | mg | 517.2 | 400 | |
| P | mg | 4,231.3 | 550 | 4,000 |
| K | mg | 4,967.9 | 3,500 | 7,000 |
| Na | mg | 799.3 | 200 | 2,300 |
| Zn | mg | 15.5 | 9.5 | 40 |
| Cu | mg | 1.1 | 0.09 | 10 |
| Mn | mg | 1.1 | 2.05 | 11 |
| Se | µg | 197.5 | 55 | 400 |
| Vitamin C | mg | 0 | 0.9 | 4,000 |
| Thiamin | mg | 2.2 | 0.88 | |
| Riboflavin | mg | 1.9 | 1.2 | |
| Niacin | mg | 47 | 12 | 500 |
| Pantothenic acid | mg | 11.8 | 5 | 20 |
| Vitamin B-6 | mg | 4.7 | 1.3 | 500 |
| Folate | µg | 235.1 | 166 | 1,000 |
| Vitamin B12 | µg | 15.7 | 0.9 | 25 |
| Vitamin A | µg | 564.2 | 500 | 8,000 |
| Vitamin E | mg | 0 | 10 | 1,000 |
| Vitamin D | µg | 188.1 | 3.8 | 15,000 |
| Vitamin K | µg | 1.6 | 70 | |
| Omega 3 | g | 2.9 | 1.35 | 5 |
| Omega 6 | g | 4.3 | 10 | 23 |
| Folic acid | µg | 0 | | |

In the table above, the phosphorus intake provided by a 2,100 kcal/day diet of only white fish is labeled as excessive. There is very little evidence about the consequences of such high levels of consumption, but exceeding the upper recommended intake is probably not harmful. Manganese, omega-6 fatty acid, and vitamins C, E, and K are listed as deficient in this diet. Given that some groups have historically relied on seafood, I suspect that these deficiencies would not prove fatal on a seafood-based diet, but I have not yet investigated this claim thoroughly.

# 8.0 Large-scale mushroom production

Here I calculate the input requirements for large-scale mushroom production, extrapolating from current cultivation practices. Currently, the US produces an average of ~345 kt of mushrooms per year. Of these, 335 kt, or 97%, are Agaricus mushrooms, which include white button, Crimini, and Portobello. These mushrooms typically have ~220 kcal/kg, so current US annual production only provides enough calories to feed ~100,000 people for one year. Modern production methods require too much energy, nitrogen fertilizer, and indoor cultivation space to scale to meet all US calorie needs, even if there was enough organic matter to serve as a suitable substrate.

A USDA-sponsored study found that mushroom farmers use ~6.9 kWh/kg of mushroom produced. Since providing all US calories with 220 kcal/kg mushrooms would take ~1.15 Gt of mushrooms per year, this level of energy intensity would require almost 2x the amount of electricity the US currently produces. Thus if production was not made more efficient, large-scale mushroom cultivation would probably be out of reach for the US based on energy needs alone.

High mushroom yields require 29.3 to 39.1 kg/m² of dry compost. Since the current production of 335 kt of Agaricus mushrooms takes up ~1,088 ha for an annual yield of about 28.3 kg/m², scaling up mushroom production to meet all US calorie needs requires a little over 4 million ha and around 1.2 to 1.6 Gt of dry compost. However, getting an estimate for total compost amount requires multiplying these figures by ~3.3 to account for the recommended 68–72% moisture content for compost, and then further multiplied by 3.3 to 4 to account for the number of harvests per year (discussed more below). Thus on a dry matter basis, meeting all US calorie needs for a year with mushrooms requires ~4.0 to 6.4 Gt of compost, or 13.0 to 21.1 Gt on a wet matter basis. The US has ~24 Gt of tree biomass. Trees represent ~70% of terrestrial plant biomass, so the US should have a little over 34 Gt of plant biomass. Thus the amount of organic matter needed to grow one year's worth of mushrooms is a substantial share of all plant matter in the US (12–19% on a dry weight basis). It is recommended to supplement compost with 2.0 to 2.4% nitrogen fertilizer, by mass. I am not sure if this is supposed to be on a dry or wet weight basis with respect to the fertilizer, but either way this suggests a need for far more N fertilizer than the US currently makes. If the recommended amount is supposed to refer to a percent of dry mass compost, then the US would need 79 to 153 Mt of N fertilizer, but currently only produces ~12 Mt of N fertilizer. Worse, ammonia, which makes up ~⅓ of US N fertilizer production, inhibits mushroom growth if concentrations in the compost are >0.07%. If the US has to use mostly non-ammonia nitrogen fertilizer, then suitable fertilizer supplies may be substantially lower than even 12 Mt. If the recommended N fertilizer concentration is supposed to refer to the percent of wet compost mass, then the amount needed would be 261 to 506 Mt.

Putting aside the energy, fertilizer, and compost bottlenecks to scaling up conventional mushroom cultivation, space would also be limiting, at least without significant new construction. The US has ~3.2 million ha of building floor space, 60% of which is residential. Together, this is only ~80% of the 4 million ha needed to grow mushrooms given the current US average yield. Also, much of that space would presumably need to be used for other purposes, and may not be suitable.

The calculation for the number of harvests alluded to above comes from a description of the mushroom production process, which lays out that spawn production, compost preparation, and the mushroom growth cycle mean that there are 57–68 days before harvests can start, and that harvests last for 35–42 days. There are ~4 92–day cycles in a year and ~3.3 110-day cycles, and so there are 3.3 to 4 harvests per year. During the 35–42 day harvests, mushrooms are harvested for 3–5 days, the compost is allowed to grow more mushrooms for several days, and then harvests are recommenced. This repeats until the compost no longer produces an appreciable amount of mushrooms. While this process usually lasts 35–42 days, it can go for up to 150 days.

## 8.1 Scaling up mushroom production

From above, it seems that large-scale mushroom cultivation, using high intensity modern production practices, is unlikely to be a good candidate for scaling to feed the US population. However, less intensive cultivation practices could be used in cases where dead plant biomass is accessible. I therefore wanted to get a sense of how quickly mushroom production could scale given existing mushroom supplies. From a quick review, I couldn't easily find especially clear data on this question, so what follows is a very rough BOTEC, and shouldn't be read as definitive.

Commercially, mushrooms are propagated using spawn, which consists of a substrate (usually grain) that has been colonized by mushroom mycelia. The colonization is kicked off by inoculating the substrate with pure mushroom culture. This can come from mushroom mycelia grown on an agar plate, but can also be a part of a harvested mushroom. This process is different from natural mushroom growth, which relies on spores. But these germinate too unreliably, and are too difficult to collect and then disperse, to be used for cultivation.

I wanted to see how much mushroom mass could be grown from spawn created with the amount of mushrooms harvested in a given typical harvest of mushrooms in the US. I started with the recommendation that compost used to grow mushrooms have an amount of spawn equal to ~2% of the compost mass. For mushrooms grown on wood, I haven't yet found an estimate for what fraction of wood mass should come from spawn, but I'll for now assume that it is the same percent, and will discuss this below. First, going with the 2% figure and the mushroom yields per unit of compost above (around 0.18 to 0.29 kg mushroom/kg compost, so that each person needs 12 to 19.4 tonnes of compost to grow their annual calorie needs), then one person would need ~240 to 387 kg of spawn per person per year, or ~80 to 129 Mt for the whole US By weight, spawn is ~63% grain or other substrate and 37% mycelia. Thus the US would need to use ~50 to 80 Mt of grain or some other substrate. For context, the US produces ~432 Mt of grains each year and has ~64 Mt stored, and harvests ~310 Mt of wood each year. Thus even getting the amount of substrate needed to produce spawn, let alone create a suitable growing substrate, would be challenging, taking ~16% to 26% of all the wood that the US currently harvests. One site recommended using the mushroom culture of a full 60 mm diameter petri dish to inoculate 400 to 450 g of substrate. Given a depth of ~3 mm in a petri dish mushroom culture, and a mushroom density of ~0.55 g/cm3, this suggests that ~18.7 g of mushroom was used to inoculate 400 to 450 g of substrate. Going with the 63%/37% breakdown of spawn weight between substrate and mycelia, this suggests that 1 g of mushroom is enough to make 34.3 to 38.6 g of spawn. Scaling this up, the 77.1 to 93.4 kt of agaricus mushrooms harvested in a normal season (given current annual production from above and the 3.3 to 4 harvests per year) could make ~2.6 to 3.6 Mt of spawn. This could "seed" ~110.2 to 180.2 Mt of compost. Given the yields from above, this could produce 19.8 to 52.3 Mt of mushrooms, or enough to provide ~1.7 to 4.5% of US calories. This would represent a ~59 to 156x increase in US annual mushroom production. Since this is a lot less than the amount needed to feed the whole US, the energy and fertilizer requirements wouldn't be beyond US capabilities, but it probably wouldn't be worth the effort given the limited number of calories produced. I should note that I've generously assumed that all harvested mushrooms could be used to inoculate a substrate, when in fact not all parts of a mushroom are capable of vegetative propagation.

If the US did try to grow more mushrooms, it could repeat the propagation process over two harvests. Given the assumptions from above, if the process was repeated and cultivation was not limited by input availability (e.g., limited N fertilizer, compost), then the US could turn the 19.8 to 52.3 Mt of mushrooms into ~680 to 1792 Mt of spawn, enough to grow 5.2 to 26.0 Gt of mushrooms, 4.5 to 22.5 times more than enough to cover all of the calorie needs for 332 million Americans. As covered above, there are many limiting factors on expanding mushroom cultivation this much, at least if the US were to rely on conventional compost-based cultivation practices used for growing Agaricus mushrooms. There are lower-tech alternatives that may be able to scale without encountering the same limitations.

Below, I found that growing enough oyster mushrooms (one of the main types of mushrooms that grow on wood that is cultivated in the US) to meet one person's annual calorie needs requires ~5.1 to 9.6 tonnes of (soft) wood. Feeding the US with oyster mushrooms would therefore take ~1.7 to 3.2 Gt of wood, or ~5.5 to 10.3x what the US currently harvests each year. This means that if current wood harvest rates couldn't be increased, then using all harvested wood to grow oyster mushrooms for a year would produce just ~10% to 18% of US calories. Significantly more wood appears necessary to grow shiitake mushrooms (below) than oyster mushrooms, making scaling up even harder for shiitakes, which are the second most widely cultivated wood-growing mushrooms in the US Importantly, I haven't found evidence that agaricus mushrooms (the overwhelmingly dominant mushrooms in the US) can grow on wood. Wood can be turned into compost, although it needs to be broken into small volumes like wood chips. Turning the wood into compost allows non cellulolytic mushrooms to use it as a growth substrate. As covered above, however, growing agaricus mushroom on compost is generally an intensive process. Agaricus mushrooms may not be able to be grown with low-tech techniques, like simply inoculating logs with spawn, as works for shiitake and oyster mushrooms.

If the more scalable, low tech cultivation processes can only be used for these specialty mushrooms, then the US would be starting with far fewer mushrooms for the purposes of inoculating a substrate to create spawn. However, mushroom propagation is so fast that it would still only take 3 harvests to scale up mushrooms to make enough to feed 332 million Americans for a year. Per year, the US produces an average of just ~3.5 kt of shiitake mushrooms and ~3.2 kt of oyster mushrooms. Given the calorie content of shiitakes and oyster mushrooms, this is enough to provide 2,100 kcal per day for a year to just 1500 people from shiitakes and 1900 people from oyster mushrooms. Annual production for shiitakes would therefore have to increase by a factor ~216,000, and annual oyster mushroom production would have to increase by a factor of ~173,000. To see how much starting mushroom culture might be available, I divide the annual production by the number of harvests each year. This source on oyster mushroom production suggests that the entire substrate preparation, spawn colonization, and mushroom growth process takes 46 to 65 days, suggesting that there are 5.6 to 7.9 harvests per year. This means that during each harvest, ~403 to 570 tonnes of oyster mushrooms are produced, given current annual production levels. The USDA finds that most growers have 2–3 shiitake mushroom harvests per year, so the per-harvest production should be ~1,150 to 1,730 tonnes. Going with the assumptions above, the amount of spawn that could be produced by the amount of oyster mushrooms harvested at one harvest is ~13.8 to 22.0 kt. If the same amount of spawn is needed per mass of substrate (in this case, logs), then that would be enough to inoculate 576 to 1,100 kt of wood. Given the yields of oyster mushrooms grown on wood that I'd found above (which are also discussed below and based on this study), this would produce ~130 to 440 kt of oyster mushrooms, or only enough for ~0.02% to ~0.08% of US annual calorie needs. If repeated, the process of turning oyster mushrooms into spawn for more oyster mushrooms could produce 21.0 to 169.8 Mt of mushrooms after a second harvest, enough to meet 3.8 to 30.8% of US annual calorie needs. A third round of doing this would provide more than enough mushrooms to feed the whole US for a year. Roughly the same picture holds for shiitakes, although much more wood would be needed. Three harvests for oysters would take 135 to 195 days, and three harvests for shiitakes would take 1 to 1.5 years. I suspect that shiitake mushroom production could be sped up some amount, bringing the time down.

The process laid out here is a massive simplification of the steps involved in mushroom production, and rests on several major assumptions. As a rough first pass, however, this work suggests that the rate of mushroom growth should allow for a massive scale up of mushrooms, enough for them to make up a major portion of US calories, before existing US food reserves would run out (~15 months if not supplemented with stopgaps). The bigger issue would be the extreme inefficiency with which mushrooms turn wood into calories, compared turning wood into calories through lignocellulosic sugar production (appendix [6.0](#)), or using lignocellulosic sugar to grow fungal or bacterial SCP (appendix [15.0](#)).[31]

## 8.2 Mushroom nutritional value

The table below shows the nutritional profile of a 2,100 kcal/day diet made of only one of three types of mushrooms. Data for mushroom nutrients comes from the [USDA](#) unless otherwise noted, and the min and max daily intakes come from this ALLFED [paper](#). I coded nutrients in green when they were provided in a safe amount, yellow when below the minimum, red when above the maximum and possibly dangerous, and blue when above the maximum but the maximum does not represent a dangerous level.

| Nutrient | Unit | Shiitake | Oyster | White (Agaricus) | Min daily intake | Max daily intake |
|---|---|---|---|---|---|---|
| Protein | g | 138.4 | 210.6 | 295 | 46 | 400 |
| Fat | g | 30.3 | 26.1 | 32.5 | 23 | 200 |
| Omega-3 | g | 0 | | 0.1 | 1.35 | 5 |
| Omega-6 | g | 8 | 7.8 | 15.3 | 10 | 23 |
| Carbohydrates | g | 419.4 | 387.5 | 311.2 | 50 | 500 |
| Ca | mg | 123.5 | 190.9 | 286.4 | 50 | 16500 |
| Fe | mg | 25.3 | 84.6 | 47.7 | 10 | 120 |
| Mg | mg | 1,235.3 | 1,145.5 | 859.1 | 400 | |
| P | mg | 6,917.6 | 7,636.4 | 8,209.1 | 550 | 4000 |
| K | mg | 18,776.5 | 26,727.3 | 30,354.5 | 3500 | 7000 |
| Na | mg | 555.9 | 1,145.5 | 477.3 | 200 | 2300 |
| Zn | mg | 63.6 | 49 | 49.6 | 9.5 | 40 |
| Cu | mg | 8.8 | 15.5 | 30.4 | 0.09 | 10 |
| Mn | mg | 14.2 | 7.2 | 4.5 | 2.05 | 11 |
| Se | μg | 352.1 | 165.5 | 887.7 | 55 | 400 |
| Vitamin C | mg | 19 | 0 | 200.5 | 0.9 | 4000 |
| Thiamin | mg | 0.9 | 8 | 7.7 | 0.88 | |
| Riboflavin | mg | 13.4 | 22.2 | 38.4 | 1.2 | |
| Niacin | mg | 239.6 | 315.6 | 344.6 | 12 | 500 |
| Pantothenic acid | mg | 92.6 | 82.1 | 143.2 | 5 | 20 |

---

[31] Using sugar as a substrate for fungal growth is considered [submerged fermentation](#), while growing mushrooms on wood or compost is [solid state fermentation](#), which is generally less efficient

| Vitamin B6 | mg | 18.1 | 7 | 9.9 | 1.3 | 500 |
| Folate | µg | 802.9 | 2,418.2 | 1,622.7 | 166 | 1000 |
| Vitamin B12 | µg | 26.3 | 0 | 3.8 | 0.9 | 25 |
| Vitamin A | µg | 0 | 214.3 | 768.8 | 500 | 8000 |
| Vitamin E | mg | 0 | 0 | 9.5 | 10 | 1000 |
| Vitamin D | µg | 24.7 | 44.5 | 19.1 | 3.8 | 15000 |
| Vitamin K | µg | 0 | 0 | 95.5 | 70 | |

I find that the most nutritious mushroom is actually white mushroom. These mushrooms may be less easily cultivated than the others since they can't be grown on wood, and so survivors of a societal collapse would have to create compost to grow them. If they were able to do that, then eating a diet of white mushrooms would provide ~3x the upper recommended safe amount of copper, and a very high but probably acceptable amount of protein. See the note on a high-protein diet (12.7) and on copper toxicity (12.8) below.

The other two mushroom species, especially shiitake, have much less protein, but are also less nutritious overall. Shiitake also has the benefit of having less than the upper limit of copper, but as I review in appendix 12.8, the evidence base for the danger of 30 mg of copper/day is mixed. White mushrooms appear to only be deficient in omega-3 fatty acids and vitamin E, but a 2,100 kcal/day diet provides 95% of the minimum amount of vitamin E, so I suspect that this is a tolerable deficiency, and eating a bit more than 2,100 kcal would provide an adequate amount.

# 9.0 Alternative food options without plant biomass

There are three broad approaches to making food without outdoor agriculture: indoor agriculture, synthesizing nutrients abiotically, and growing food on nonagricultural substrates. The table below summarizes the main findings from a quick investigation of different routes to food production (also covered in the main text above), with more detail provided in the subsections of this appendix. I evaluated the potential of each food source to act as the sole source of calories for the US, even though most sources evaluated wouldn't provide anywhere near a complete diet. The valuations should be understood as a first-pass assessment of scale potential rather than a literal suggestion that a post-disaster diet be made of a single food source. The options evaluated are those that do not require any plant-derived material, but some scenarios may allow access to dead plant biomass. The options opened up by this are discussed in appendix 15.0.

| Synthetic fat | All proposed processes require glycerol, which in the US comes from biofuel production. Since this production requires crops, and since the disasters considered here would force humans to directly consume crops instead of making biofuels, biofuel-derived glycerol would be inaccessible.<br><br>The US could make enough synthetic glycerol from its current annual |
| | production of propylene to then make enough synthetic fat to meet its calorie needs, but it would have to increase its annual chlorine production by ~2x.<br><br>Going with historical precedent, using all US coal could produce 24% to 65% of US calories (if not bottlenecked by the glycerol supply), but the US has few facilities to make coal into a liquid fuel or to separate out Fischer-Tropsch wax from this liquid. Modern production methods find higher yields and allow other feedstock, like natural gas, to be used. However, production would still require synthetic glycerol, which is a likely bottleneck. |
| Synthetic carbohydrates | Sugar has been produced from $CO_2$ through electrosynthesis, although only at an experimental scale. There are several proposed processes, but the one that is likely most feasible involves first reacting $CO_2$ with water to make formaldehyde, and then reacting formaldehyde with calcium hydroxide and water to make glucose. Assuming a maximum theoretical efficiency of the intermediate steps, the electrosynthesis of glucose from $CO_2$ would require ~a third of current US annual electricity production to make a year's worth of calories for all Americans. Accounting for likely inefficiencies, production would need ~1.7x the amount of electricity currently produced in the US. |
| Synthetic amino acids | Only one of the nine essential amino acids that humans need, methionine, has been synthesized abiotically. While the US already produces ~150 kt of methionine per year from methanethiol and acrolein (made from methanol and propene, respectively), achieving the ~260x increase in production needed to meet all US calories would require a ~100x increase in methanethiol production and a ~125x increase in acrolein. This in turn could only be done by increasing the production of chemicals used to make these inputs, like increasing methanol production by almost 4x. |
| Crops grown with artificial light/in greenhouses | To grow enough food with fully indoor ag, the US would have to use something like 7x its current total energy production.<br><br>To grow enough food with greenhouses that have at least some filtration system to keep out a bio threat, the US would have to use ~124% of its current electricity.<br><br>There'd also be various material constraints that I didn't look into, like light bulbs and cover material for greenhouses. |
| Microalgae | If grown with artificial light, would need ~1.7x current US electricity |
| | production to make all US calories. |
| | Would need less electricity to grow outdoors (~5%), but this may not be biosecure, depending on the threat. |
| | Could be grown in transparent tubes that allow growth on sunlight but also protect from the outdoors, but the US does not currently make enough plastic or glass to make enough of these. |
| | In all cases, production may be nutrient limited, although more optimistic assumptions about nutrient uptake and recycling efficiency and microalgae mineral content suggest that microalgae may be a viable option for producing all US calories. |
| Microbes grown on methanol | The US would need to ~20x current annual methanol production to make enough to feed all Americans. |
| Microbes grown on $CO_2$ and $H_2$ | The US would have to ~2.7x the amount of $CO_2$ it captures to make enough SCP (assuming that all of its current capture could be diverted to food production, even though some is used for enhanced oil recovery). |
| | The US would also have to ~3x $H_2$ production. |
| | Doing this appears possible given the current availability of inputs to $H_2$ production and $CO_2$ capture or production. But the need to scale up these industries — on top of scaling up microbial food production — probably makes this option slightly worse than others considered here, where the relevant feedstocks are already available in sufficient quantities in the US. |
| Microbes grown on diesel | If the US used ~67% to 100% of its current annual production of diesel, it could possibly make enough SCP to feed all Americans, but there's extremely little information on this process (including actual evidence that SCP is safe for human consumption), so this option is not investigated further in this report. |
| Microbes grown on natural gas | Using only ~13% of current US natural gas production, the US could feed itself on MOB SCP. |

Note that in investigating highly scalable options, I generally was looking for what I call "easy wins": cases where the major materials and energy inputs are already available in large enough amounts for a single food source to meet all of the US's calorie needs. I later investigated the equipment needs and nutritional quality of different food sources, but the limits of my initial search process still meant that I was not looking for an optimal solution, and am likely to have missed other food production options. My findings should be taken more as an existence proof that there are promising options for food production without agriculture, rather than as being prescriptive about a particular approach for responding to an agricultural disaster. For a brief discussion of the limits of my approach, see appendix 10.0.

## 9.1 Nutrient synthesis

### 9.1.1 Synthetic fat production

Synthetic fat production, as historically achieved, requires wax from fossil fuels and glycerol. The wax that was traditionally used was produced through the Fischer-Tropsch (FT) process, used to process coal or natural gas. ALLFED has proposed using paraffin wax instead of FT wax. Paraffin wax occurs naturally in petroleum, and is sometimes removed during the oil refining process, or left in in cases where the oil has a low wax content. About 70% of glycerol worldwide and all glycerol made in the US is produced as a byproduct of biodiesel production. In a loss of agriculture scenario, synthetic glycerol from propylene, a propane derivative, would have to be used.

An old NASA report finds that coals can be made into a margarine-like food with a mass conversion efficiency of 1.6–4.3%. The US currently produces ~535 Mt of coal each year. If the product discussed in NASA's report has the same caloric content as margarine, then it should have 7170 kcal/kg. Given these conversion efficiencies and calorie density, the current US annual coal production could provide between 24% to 65% of US calories, if all coal was used for making food and there was no other limiting factor. Given current mining rates, the US has >400 years worth of coal that are economically accessible, so even at elevated production rates, supplies should last a while.

If the US instead used paraffin wax from oil as a feedstock for synthetic fat production, it could probably make about 36 Mt of wax per year from its current annual oil production. This figure is based on a literature review that reports paraffin wax abundances in different oil fields. I took a weighted average of these abundances based on the Energy Information Agency's (EIA) list of oil production by state to estimate that US domestic oil has a typical wax abundance of ~5% by weight. Since producing a kg of synthetic fat requires about 1.43 kg of wax (based on information provided by Savor-it), this amount of annual wax production would be enough to make fat equal to around ~60% of US annual calorie needs.

Savor-it claims higher yields from coal or natural gas (as feedstocks to make FT wax) than what was achieved historically. With a total mass yield from coal and natural gas of 0.59 and 0.34 respectively (as stated by Savor-it), the US could meet all of its yearly calorie needs with just ~22% of its coal or ~8% of its natural gas annual production. The company also finds that producing a kg of synthetic fat requires ~0.15 kg of glycerol. This means that the US would have to make ~6.1 Mt of synthetic glycerol. Given the inputs to synthetic glycerol production, reviewed by ALLFED, this would require ~52% of current US annual production of caustic soda and ~2.1x current US annual production of chlorine. In addition, the FT process, needed to make wax for fat production, is not widely used in the US. One current application of this process is the production of syngas to make $H_2$ gas, but this is only used to make about 1 Mt of $H_2$ gas per year. Given the yield of 1 tonne of $H_2$ per 3.6 tonnes of natural gas, this suggests that only about 3.6 Mt of natural gas are used in an FT process in the US each year. This is just ~5% of the natural gas that would be needed to produce enough synthetic fat to meet US calorie needs based on the yields claimed by Savor-it.

The overall picture appears to be that significant production of synthetic fat in the US would require multiple industrial processes to be scaled up simultaneously – increased chlorine production, synthetic glycerol production, FT processing of natural gas or coal or an increase in annual oil production, and synthetic fat production. This makes this option appear like a poor choice as a major food source.

### 9.1.2 Glucose synthesis from $CO_2$

To my knowledge, carbohydrates have not been abiotically synthesized at an industrial scale, but there are several proposed mechanisms for production of glucose from $CO_2$, a few of which have been achieved in laboratory settings. The most promising route appears to be to synthesize formaldehyde from $CO_2$ and water (with the hydrogen in the water first being split into protons and electrons through an anodic reaction), and then to use the formose reaction to produce glucose from the formaldehyde.[32] A review of the energy efficiency of this electrosynthesis process estimated that an efficiency of ~20% may be achievable. This was also the maximum conversion efficiency estimated by ALLFED in their review of this technology as a possible food source for space missions. Importantly, ALLFED found that a ~4% conversion efficiency may be more realistic, and the other review article similarly noted that much lower conversion efficiencies were probably more likely in practice.

The US produces ~4,300 tWh of electricity each year. This is equivalent to about $3.7 \cdot 10^{15}$ kcals. Since feeding everyone in the US 2,100 kcal/day for a year requires ~$2.5 \cdot 10^{14}$ kcals (not accounting for waste), a 20% energy efficient process would require the use of ~34% the amount of electricity currently produced in the US to feed all 332 million Americans. This reflects a maximum energy conversion efficiency, and a more realistic 4% efficiency would require ~172% of all US annual electricity production to meet US calorie needs. This means that, while carbohydrate electrosynthesis may not be energetically out of reach as a large-scale food source, it would require a conversion process that operates near maximum theoretical efficiency.

However, even if carbohydrate synthesis could be done at scale without exceeding overall US electricity production, the process may be limited by various material input requirements.

Glucose is ~7% hydrogen by mass, and the proposed electrosynthesis processes use hydrogen gas ($H_2$) as the hydrogen source. Since 64 Mt of carbohydrates would be needed to meet all US calorie needs (assuming no food waste) and there is a ~30% $H_2$ selectivity in glucose electrosynthesis, making enough sugar to feed the US would require ~14 Mt of $H_2$ gas per year. This is ~140% of current US annual hydrogen production. This level of increase is probably achievable in a short period of time, but may pose a scale up challenge to synthetic carbohydrate production.

Another input requirement is calcium hydroxide, which helps catalyze formaldehyde production from $CO_2$. Assuming 60% formaldehyde selectivity when synthesizing sugar, and using the reported concentrations of $Ca(OH)_2$ and formaldehyde from the same study, making enough sugar to meet US calorie needs would require almost 177 Mt of $Ca(OH)_2$ per year, well beyond the current annual production of ~3 Mt.

Another barrier to large-scale production might be the capture of $CO_2$. The previously mentioned study of sugar electrosynthesis methods was unable to experimentally achieve a selectivity of $CO_2$ above 3% when making glycolaldehyde, an intermediate in the production of glucose. At this low selection efficiency, making 64 Mt of glucose (which is 40% carbon by mass) would take over 3 Gt of $CO_2$. This is over half of all US $CO_2$ emissions (5 Gt), and it is nearly 20x more than the ~230 Mt of $CO_2$ that the US currently captures each year. The authors report other methods from the literature that might have significantly higher $CO_2$ selectivity, the highest of which (~63%) would put total $CO_2$ requirements at an achievable ~150 Mt/y. However, they were not able to replicate these methods. This makes me skeptical of carbohydrate electrosynthesis methods, but with some refinement and increased reproducibility, it seems possible that these could be a viable large-scale food production method.

---

[32] Catalyzing the formose reaction also requires small amounts of glycolaldehyde, which is also produced from $CO_2$, but less efficiently

While glucose is the simplest carbohydrate, other simple edible carbon compounds may also be produced from $CO_2$. In ALLFED's review, they looked at the potential use of glycerol synthesis as a food supply for space exploration and found that ~31 to 148 kWh are needed to produce 1 kg of glycerol. Since glycerol has an energy content of ~4 kcal/g, meeting all US calorie needs would take a little under 64 Mt of glycerol per year (same as sugar), ignoring food waste. ALLFED's estimates suggest that producing this much glycerol would require ~46% to 219% of US annual electricity. This again shows that this synthesis route may not be energetically out of reach for the US, but it would still require a massive fraction of all current electricity, and seems less desirable than the microbial food production options reviewed in appendix 9.2 Microbial food grown on non-plant substrates.

### 9.1.3 Protein synthesis from methanol and propene

Only one of the nine essential amino acids, methionine, required by humans is currently synthesized abiotically. Methionine is made from methanethiol, a gas made from methanol, and acrolein, a liquid made from the petrochemical propene (also known as propylene). The US currently produces ~246 kt of methionine each year, enough to provide ~0.4% of US calories, assuming no waste. If methionine could somehow be the country's only food source, its annual production would have to increase by ~260x.

I estimate that producing one kg of methionine requires around 0.23 kg of methanethiol. This estimate comes from the fact that China uses 70 kt of methanethiol to make 300 kt of methionine per year. Currently, the US makes ~150 kt of methanethiol per year, so meeting all US calorie needs with methionine production would require an increase in methanethiol production by ~100x. Methanethiol is made from methanol, with a max yield of ~92%. This means that ~0.25 kg of methanol are needed per kg of methionine. Since the US produces about 5.7 Mt of methanol, it would have to increase its production of methanol by ~3x to make enough methionine to meet all US calorie needs.

Globally, about 90% of acrolein is used to make methionine. Since annual production of acrolein is about 483 kt, and global methionine production is ~1.6 Mt, I infer about 0.27 kg of acrolein are needed to produce 1 kg of methionine. This means that the US would need ~17 Mt to make enough methionine to feed its entire population. Current US production of acrolein is only ~170 kt, so annual production would have to increase by ~125x.

### 9.1.4 Note on electricity-driven processes

Beyond using electrosynthesis to make glucose from $CO_2$, several other options for using electrosynthesis to make food have been proposed. As noted above, US annual calorie requirements are ~2.5 · $10^{14}$ kcals, and electricity production is equivalent to 3.7 · $10^{15}$ kcals per year. Electricity-to-food techniques therefore need a fairly high energy conversion rate to work, and this paper finds a conversion rate of 4–8% depending on the metabolic process used, so slightly higher than the 3% found for $CO_2 \rightarrow$ glucose production from above. Assuming the upper end of that range (8%), the US would need to use 86% of its electricity to generate enough calories. Therefore, I am generally skeptical of electrosynthesis methods of calorie production, but I haven't looked into them in much depth yet.

## 9.2 Microbial food grown on non-plant substrates

Here I review a few carbon sources for microbial food production that are not made from plants. The three substrates that I look at — natural gas, $CO_2$, and methanol — have all been used to produce single-cell protein (SCP) at least on a pilot scale, and methane and methanol have been used at an industrial scale. I could therefore find a fair bit of information about these processes. Other substrates, like peat, diesel, plastics, paraffin wax, and crude oil, have also been proposed and studied at a lab scale, but little information is available. I think that some of these less well studied substrates have some potential to be useful for societal collapse scenarios because they likely require less complicated and bespoke equipment than natural gas, $CO_2$, or methanol fermentation. I've included a brief discussion of them in a separate section, appendix 18.4.

### 9.2.1 SCP from methanol

In the 1980s, the UK's Imperial Chemical Industries (ICI) had a large methanol-fed SCP facility, with a then state of the art 500,000 L fermenter. ICI shut down its SCP production, which was targeted at the animal feed market, as economic conditions changed. However, this provides good precedent for large scale methylotrophic SCP production.

Producing one kg of SCP from methanol requires a little under 1.8 kg of methanol. If methanol-fed SCP had the same kcal/kg content as MOB SCP such that the same 73 Mt/year was needed to meet US calorie requirements, the US would need to use about 131 Mt of methanol annually. The US currently makes about 5.7 Mt per year, so this would require a ~23x increase, making this an unlikely source for large scale calorie production.

### 9.2.2 SCP from $CO_2$

$CO_2$ can also act as a carbon source for SCP cultivation, usually by growing hydrogen-oxidizing bacteria (HOB)[33]. There is relatively little information on this route for SCP production, so I base this section mostly on one life cycle assessment (LCA) paper. Encouragingly, however, an SCP company using captured $CO_2$ was recently approved to enter the market in Singapore, and is waiting approval by US regulators.

The paper found that the microbial mass was 70% crude protein and 10% lipids, suggesting that it probably has ~3,700 calories/kg. This means that if it was the only food source, a person would consume ~207.2 kg per year, and the US as a whole would need to produce ~67 Mt per year (not accounting for waste). The feedstock used in the study is captured $CO_2$. Given the stoichiometry provided in the paper, the US would need about 121 Mt of $CO_2$ per year. Currently, the US captures ~46.7 Mt of $CO_2$ per year, or ~39% of the $CO_2$ needed to make enough $CO_2$-fed microbial food, if all of the currently captured $CO_2$ could be used for SCP cultivation.

Hydrogen gas is also required for $CO_2$-fed microbial food production, and the US would similarly have to increase production of this input to meet all US calorie needs with this food option. The US currently produces ~10 Mt of $H_2$ gas per year. About 60% of this production occurs at oil refineries, 30% comes from ammonia synthesis, and around 10% comes from the production of syngas — a mixture of CO and $H_2$ — from natural gas. Since making a tonne of $CO_2$-fed microbial food requires ~0.44 tonnes of $H_2$ gas, meeting all US kcal needs with this food would require 3x current $H_2$ production. This presents another potential bottleneck to production. Producing all US calories from HOB SCP therefore requires increasing $CO_2$ capture and $H_2$ production substantially. The table below shows the amount of different inputs to $H_2$ production that could be used to achieve this increase, as a fraction of current US annual supply.

---

[33] One company, Solar Foods, uses species of Xanthobacter, but other species can be used as well.

| $H_2$ production source | Conversion rate (t or kWh source/t $H_2$) | % US source needed to make enough $H_2$ |
|---|---|---|
| NG | 3.6 | 12.1 |
| Coal | 7.6 | 42.7 |
| Electricity | 58,800 | 41.4 |

The table shows that the best feedstock for $H_2$ gas production in the US is probably natural gas, and that a similar amount would have to be used as would be needed for methane-fed microbial food (9.2.3). Given that additional steps are needed to produce the feedstock needed in this case, I evaluate that $CO_2$-fed SCP is likely slightly worse as an option for large-scale food production compared to natural gas-fed SCP, but both options may be suitable.

### 9.2.3 SCP from methane

Methane can act as a carbon source for many species of microbes, called methanotrophs or methane-oxidizing bacteria (MOB).[34] Some methanotrophs are already used as food and feed, generally as SCP. Natural gas is ~90% methane and is the most abundant source of methane that could be used to cultivate MOB SCP.

Annual US natural gas production is around 882 Mt. Going with the ~95% methane content from above, this means that US annual methane availability is ~794 Mt. However, most sources report methanotrophic yields relative to natural gas, so the relevant figure for determining US feedstock availability is the 882 Mt of annual natural gas production.

One company producing methane-fed SCP, Unibio, shows that their product has an energy density of ~3,892 kcal/kg. Given this calorie content, if MOB SCP was the US's only food source, the country would need about 64 Mt of MOB SCP per year, without accounting for waste. With 13% waste, this rises to about 73 Mt. Unibio uses ~1.52 kg of natural gas to make 1 kg of MOB SCP, so making 73 Mt of MOB SCP requires a little under 112 Mt of natural gas, or ~13% of annual US natural gas production.

For a few calculations of the inputs required to make microbial food grown on natural gas, see tab 10 of this Google Sheet.

---

[34] The company Unibio currently produces MOB SCP for livestock feed using *Methylococcus capsulatus*, but many other species could be used as well.

## 9.3 Indoor agriculture

### 9.3.1 Indoor crop cultivation

One of the most straightforward ways to try to decouple food supplies from outdoor plant growth is to grow plants indoors. This could be done either in buildings with opaque walls and ceilings so that the only light source is artificial, or in structures with transparent walls that let in sunlight, forming a greenhouse. The advantage of the former, which I'll call indoor agriculture, is greater environmental control, greater ability to exclude pests or pathogens, and the ability to have multistory farms to cut down on space and construction requirements. Conversely, greenhouses require less non-solar energy for plant growth and are likely easier to construct. In both cases, however, it appears that the energy requirements for crop cultivation are probably prohibitively high for large scale food production.

For greenhouses, I used a [study](#) that modeled resource use in greenhouses in different environments and chose the most favorable energy use finding. The greenhouses modeled in the study had HVAC systems, dehumidifiers, and high internal $CO_2$ concentrations. In the most favorable conditions modeled by the study, the authors found a kWh/kg of dry weight use of ~70. Looking at one staple crop that is widely grown in the US, corn, feeding all Americans would require a little under 70 Mt of food per year, ignoring waste. If it took ~70 kWh of electricity per kg of corn produced, this would require ~4,880 tWh of electricity, or ~114% of what the US [currently produces](#) each year.

To estimate the energy that would be used for indoor agriculture, I drew on a [study](#) that modeled the hypothetical attainable wheat yield in an optimized indoor growing facility. Wheat has a calorie density of 3,320 calories/kg dry weight, so the US would need to produce almost 77 Mt per year (~71% more than it [produced](#) in 2021) if it was the only food source. The study found a kWh/dry kg requirement of 393.3 which, if scaled to 77 Mt/year, would mean ~7x the current US [production](#) of electricity.

In addition to requiring too much electricity to be used at a large scale, indoor agriculture and high greenhouses would also need large amounts of materials that are not currently produced in sufficient quantities in the US, putting this production option further out of reach.

### 9.3.2 Microalgae

Microalgae, such as cyanobacteria or green algae, convert sunlight to biomass with an efficiency [3–4x](#) higher than that of most terrestrial crops. They also have a harvest index, the fraction of biomass that is edible, near 1, whereas [major crops](#) have harvest indices maxing out near 0.5, except for tubers which can be over 0.8. This, and the relative nutritional completeness of some microalgae species, makes these plausible alternative food candidates. While growth on artificial light appears to be far too energy intensive for large-scale production, microalgae can be grown fairly efficiently in transparent tubes that let in sunlight. A review of the literature shows that under some assumptions, current US nutrient fertilizer production is sufficient to grow enough microalgae to meet all US calorie needs, but more pessimistic estimates suggest that the US would have to substantially increase its annual fertilizer production. In all cases, it appears that the US would need to increase its production of one or more materials needed to construct microalgae cultivation facilities. Given my criteria of excluding food options that require more materials to produce than the the US currently makes in a year, microalgae are likely a poor choice for large scale production. However, at least under the more optimistic assumptions about nutrient uptake and recycling efficiency that I found in the literature, it seems that microalgae could be a plausible contender for large-scale food production that is not reliant on traditional agriculture.

Microalgae species can be grown outdoors, often in artificial ponds with circulation to ensure high exposure to light and nutrients. However, these growing environments may not be secure in some of the disasters considered in this report. One option would be to grow algae fully indoors, using artificial light. Despite being more efficient at turning light into edible calories than land crops, however, using artificial light to grow algae is still too energy intensive to be feasible at a large scale, given current US energy supplies. Even with a 4% conversion efficiency from light energy to chemical energy, and ignoring all losses to heat and other energy requirements for cultivating algae indoors, providing enough artificial light to produce enough algae feed the US would require using ~1.7x more electricity than the US currently produces,[35] so this approach appears unreachable in the near future.

Algae can be grown in closed, transparent tubes called photobioreactors (PBRs) that may provide protection from outdoor threats but still allow algae to grow on natural light so that less electricity is needed for cultivation. There are several different types of PBRs that have been studied, but all show that the electricity needed to produce enough algae to meet US calorie needs is probably a feasible amount (<20%). Studies differ substantially in the nutrient requirements that they model as necessary for efficient growth. Some of the variation comes from differences in the species studied, as some species have a higher mineral content as a percentage of their biomass. Some of the variation is also due to different assumptions about nutrient uptake and recycling efficiencies.

The two main sources that I draw on for this section, mostly because of their completeness, are a life cycle assessment (LCA) of *Spirulina* (a cyanobacteria) production in China and a techno-economic assessment (TEA) of *Scenedesmus* (a green algae) production published by the US National Renewable Energy Laboratory (NREL). The table below shows the differences in biomass composition of the species evaluated in the two studies and the differences in assumed nutrient uptake efficiencies. An uptake efficiency of 30% means that for every kg of a nutrient added to an algae PBR, only 0.3 kg are used and 0.7 kg are lost. For uptake efficiency, the LCA of spirulina production provided two estimates, one which matches present processes and one that the authors believe could be achieved through further refinement and optimization of algae cultivation practices. I show a range to reflect this. For percent biomass composition by different nutrients for *Spirulina*, I found a source that provided typical ranges for cyanobacteria, so I show a range. For *Scenedesmus* I found a source that provided point estimates, so I do not show a range.

| Nutrient | *Spirulina %* *biomass* | *Scenedesmus* % biomass | Uptake efficiency assumed in LCA (%) | Uptake efficiency assumed by NREL (%) |
|---|---|---|---|---|
| Nitrogen | 9.6 | 1.8 | 28 to 70 | 83 |
| Phosphorus | 0.3 to 1.2 | 0.2 | 3 to 30 | 83 |
| Potassium | 1.1 to 2.4 | 0.6 | 12 to 27 | 83 |

---

[35] One kcal is equal to ~1.16 watt hour (Wh). Since feeding the whole US, ignoring waste, requires ~2.5 * $10^{14}$ kcals per year, this means that feeding the whole US requires an amount of energy equal to ~296 terawatt hours (tWh). The US currently produces ~4,300 tWh of electricity, so a 4% conversion of electricity to kcals in the form of microalgae means that feeding everyone in the US for a year would take ~1.7x the amount of electricity that the US currently produces each year.

These differences in organism and uptake efficiency assumptions mean that the two studies show very different total nutrient requirements for algae cultivation at the level needed to feed the whole US. Staying focused on these three major nutrients, the table below shows the percent of current annual US nutrient production needed to make enough of *Spirulina* or *Scenedesmus* to meet all US calorie needs, based on the same two studies. For *Spirulina*, I went with the high uptake efficiency provided by LCA authors as their estimate for what could be achieved in an optimized process. Data on US nutrient production comes from [FAO](#).

| Nutrient | % of US current annual production needed for Spirulina production | % of US current annual production needed for Scenedesmus production |
|---|---|---|
| Nitrogen | 81.6 | 9.8 |
| Phosphorus | 131.7 | 9.2 |
| Potassium | 3,472.0 | 199.1 |

In both cases, the US would have to increase its potassium production, but going with the more pessimistic estimates from the LCA on *Spirulina* production, the US would need a nearly 35x increase in annual nutrient potassium production, which may pose a serious bottleneck to rapidly scaling up production.

The NREL study reviews several PBR designs. The authors describe the material and energy requirements per unit area and volume of cultivation space. Combining this with their yield estimates, the tables below show how much of different inputs would be needed to make enough microalgae to meet all US calorie needs, as a percent of current US annual supply of each input.

Tubular PBR

| % Electricity | % low density polyethylene | % Steel (for pipes) |
|---|---|---|
| 3.6 | 566.1 | 7.8 |

Helical PBR

| % Electricity | % Glass | % Steel (for pipes) |
|---|---|---|
| 19.0 | 3,933.8 | 6.3 |

Flat panel PBR

| % Electricity | % low density polyethylene | % Steel (for pipes) |
|---|---|---|
| 5.3 | 214.8 | 7.0 |

For all designs described by NREL, the US would have to use more PBR construction material — either glass or polyethylene material — than it currently produces in a year. A several-fold increase in polyethylene production may well be achievable, but combined with the uncertainty about possibly needing substantial nutrient production increases to make microalgae cultivation feasible, I conclude that microalgae may be a poor choice as a major food source. However, going with the more optimistic estimates for nutrient requirements, they appear to be a plausible contender for a nonagricultural food production option that could scale quickly.

# 10.0 Limits to an "easy wins" approach to evaluating food production options

In the previous section, I detailed some of the food production strategies that I considered in this report, and showed why I think most are not a good choice for scaling up quickly in response to a massive loss of agricultural production potential. For a quick assessment of a technology's potential, I looked at the major material and energy requirements for producing enough of one food to meet all US calorie needs. I call this approach "easy wins".

My approach to looking for highly scalable food options led me to focus this report on natural gas fermentation. This is because the US only needs a fraction (~10%) of its current annual natural gas supply to meet its yearly calorie needs, and it doesn't have to use too high a proportion of its electricity in the production process (~6%). I stand by this approach and my findings as a useful first pass assessment. However, my findings should be taken as providing a plausible proof of concept that there are food production options in a loss of agriculture scenario, rather than as a prescription for an optimal response to any specific disaster. Compared to a fuller account of how societies might respond to a severe attack on outdoor crop production, the approach that I've taken in this report has three main limitations, since this report does not thoroughly study:

1. Trade-offs between needing to increase the supply of the main feedstocks for food options and needing to increase the equipment supplies needed for the food
2. Possible combinations of foods that could minimize nutritional issues, production start-up time, costs, and material requirements
3. Different threat scenarios

To give an example of the first limitation: While promising, fermentation using natural gas as a feedstock would require massive amounts of bespoke equipment, and rapid scale-up could therefore face significant manufacturing bottlenecks. By comparison, even though there technically is not enough nutrients currently produced in the US to meet all its caloric needs with cultivated algae, the equipment needed to produce that algae may be more easily manufactured. Therefore, in at least some scenarios, algae may be more promising than natural gas fermentation.

As a proof of concept of the ability to produce food without agriculture, I focused on the ability to meet all calorie requirements with just one food product. However, even if most options cannot readily scale to meet all US calorie needs, producing small amounts of different foods might make a response more robust to errors in the production process of one type of food. Having a portfolio of different foods might also reduce bottlenecks in scaling up food production from a single food technology by reducing the need for some hard to produce pieces of equipment. Access to multiple foods may also make meeting nutrition needs easier. On the flip side, economies of scale might favor repeating a few manufacturing and construction processes many times, so it shouldn't be taken as obvious that more foods are better, but a more thorough investigation of a response to a loss of agriculture would examine an optimal portfolio of foods. I did a very quick investigation of how access to different foods may improve diet quality (appendix [12.0](#)), and I also looked briefly into how natural gas-fed microbial food might be supplemented with sugar and or sugar-fed fungal or bacterial foods in scenarios where a society has access to dead plant biomass (appendix [15.0](#)), but these investigations were not the focus of this report.

This report assumes a very broad loss of agriculture scenario, mentioning only a few variations — e.g., access or no access to dead plant biomass, collapse or preservation of aquatic ecosystems, different timings for an attack. However, I have not considered how different scenarios would impact an optimal response. This is outside the scope of this report, but worth addressing in future work.

## 11.0 Methane-fed microbial food nutrition

The table below shows the nutritional profile of SCP made through natural gas fermentation.[36] The data on nutrition come from [Unibio](#), and the minimum and maximum safe consumption estimates come from this ALLFED [paper](#). I have labeled nutrients in green when they are within the acceptable range, yellow when they are below, red when they are above, and have left many nutrients uncolored for which data was not provided. It is possible that some of these are actually provided in sufficient quantities, but for now I was unable to find data on these categories.

| Nutrient | Unit | Amount/2,100 kcal | Min | Max |
|---|---|---|---|---|
| Protein | g | 380.9 | 46 | 400 |
| Fat total | g | 52.9 | 35 | 200 |
| Saturated fats | g | | 0 | 200 |
| Omega-3 (ALA) | g | | 1.35 | 5 |
| Omega-6 (LA) | g | | 10 | 23 |
| Fiber | g | 3.7 | 25 | 500 |
| Vitamin A | µg | 161.9[37] | 500 | 8,000 |
| Vitamin E | mg | 2.7 | 10 | 1000 |
| Vitamin D | µg | | 3.8 | 15,000 |
| Vitamin C | mg | | 28 | 4,000 |
| Thiamine (B1) | mg | 6.5 | 0.88 | |
| Riboflavin (B2) | mg | 39.4 | 1.2 | |
| Niacin (B3) | mg | | 12 | 3,000 |
| Vitamin B6 | mg | | 1.3 | 500 |

---

[36] The species used for the product is *Methylococcus capsulatus*.

[37] This may well be an overestimate. The Unibio data shows that there is < 1 IU of vitamin A/g of SCP. 1 IU/g translates to 300 µg/kg. It is possible that vitamin A levels are actually much lower, or even absent and the reported "< 1 IU" just means that the level is below the minimum level for detection.

| | | | | |
|---|---|---|---|---|
| Vitamin B12 | µg | | 0.9 | |
| Vitamin K | µg | | 70 | |
| Folate (B9) | µg | | 160 | |
| P. acid (B5) | mg | | 5 | |
| Calcium | mg | 2,536 | 500 | 16,500 |
| Iron | mg | 116.5 | 13 | 1,240 |
| Magnesium | mg | 1,133.1 | 325 | |
| Phosphorus | mg | 5,611.5 | 550 | |
| Potassium | mg | 1,996.4 | 3,500 | |
| Sodium | mg | 485.6 | 2,000 | |
| Zinc | mg | 9.2 | 9.5 | 40 |
| Copper | mg | 49.1 | 0.9 | 10 |
| Manganese | mg | 0.5 | 2.05 | 11 |
| Selenium | µg | 9.2 | 55 | 400 |
| Iodine | µg | | 150 | 3,000 |

The table above shows that a diet consisting of only methane-fed SCP, going with the products that are currently produced, would provide a possibly dangerous amount of copper, and would be deficient in essential fatty acids and several necessary vitamins and minerals. Appendix 12.0 covers how a predominantly methane-fed SCP diet could be improved, either through changing the gas fermentation product or supplementing with other food options. Several of the deficiencies shown above are probably fairly easy to address without producing other foods. Vitamin D can generally be synthesized by the body with adequate exposure to sunlight, and the minerals (zinc, manganese, selenium and iodine) can be provided through mineral supplements that do not require any biological processes, so these are not covered in the following section.

One issue not listed above is excess nucleic acid content. By weight, Unibio's SCP product is ~7.2% nucleic acid (DNA and RNA). This poses a health risk since a high consumption of nucleic acids produces uric acid, a precursor to gout and kidney stones. Nucleic acids are not generally removed from SCP products when it is used as livestock feed, because fish and pigs can process uric acid and suffer no adverse health effects from high nucleic acid consumption. For direct human consumption, nucleic acid levels should be reduced. Daily consumption above 4 g is not recommended, however higher consumptions have not been studied, and this upper bound may be set conservatively[38]. Given the calorie density of Unibio's product, this means that nucleic acid content should not exceed ~0.7% of mass. There are several methods for reducing nucleic acid content in SCP. Since this is a problem with many microbial foods, techniques for nucleic acid reduction have been used for decades by companies like Quorn that have long delivered safe products to consumers that initially contain similarly high levels of nucleic acid. In my account of the equipment and energy needed for SCP production, I assumed that a heat treatment process is used to bring nucleic acid levels to a safe range.

---

[38] Meat and fish generally have an NA content of ~0.15 to 0.8% by mass. At the upper end of this range, a maximum of ~500 g could be consumed before exceeding the recommended maximum of 4 g of nucleic acids. Meat and fish range from ~1000 to 2500 kcal/kg (eg white fish is 1340 kcal/kg and chicken is 1070 while pork is 2630). This means that eating > about 500 to 1250 kcal of meat or fish per / day could put someone over the upper recommended nucleic acid consumption, if the product was 0.8% nucleic acid. Certainly there is precedent for populations eating this much or more meat per day, but it's possible that it was generally lower in nucleic acid content.

# 12.0 Options for improving or supplementing methane-fed microbial food diet

A diet made of only methane-fed microbial food, as it is currently produced, would have several important nutritional deficiencies and potential toxicity. I believe that these issues can likely be resolved, especially with some pre-disaster preparation and research. Addressing the nutritional challenges of a diet could likely be done through a mix of supplementing the diet with other food sources that could still be produced in disaster, and changing the nutritional profile of methanotrophic bacteria.

We have reason to expect MOB SCP products could be improved for human consumption. While producers have investigated options for human consumption, and one human-edible HOB SCP product is currently on the market (in Singapore), the vast majority of SCP is used for high protein supplements for livestock feed. Since little effort has gone into making methanotrophic bacteria a complete source of human nutrition, I expect there to be a fair amount of room for improving their nutritional quality. This is especially true for methanotrophs since there are varieties that can produce many different vitamins, and methanotrophs have been investigated as cell factories for various products.

I also do not expect methanotrophic bacteria to be the only food source in a post disaster diet. The exact portfolio of foods available will depend on the nature of the threat, and a thorough assessment of the optimal food selection – e.g. to minimize nutritional issues, costs, equipment or material bottlenecks, etc – is beyond the scope of this report (see appendix 10.0 for a discussion of the limitations of the approach taken in this report). However, below I suggest a few options for supplementation, and suspect that others may be possible.

My quick investigation of options for improving the nutrition of a post disaster diet suggest that there are very likely ways to make a nutritionally complete diet given the constraints of the disasters considered in this report. The key issue will be whether these can be scaled up in time. This depends on both how long it takes to start producing food — for bulk calories and for complete nutrition — and how long before nutritional deficiencies from a potentially inadequate diet begin to cause severe health problems. It is possible that bulk calorie production can be ramped up to meet US calorie needs in time, as my other work presented here suggests, but that additional time will be needed to make a more nutritional mix of foods. The time budget that the US would have to begin producing a nutritious diet depends on how long it can survive on a diet with nutritional deficiencies and possible toxicity. Getting a clear sense for how long a population can survive on an incomplete diet is an important area for future research.

It should be noted that all of the nutrition findings here are essentially theoretical. While a nontoxic diet that provides an adequate amount of all essential nutrients should be safe, any diet after a total loss of agriculture would be quite unprecedented, and there remains a chance that it could pose unanticipated challenges.

## 12.1 Essential fatty acids

ALLFED's paper finds that humans need 1.35 g of omega-3 and 10 g of omega-6 per day. These nutrients are available in many plant and animal sources, but could be scarce in a diet of microbial food. A cursory review of the literature suggests the consequences of deficiency are fairly severe, though perhaps not fatal. For now, I'll assume that it is necessary to meet these minimum requirements.

 In Unibio's nutritional description of their SCP, they do not include any C18:2 (omega 6) or C18:3 (omega 3). They do, however, list 2.4% of its fatty acid content as being "other" from the types shown. Even if optimistically all of that 2.4% was one or the other essential fatty acids, though, this would only provide ~1.24 g/day in a 2,100 kcal diet. I haven't found much information on the fatty acid content of other methanotrophs yet, and I suspect little work has gone into identifying if any strains do currently or could be engineered/selected to accumulate more essential fatty acids. This paper finds a concentration of omega-6 in methanotrophs to be ~0.6% of cell dry weight.[39] If this level of production could be replicated in an SCP, then this would still only provide ~3.2 g of the 10 g of required omega-6/day, assuming the same kcal/kg content as Unibio's product. I couldn't find any mention of omega-3 in any methanotroph.

After pre-disaster food reserves have run out, the two main ways to get essential fatty acids in a total loss of agriculture disaster would probably be cultivating oleaginous microbes or growing photosynthetic organisms in biosecure facilities, like photobioreactors, depending on the scenario. Below I review a few options in these categories. I find that cultivating some aquatic plants may be the best overall option, but in scenarios where dead plant biomass can be accessed and turned into sugar, growing fungal foods may also be promising.

In scenarios where the threat only affects terrestrial plants — which is also necessary for fish to be a viable food option — it may be possible to supplement a diet with aquatic plants grown outdoors. In cases where the threat precludes this, these organisms may still be cultivable in more biosecure settings, but this would require more material and energy inputs. One option would be to grow seaweed. However, one example species, wakame, doesn't have enough omega-6 or omega-3 to provide these in adequate amounts even if it made up all of a 2,100 kcal/day diet. Similarly, cyanobacteria species like *Spirulina* and *Chlorella* do not have enough essential fatty acids as a fraction of their biomass to be great sources of essential fatty acids.[40] Lemnaceae, colloquially called "duckweed," have much higher essential fatty acid content. Given the density of omega-3 fatty acids in its biomass, eating between 43 to 72 g of Lemnaceae per day (79–105 kcal/day) would provide the minimum daily amount of omega-3. If Lemnaceae can be grown with a similar efficiency as the *Spirulina* (cyanobacteria) investigated in 9.3.2, then producing 75 g per person per day would only take ~12% of the resources that I estimated were needed to feed the US with *Spirulina* grown in photobioreactors.[41] I have not investigated the other inputs to lemnaceae cultivation, but from a high-level first pass, this may be the most appealing option for providing enough essential fatty acids through dietary supplementation.

---

[39] This is based on the concentration of phospholipid ester-linked fatty acids (PLFAs) of 213.26 μmol/g-cell dry weight, this table of the molecular weight of different fatty acids, and table 1 in the paper that shows that omega-6 fatty acid is ~10% of the total PLFA content in the methanotrophs studied.

[40] A 2,100 kcal/day diet of *Spirulina* would provide only ~90% of the minimum omega-6 intake, just getting enough omega-3 requires only 476 kcal. For *Chlorella*, ~1,618 kcal/day are needed to get 10g of omega-6, but just 289 kcal is enough to provide the minimum daily dose of 1.35 g of omega-3.

[41] 12% of the resources required for *Spirulina* production (production at the level needed to meet all US calorie needs) is equal to ~417% of current US annual nutrient potassium production, ~16% of current US annual nutrient phosphorus production, ~10% of current US annual nitrogen fertilizer production, and between ~0.4% and ~2.3% of US electricity. Fewer nutrient fertilizers would be needed under more optimistic nutrient uptake and recycling assumptions, such as presented by NREL (9.3.2).

Another promising option for essential fatty acid supplementation is fungal SCP, probably grown on sugar. A model fungal SCP is Quorn, a meat alternative that has been produced from the fungus *Fusarium venenatum* for decades. Even though it's not especially high in fat overall (only ~32% by kcal), Quorn has about 0.4 g of omega-3 in an 85 kcal serving, so meeting min omega-3 consumption only takes ~289 kcal/day. Quorn also has about 1.4 g of other poly unsaturated fats (PUFAs) in an 85 kcal serving. This means that ~607 kcal/day of Quorn is needed to provide 10 g of omega-6 fatty acid. Note that the other PUFAs could include omega-9, which is not an essential fatty acid. In that case, more Quorn would be needed to meet the minimum daily omega-6 requirement.[42] Quorn has 3,400 kcal/kg, and fungal SCP typically has a conversion rate of 0.31 to 0.50 g/g substrate (i.e., glucose) and a growth rate of 0.74 to 2.33 g/L/h. The following table is based on these values and shows how much glucose is needed per year to produce enough fungal SCP to provide either only enough omega-3 or both enough omega-3 and omega-6. The table also shows how much fermenter volume would be needed in both cases.

| Kcal/person/day | Mt of glucose/y | Million m3 of fermenter space |
| --- | --- | --- |
| 289 | 20.6 to 33.2 | 0.5 to 1.6 |
| 607 | 43.3 to 69.8 | 1.1 to 3.3 |

These values suggest that if the US has access to woody biomass, it could likely make enough fungal SCP to meet its essential fatty acid needs. If the US does not have access to woody mass, then this option is likely not feasible, though there could be some production of glucose through electrosynthesis (9.1.2). As covered in appendix 6.0, the US could probably make ~33.2 Mt/y of lignocellulosic sugar by retrofitting all breweries, ethanol plants, and pulp and paper mills. So even without making any new facilities, the US could probably produce enough sugar to grow enough fungal SCP to meet all of its omega-3 needs. Making more lignocellulosic sugar would require building new facilities, which would take more time and cost more. I've estimated that making enough methanotrophic SCP requires ~3.5 million m$^3$ of fermenter volume to produce all US calories. This represents a similar calorie production per unit volume as the higher end of volumetric yield given above, so if the US can make enough fermentation capacity to meet its calorie needs with MOB SCP, it can likely make the capacity needed meet its omega-3 fatty acid needs from fungal SCP.

There are several other options for providing essential fatty acids through dietary supplementation, but these generally seem inferior to the options provided above.

Overall, it seems that producing enough essential fatty acids is quite challenging. I'm optimistic that more research could solve this problem. However, it seems possible that such research would take a long time, such that it may be difficult to find a solution quickly in the midst of a disaster without prior investigation. It is possible that I'm missing some easy solution, such as a methanotroph strain that produces a lot of essential fatty acids and is also suitable for efficient, large-scale cultivation. If it turns out that people can go without the minimum essential fatty acids for several years without extreme health effects, then I expect that a solution could be found even without pre-disaster preparation.

---

[42] As a separate note, Quorn also has vitamins B5 and B6, which are absent, or at least not mentioned, in methanotrophic SCP, although to get enough B5, someone would need to get most of their kcals (1700/day) from Quorn, and getting enough B6 requires 884 kcal/day.

I also suspect that society may have more time to figure out and scale up an efficient method for essential fatty acid production than it has to scale up alternative food production as a whole. This is because there is a fair amount of essential fatty acids available in stored food. For example, soy has some omega-6 and omega-3, and it represents ~22% of the stored crop calories in the US.[43] In September, the amount of soy stored in the US (~7 Mt) and the amount of soybean oil (~957 kt) contains ~1 year's worth of omega-6 fatty acid, although only ~1.5 months of omega-3. If there is good coordination, soy could be saved during the early part of the disaster (especially if protein and fat needs could be met through animal product consumption), and then eaten slowly later on in the disaster, potentially giving the US extra time to ramp up the options discussed above. Soybeans could also be processed to extract soybean oil, so the carbohydrates and protein could be eaten earlier and the fats saved for later. If fishing is possible, this could be another source of essential fatty acids. That said, fish has to be consumed in fairly large quantities to provide all of a person's essential fatty acids. One needs to eat ~650 kcal of salmon and ~989 kcal of "mixed species white fish" to get 1.35 g omega-3, so I don't expect this to buy much time.

## 12.2 Vitamin A

As noted above, Unibio reports that their SCP product contains "<1 IU/g" of vitamin A. 1 IU/g translates to 300 µg/kg, or ~162 µg/day on a 2,100 kcal diet. However, it's unclear if Unibio's product actually has that much vitamin A. Even if it does, this is still only about ⅓ of the minimum 500 µg daily intake, so this is an important deficit to address. It seems that some methanotrophs produce extremely high amounts of carotenoid, a compound that the human body can turn into vitamin A. A team inserted a gene cluster that massively increased carotenoid production in some methanotrophs — although not the ones used for SCP production — which suggests a path to genetically modifying SCP-yielding methanotrophs to produce more vitamin A. The team increased carotenoid production by up to 20x, resulting in a concentration of carotenoids at 1 g/kg of dry cell mass. Since minimum vitamin A consumption only needs to be 500 µg (0.0005 g), it seems that even if the methanotrophs used for SCP can't be edited/selected to produce so much carotenoid, the US could probably produce a lot as a supplement just by cultivating the strains identified by the researchers. Since the bacteria are also methanotrophs, the infrastructure that the US would build to make SCP would likely be usable to cultivate these organisms.

## 12.3 Vitamin E

Going with Unibio's figures, a 2,100 kcal/day diet of MOB SCP only provides ~27% of the recommended daily intake of vitamin E. I haven't yet found cases where methanotrophs produce large amounts of vitamin E, but it is possible that, as with the case of vitamin A, some strains could be engineered for higher vitamin E production. More research is needed on this front.

## 12.4 Vitamin C

I haven't found any evidence of vitamin C being synthesized by methanotrophs, so it would probably have to come from other sources.

---

[43] In reserves in September.

Thankfully, vitamin C is already produced synthetically in large amounts. The world produces ~95 kt of vitamin C per year. Of this, 80% is made in China. I haven't yet found an estimate for US production, but to provide 28 mg/person/day (the minimum amount), the US would only have to produce ~3.6% of current global supplies. Since the US, along with India, is listed as one of the next top producers after China, it seems reasonably likely that the US already makes this much. To continue doing so in a disaster, the US would need a supply of glucose, but not very much.

Vitamin C production involves the following reactions, with the associated yields.

- Glucose → Sorbitol (~98% yield)
- Sorbitol → Sorbose (~53% yield, calculated from this paper)
- Sorbose → 2-Keto-L-gulonic acid (~50% yield)
- 2-Keto-L-gulonic acid → vitamin C (~90% yield, calculated from this paper)

Total yield from glucose is ~23.6%. This yield suggests that ~14.4 kt of glucose is needed to make the ~3,393 tonnes of vitamin C that the US needs per year. This is far less than the ~33.2 Mt/y of lignocellulosic sugar that I estimate could be made if all brewing, ethanol, and pulp and paper facilities were repurposed. In fact, the US has ~578 kt of stored sugar in reserves. While this only represents ~3 days of kcals for the whole US, it is in theory enough to provide ~40 years worth of vitamin C (assuming no spoilage).

## 12.5 Vitamin K and B Vitamins

Unibio's data did not provide any information on Vitamin K or B vitamins, except for B1, B2, and B12. Based on this and a cursory review of the literature, they may be totally absent from methanotrophs. This source does say that Calysta's FeedKind product (SCP for fish and livestock) contains high levels of "B vitamins", but isn't specific on amounts of types, so may be referring to just B1, B2, and B12. Similarly, no information is provided about the presence of vitamin K.

A future research direction would be to review the existing microbial production routes for vitamin K and various B vitamins (like vitamin B5). For example, one team recently demonstrated B5 production via a heterologous pathway in baker's yeast. Similarly, many organisms have been used to produce vitamin K, although glucose, sucrose, glycerol, starch, and other substrates would be needed to grow them. It is possible that scaling up the production of these vitamins poses a critical challenge to survival, and more research in this area is needed.

## 12.6 Fiber

If there is wood, it seems possible to make soluble fiber dietary supplements. In scenarios without wood, getting the recommended 25 g of daily fiber may be quite hard. There are reports of people surviving exclusively on animal products, which provide no fiber. While far from ideal, these diets appear to have kept people alive for some time, suggesting that people could survive on the meager ~3.6 g of fiber provided each day on 2,100 kcal of SCP. Another option may be using plastic as a replacement for dietary fiber. This approach has been trialed with human subjects before and appears safe, but is fairly understudied.

## 12.7 Note on a high-protein diet

Current methanotrophic microbial foods are extremely high in protein. A diet made of only MOB SCP would provide over 380 g of protein per day, which is well above recommendations. However, it is not outside of historical precedent for high protein diets. It seems that this may be a tolerable level of intake, but there are also several ways to reduce protein intake, both through supplementing a diet with less proteinaceous foods, and by changing the methanotrophic product.

It is not recommended that long-term protein intake exceed about 2 g of protein/kg of body mass, although healthy adults are able to tolerate up to 3.5 g/kg of body weight. I calculated that if US adults reduced their average BMI to 18.5 (lower limit of what is considered healthy), then on average, across men and women, body weight for US adults would be just 53.5 kg. This would put the upper tolerable protein intake at only ~187 g/d. However, some populations, such as the Inuit, have historically had extremely high protein intakes. I have not investigated long-term health outcomes for this population, but this at least suggests that high protein intake is unlikely to be fatal in the short run. One study of historic Inuit diets found that men had an average daily protein intake of 370 g, while women ate an average of 280 g. If 3.5 g of protein/kg of bodyweight were the upper limit, this would suggest that men had an average body weight of over 100 kg, when they were instead found to have an average body weight of ~63 kg. However, their diet was quite different from a 2,100 kcal/day diet of MOB SCP, with higher total kcal/day consumption, meaning that protein was a lower percent of total intake. I suspect that while a high-protein microbial food diet is probably survivable, it likely would lead to various adverse health outcomes for the population, so reducing protein intake would be desirable.

One way to reduce protein content in methanotrophic microbial food may be to reduce the amount of nitrogen used in the growth media. N starvation appears to have a large impact on methanotroph macronutrient production. A group studying the production of PHB (a lipid that can be used as a plastic) increased lipid concentration in cell mass ~2.7x by reducing N levels in the growth medium, and this didn't significantly change the doubling time. As a toy example, if lipid production in Unibio's MOB SCP (appendix 11.0) was doubled from 9.8% to 19.6% of total biomass, and protein decreased by 9.8% down to 60.8%, then total protein on a 2,100 kcal of SCP/day diet would be only ~283 g. This of course assumes a 1-1 trade-off in protein and lipid content, which is likely unrealistic. Still, this example shows a potential route to improving the nutrient content of methanotrophs.

Another option could be to choose methanotrophs with lower protein content. Importantly, existing SCP companies seek to maximize protein levels. Calysta and Unibio report basically the same % protein for their products, 71% and 70.6%, respectively. However, most methanotrophs are naturally 50–65% protein, so one could likely attain a more balanced macronutrient profile by cultivating different methanotrophs. At 50% protein content, daily protein consumption could well fall below 263 g.[44] Methanotrophs investigated as potential sources of oil for biodiesel production have higher lipid content, such as a strain described here that is ~22% fat (as opposed to 9.8% for Unibio and 8% for Calysta), so there may already be precedent for growing other methanotrophs at a large scale.

Finally, protein consumption could also be reduced by pairing MOB foods with other food sources, especially carbohydrates and fats, such as sugars from lignocellulosic biomass or synthetic fats. As discussed elsewhere (appendix 9.0), scaling up these other foods presents real challenges, but even getting 15% of daily kcals (315 kcal/day) from a food source that is either all fat or all carbohydrate would reduce an individual's daily protein intake from 380 g to 323 g.

---

[44] The exact number of g of protein this would imply on a 2,100 kcal/day diet depends on the other macronutrients, but since fat is more calorie dense, having protein make up 50% of the biomass would mean that it is <50% of the kcal content (unless a lot of the biomass is inedible), so the g of protein per day would be <263 (since this is 1050 kcal).

## 12.8 Note on copper toxicity

The SCP produced by Unibio would, if eaten at a level of 2,100 kcal/day, provide about 5x the upper recommended level of copper (10 mg/day). From a brief review, it seems possible that this level is set conservatively and could in fact be exceeded without incurring extreme harms. It may also be possible to produce methanotrophic microbial foods with much lower copper levels.

At least one paper recommends raising the safe upper limit of copper consumption from food to 49 mg/day for adults, which is right at the level provided on a 2,100 kcal/day diet of MOB SCP. The paper uses a mix of case studies of people eating high levels of copper without showing any adverse effects for at least several years, and animal studies (particularly rhesus and capuchin monkeys, but also rodents and others) that also did not measure any adverse effects from significantly higher levels of Cu consumption.[45] One cited study gave participants 10 mg of Cu (in capsules) per day (in addition to their normal Cu consumption) and found that the participants did not have any increase in Cu serum levels, suggesting that the body can maintain Cu levels reasonably, at least within some limits. In the present report, I set average caloric intake to 2,100 kcal/day for US adults, assuming that average BMI would fall to the lower end of the health range at 18.5. This would bring average (across men and women) weight to 53.5 kg given average US height, so 48 mg Cu/day is almost 1 mg Cu/kg body weight. Monkeys were fed 7.5 mg Cu/kg body weight for three years without any observable effects. It is worth noting, however, that the CDC report on Cu toxicity finds that the lowest observed adverse effect level (LOAEL) is generally much lower in humans than in animal models. I suspect that much of this is because the first reported adverse effects are discomfort, which may be hard to observe in nonhuman animals.

Importantly, the few case studies of high Cu consumption appear somewhat inconsistent. For example, one case highlights an otherwise healthy man who ate 30 mg/day and reportedly had to receive a liver transplant as a result. Both the CDC and the National Academies have reviewed Cu toxicity, but neither cited many cases of people consuming much beyond the recommended 10 mg/day. The CDC does note that people who consume large amounts of shellfish may expect to eat ~16.5 mg Cu/day on average, and that this population doesn't display any particular signs of ill health as a result. There was also a case of three Massachusetts towns whose water had 8.5 to 8.8 mg of Cu/L for at least two years, and researchers following the town did not report any Cu-related health problems in the population, even among young children. The daily copper intake that this implies is hard to say, since it depends on water consumption. If someone drank the average recommended 3.2 L of water per day (averaged across men and women), then they'd get around 27–28 mg of Cu/day, and Cu in water is apparently more available than Cu in food, suggesting that these towns had several times the daily upper limit without apparent ill effects. However, the EPA assumes that adults average ~2 L of water per day for the purposes of setting safety guidelines, and consumption appears to actually average closer to 1.5 L, so consumption may have been lower. It's also important to note that the National Academies report says that researchers did not follow up long enough to detect potential long-term damage caused by higher Cu intake.

Even if 10 mg of copper/day is a fairly hard cutoff, some methanotrophic bacteria accumulate copper at a low enough rate that they could be relied on as a major calorie source without exceeding this limit. This [study](#) finds a Cu content per kg of just 1.7 mg, so a 527 g serving (the amount needed to get 2,100 kcal going with Unibio's kcal content) is just 0.9 mg, which is actually lower than the recommended consumption. This suggests that some careful selection or engineering could produce a product that has safer levels of copper. The key will be identifying a strain that can still grow fast enough, and can assimilate other nutrients efficiently enough, to produce food at a similar level per unit effort as assumed throughout this report.

---

[45] The animal studies provided animals with copper intakes that were higher than are recommended for humans, both relative to the animals' bodyweight and in absolute amounts

# 13.0 Methane-fed microbial food production facilities

Above, I find that the major feedstock for methane-fed microbial food production, natural gas, is available in ample supplies to allow the US to feed itself without increasing natural gas production. However, producing enough microbial biomass to meet US calorie needs also requires many new facilities, bespoke equipment, and a nutrient-rich media for microbial fermentation. All of these have to be constructed, manufactured, or sourced quickly so food production can begin within the ~22 months that the US has before existing food reserves run out (assuming worst-case scenario timing of a disaster, but also assuming that supplies are well managed and rationed).

Here I show that we should not expect a well coordinated effort to scale up methane-fed food production to be bottlenecked by energy input requirements, materials needed for construction or the nutrient media, or manufacturing or construction capacity, given existing US infrastructure and material supplies. As a first pass, I checked these possible bottlenecks through an independent literature search. Afterward, while at Open Philanthropy, I collaborated with Synonym Bio, a financing and development platform for bio manufacturers, to make a techno-economic assessment (TEA) model of SCP production using natural gas as the carbon source. In the following sections, I'll briefly mention the approaches that I took on my own and some of my initial findings, but I will generally rely on the outputs of the [TEA model](#).

## 13.1 Facility construction and operation costs

The total cost of making and operating enough methane-fed SCP production facilities to feed the US provides a high level indication of the scale of resources that society would need to pull off a complete replacement of agricultural food sources with nonagricultural ones. For these cost estimates, I initially referenced [projections](#)[46] provided by Unibio on their proposed 100 ktpa plant, which put capital expenditures (CAPEX) at ~$251 million and annual operating expenses (OPEX) at ~$63 million. Scaled up to the number of plants needed to feed the US, this would represent a total CAPEX of ~$184 billion (~0.7% of [US GDP](#)) and an OPEX of ~$46 billion (~0.2% of US GDP). I used several other approaches to get different estimates, but overall the figures provided by the Synonym Bio TEA are probably the best estimate.

Synonym Bio's team constructed a model of an SCP plant making ~100 ktpa of food-grade product where natural gas was the carbon source.[47] The model incorporates a complete process flow diagram for SCP production, and from this provides a list of all equipment needed for a full plant (see my discussion of equipment requirements in appendix 13.5). Synonym's team then contacted US vendors for each piece of equipment, or used their existing database, to get cost estimates for equipment, scaled to the size appropriate for a 100 ktpa facility. Adding these costs together gives a total purchased equipment cost estimate. This can then be used to estimate the overall cost of a facility through a Lang Factor approach, where data on the CAPEX breakdown from other similar plants provides a typical ratio of purchased equipment costs to other costs (e.g., for a plant that requires $X in purchased equipment, it will typically cost $0.25X in electrical installations, $0.40X in piping, etc.). When the approach taken by the Synonym Bio team has been applied to other projects that were then actually built, cost estimates have proven to be accurate within roughly a ±30% range. In this case, the model shows a total CAPEX of ~$752 million per plant. Accounting for plant downtime, product losses during recovery and processing, and other such factors, the model plant would produce a little over 104 kt of SCP product a year, meaning that the US would need about 704 such plants to meet its calorie needs, given the calorie content of current SCP products (~3,982 kcal/kg, according to Unibio). This puts the total cost at ~$530 billion, or ~1.9% of US GDP.

---

[46] These data were also used by ALLFED in a [paper](#) with a similar focus as this report.
[47] Synonym also developed a model for a $CO_2$- and $H_2$-fed SCP plant since the technologies — and the models — are quite similar. As discussed elsewhere, though, I find that $CO_2$ is a less available carbon source for large-scale microbial food production.

The Synonym TEA shows an annual OPEX for a model plant of ~$184 million. This figure is based on the cost of different inputs, such as electricity and nutrients, which in turn are determined by various operating parameters for the plant, such as microbial growth stoichiometry, metabolic heat production and fermenter chiller energy needs, etc. The OPEX also assumes a certain amount of labor, maintenance and repairs, and other costs that are based on the requirements of reference facilities. Multiplying the OPEX across all 704 facilities that the US would need to meet its calorie requirements provides an annual OPEX of ~$130 billion, or ~0.5% of US GDP.

In the tables below, I show the contribution of different cost drivers for CAPEX and OPEX for 704 large SCP production plants, and compare these to the size of some relevant reference industries in the US to provide a sense of scale of the resources required for SCP facility construction and operation. Note that I do not show all component costs, just a few select components. More granular comparisons of CAPEX and OPEX requirements to current US spending on reference industries are made in follow sections.

CAPEX

| Cost Component | Cost (billion $) | Reference Industry | Reference Industry Size (billion $) | Cost of equipment or service as a % of reference industry annual sales |
|---|---|---|---|---|
| Total CAPEX | 529.8 | CAPEX of US manufacturers | 179.1 | 295.8 |
| Purchased Equipment | 116.5 | (Sales of) Fabricated metal product manufacturers | 343.2 | 33.9 |
| Construction | 179.4 | (Sales of) Nonresidential building construction sector | 458.5 | 39.1 |
| Engineering | 33.1 | Engineering services | 367.8 | 9 |

OPEX

| Cost Component | Cost (billion $) | Reference Industry | Reference Industry Size (billion $) | Cost of equipment or service as a % of reference industry annual sales |
|---|---|---|---|---|
| Total OPEX | 129.6 | OPEX of chemical manufacturers | 64.5[48] | 200.9 |
| Media nutrients | 27.2 | Fertilizer sales | 26.2 | 103.8 |
| Wages | 6 | Wages of chemical plant workers | 56.1 | 10.7 |
| Maintenance | 6.3 | Maintenance expenditures by chemical manufacturers | 8.5 | 74.1 |

## 13.2 Electricity

One reason that large-scale food production without agriculture is challenging is the need for high levels of energy inputs. Above I found that this was the main bottleneck for options like indoor agriculture. For methane-fed microbial food, the energy for biomass production comes from natural gas, which appears to be in adequate supply in the US to make enough food to meet US calorie requirements. Thus this process is a better option for rapid deployment than food options that use electricity as the main input. A substantial amount of electricity is still needed to operate the equipment required for single-cell protein (SCP) production, but the amount of electricity needed does not appear prohibitive.

As a first pass, I looked at reported literature values of methane-fed SCP production, finding that a kg of SCP was estimated to require ~11.02 kWh of electricity to produce. Given the kcal content of current methane-fed microbial food products (appendix 11.0), and assuming a waste of ~13% (appendix 2.3), providing everyone in the US with 2,100 kcal/day of SCP for a year would take ~809 tWh, or ~19% of current US annual electricity production. However, I believe that the actual electricity requirements are likely much lower based on the results of the Synonym TEA model. By selecting more efficient chiller technologies, the total electricity requirement drops to ~6% of current US electricity production. As discussed in the main text section on methane-fed microbial scale up, historic precedent suggests that this is a feasible amount of electricity to expend on a single national project, and in appendix 13.6 I discuss how economic proxies argue for this being an amount of electricity installation that could be completed within a year given existing US installation capacity.

---

[48] OPEX for food manufacturers (not food producers, i.e. farmers, but rather manufacturers of food products), another potential reference category, is ~$48.3 billion

The Synonym TEA model found that the breakdown of electricity use for SCP production is ~55% for chillers, 28% for an air separation unit (to provide pure $O_2$), and 17% for other plant operations.

## 13.3 Equipment and building materials

Given the scale of the construction required, and US reliance on foreign supplies for certain key material inputs to construction and manufacturing, I wanted to check if the US could autarkically supply the raw materials needed to build enough facilities to meet its calorie needs with natural gas-fed microbial food. Given the scope of this project and time available, I focus on stainless steel, carbon steel, structural steel, and concrete. It appears that current US annual production of all three is more than sufficient to make enough SCP facilities, but the US would have to engage in some recycling or trade to secure enough nickel and chromium for stainless steel.

When I initially looked at this, I focused on the stainless steel required for fermenters, and minerals needed for its production. This was motivated by expert interviews suggesting that stainless steel supplies may be limiting in an autarkic scenario and that fermenters were the largest source of stainless steel demand. I was unable to quickly identify many other material needs for equipment or construction, but I did do a quick calculation to show that the material needs for filtration (a step used to separate microbial biomass from nutrient media) are currently available in adequate amounts in the US. For those interested, I provide those first pass findings in a subsection, but here I focus on the outputs of Synonym's TEA model.

Based on the equipment sizing calculations and industry standards for piping and foundation requirements for processing plants, the table below shows the amount of stainless, carbon, and structural steel, and the amount of concrete, that a 100 ktpa natural gas-fed SCP production plant should require. The table also shows how much of each material would be needed for the 704 facilities required to feed the US and compares this to current US annual production. The choice to compare to annual production is slightly conservative since the US would have more than a year to make enough facilities to meet its calorie needs before reserves run out.

| Material | Amount for a 100 ktpa plant (tonnes) | Amount for the whole US (Mt) | Current US annual production (Mt) | Percent of current US annual production needed |
|---|---|---|---|---|
| Stainless steel | 1,970 | 1.4 | 2.6 | 53 |
| Carbon steel | 934 | 0.7 | 67.1 | 1 |
| Structural steel | 1741 | 1.2 | 3.2 | 39 |
| Concrete | 35,770 | 25.2 | 741.3 | 3 |

The table shows that current US production of major material inputs to equipment and construction are likely sufficient for making enough microbial food plants to meet its calorie needs. However, the production of these inputs relies on raw materials, not all of which the US produces in sufficient quantities.

The main input to concrete is lime, which the US has been a net exporter of in three of the last five years. Exports and imports represent only ~4% of domestic production, so existing US capacity should be more than sufficient.

Structural steel composition varies by grade, but is mostly iron, with trace amounts of other elements. These elements come from the addition of petroleum coke, a by-product of the petroleum industry. Since the US is a net petroleum exporter and a net iron ore exporter, it seems likely that the US has adequate domestic supplies of the inputs to structural steel production. Similarly, carbon steel is made using iron ore and coal, and the US is a net exporter of coal, so it should have adequate supplies to make enough carbon steel. Stainless steel, however, contains significant amounts of nickel and chromium. For the main grade of stainless steel used in the materials considered, nickel makes up 8–10% of the steel by weight, and chromium is 18–20%. Going with the middle values and the amount of stainless steel needed from above (1.4 Mt), the US would need ~263 kt of chromium and ~125 kt of nickel. Currently, the US only produces ~120 kt of chromium, and only through recycling, and ~18 kt of nickel.

It seems that the potential mineral bottlenecks for stainless steel production can be addressed with recycling. In the US, ~95% of the stainless steel in use is recycled once the product that it is in has reached the end of its life. At current production levels, this translates to over 60% of annual stainless steel production coming from recycled steel. Since stainless steel can be 100% recycled without loss of quality, and the US only needs ~53% of its current annual stainless steel production for SCP plants, I think the US will not need primary sources of chromium and nickel.

### 13.3.1 First pass findings on equipment material requirements

My first pass look at steel production showed the same picture as above. I wanted to briefly check domestic supplies of material used in filters to see if these presented a possible bottleneck. From what I could find, the answer appears to be no. I focused on filters mostly because the material needs for other equipment were hard to find, so I do not consider this section of my investigation complete. Notably, another type of equipment that relies on materials that could be hard to source domestically are sensors (for pH, temperature, oxygen, methane, etc.). The US is fairly reliant on imports for much of its electrical equipment, so making these specific devices could be challenging if the US is cut off from international trade, and two experts raised the supply of sensors as a possible issue in a disaster scenario. This remains an outstanding question and I haven't been able to reach clear conclusions as to whether this would be an issue.

For filters, a common design cited as useful for SCP harvesting is a rotary drum filter. This type of filter uses diatomite. Based on the material flow rate through a rotary drum filter ($100$–$2000$ L m$^{-2}$ h$^{-1}$), typical thickness of the diatomite layer ($3$ mm), and diatomite density ($2.08$ g/cm$^3$), the US would need between 57 thousand and 1.1 million tonnes of diatomite to process the needed amount of SCP. The US currently produces 830,000 tons of diatomite per year, so supplies are likely sufficient to cover this, and other materials can be used. Filter material also requires polymer yarn, which can be made from PET or nylon. Typically, cloth thickness is around $0.1$ mm, and the density of PET is $1.39$ g/cm$^3$. Given the filter rate from above, the US should need between 1,270 and 2,540 tonnes of PET to make all of the filtration equipment needed. It currently produces at least 22.4 Mt of PET per year, so supplies should be ample.

## 13.4 Media nutrients

While natural gas provides carbon, methanotrophic microbes require other nutrients to grow. If the US could not provide enough of all nutrients, this could represent a substantial bottleneck to scaling up natural gas-fed food production. In my first pass, I looked at several nutrient media recipes: one from a [patent](#) from the methanotrophic company Unibio, one from the database [MediaDive,](#) and one from a [source](#) on general methanotrophic cultivation. These recipes provide nutrient concentrations per media volume, and so do not directly show how much of each nutrient is needed to produce a given amount of microbial food. I used data on the [chemical composition](#) of Unibio's SCP product to see how much of different nutrients are "taken out" by the production of a unit of biomass. I calculated how much of different nutrients are taken out from the production of all of the SCP needed to meet US calorie needs, and added this to the nutrients in the volume of media needed to produce this much microbial biomass (the volume calculation comes from the [growth rate](#) of 4 g biomass/L h). Since this method assumes zero waste and ~100% uptake efficiency of all nutrients, it is an underestimate of the total nutrient requirements. Still, I include the findings of this first pass approach in the following subsection.

To estimate nutrient requirements per unit of biomass produced, Synonym's TEA model used standard assumptions about waste and nutrient uptake, and data from current SCP producers on biomass loss rates during downstream processing. However, due to variability in nutrient media recipes, they only looked at major nutrients: nitrogen, phosphorus, potassium, calcium, sodium, magnesium, iron, and copper. There are a handful of trace elements, like cobalt and manganese, that this does not include, but these are addressed in my first pass on nutrient requirements. The table below shows the amount of each nutrient required to make enough SCP to meet US calorie needs each year, combining information from the TEA model, the current US annual production of each nutrient, and the percent of current annual supply that would have to be used for natural gas-fed microbial food production if this was the US's only food source.

| Nutrient | Amount needed per year (Mt) | Current US annual production (Mt) | % of current annual production needed |
|---|---|---|---|
| Anhydrous ammonia | 16.4 | [5.2](#) | 315.4 |
| Potassium hydroxide | 0.5 | [0.5](#) | 100 |
| Phosphoric acid | 3.9 | [6.3](#) | 61.9 |
| Calcium chloride dihydrate | 1 | [2.4](#) | 41.7 |
| Sulfuric acid | 0.8 | [22.8](#) | 3.5 |
| Magnesium Sulfate Heptahydrate | 1 | | |
| Ferrous Sulfate Heptahydrate | 0.1 | [0.3](#) | 33.3 |
| Copper Sulfate Pentahydrate | 0.02 | [0.08](#) | 25 |
| Sodium hydroxide | 0.4 | [11.6](#) | 3.4 |

The table above shows that current US domestic production of the major nutrient inputs to methanotrophic microbial food production are sufficient to cover annual needs, with the exception of anhydrous ammonia, production of which would have to roughly triple, and magnesium sulfate, for which I could not find clear production data. However, I believe that the US is very likely capable of making enough magnesium sulfate given its current production of the two precursors to magnesium sulfate, magnesium oxide (or other magnesium compounds like magnesium carbonate) and sulfuric acid. In fact, the US may already make enough, and I just haven't found clear data on the amount produced each year. The US also makes significantly less anhydrous ammonia than it does other nitrogen fertilizers. These other fertilizers can likely be used as the nitrogen source in a nutrient media, or their production facilities can likely be repurposed to make anhydrous ammonia.

For magnesium sulfate, making enough to meet the requirements shown in the table (~1 Mt/year) requires roughly 0.8 Mt of sulfuric acid and ~0.3 Mt of magnesium oxide. The US produces 22.8 Mt of sulfuric acid (and only has to use ~3.5% of it directly for microbial food production) and ~0.4 Mt of MgO per year, so the US should be able to make enough magnesium sulfate.

The US produces only ~5.2 Mt of anhydrous ammonia per year, but makes ~27.4 Mt of other nitrogen fertilizers, for a total of ~32.6 Mt of nitrogen fertilizer. Different media recipes use various nitrogen fertilizers, suggesting some flexibility in the nutrient delivery options. For example, Unibio's patent uses potassium nitrate. However, US domestic supplies may still not be enough to make the highly protein-rich microbial foods modeled here. For a quick demonstration, the composition of Unibio's SCP product suggests that it is ~11.3% N by mass. Since anhydrous ammonia is ~82% N, the use of 16.4 Mt to make ~73.5 Mt of SCP suggests an N uptake rate of ~61% (assuming no recycling year to year). The ~32.6 Mt of nitrogen fertilizers produced in the US each year contain ~12.7 Mt of N. Given the ~61% uptake rate and the 11.3% N composition of SCP, this is enough to make about 68.6 Mt of SCP, which is only ~93% of the amount needed to meet US calorie needs. As I discuss in appendix 11.0, however, the microbial food that Unibio and others produce from methanotrophs is currently extremely proteinaceous, and so very high in nitrogen. This high protein content is likely undesirable, and there may be ways to reduce protein production in methanotrophs (12.7), which would involve lower nitrogen demand. Therefore I am inclined to believe that the existing US nitrogen production capacity is sufficient to make enough natural gas-fed food, but this is an area of mild uncertainty.

### 13.4.1 First pass findings on media nutrients

As noted above, the media nutrient data provided by the Synonym TEA is likely a better estimate of overall needs than the first pass approach that I used, which assumed zero waste and ~100% nutrient uptake efficiency. However, the TEA also misses a few trace minerals that could be important for efficient methanotroph cultivation. Therefore I've decided to include my initial findings, but stress that I think that those presented above are a better representation of the overall nutrient requirements. I should also emphasize that, as shown by my use of several different sources, nutrient media recipes vary. I also do not expect that the high protein strains currently cultivated for livestock feed will be the exact ones used in a response to a loss of agriculture scenario, at least not exclusively, so nutrient requirements will vary from those shown here.

| Nutrients from | Total annual requirements | Current annual production | Current annual production % |
|---|---|---|---|

| Unibio's patent | (tonnes) | (tonnes) | of requirements |
|---|---|---|---|
| KNO3 | 660,000 | 200,000 | 328 |
| CaCl2 | 840,000 | 2,400,000 | 35 |
| FeCl2 | 31,000 | 18,000 | 175 |
| CuCl2 | 12,000 | | |
| ZnCl2 | 2,300 | 281 | 826 |
| Na2MoO4 | 260,000 | | |
| CoCl2 | 1,000 | | |
| MnCl2 | 160 | | |
| NiCl2 | 290 | 9,716 | 3 |

| Nutrients from MediaDrive | Total annual requirements (tonnes) | Current annual production (tonnes) | Current annual production % of requirements |
|---|---|---|---|
| MgSO4 | 670,000 | | 126 |
| CaCl2 | 830,000 | 2,400,000 | 35 |
| KNO3 | 610,000 | 200,000 | 306 |
| Ammonium ferric citrate | 7 | | |
| HCl | 490,000 | 2,500,000 | 20 |
| FeCl2 | 31,000 | 18,000 | 172 |
| ZnCl2 | 2,400 | | |
| MnCl2 | 260 | | |
| H3BO3 | 11 | 290,000 | 0.004 |
| CoCl2 | 670 | | |
| CuCl2 | 12,000 | | |
| NiCl2 | 310 | 9,716 | 3 |
| Na2MoO4 | 52,000 | 4,190 | 120,000 |

| Nutrients from Methanotroph.org | Total annual requirements (tonnes) | Current annual production (tonnes) | Current annual production % of requirements |
|---|---|---|---|
| MgSO4 | 670,000 | | |
| KNO3 | 610,000 | 200,000 | 307 |
| CaCl2 | 830,000 | 2,400,000 | 35 |
| C10H12FeN2O8 | 680 | | |

| NaMo | 90 | | |
| FeSO4 | 38,000 | 18,000 | 211 |
| ZnSO4 | 3,100 | | |
| MnCl2 | 160 | | |
| CoCl2 | 530 | | |
| NiCl2 | 290 | 9,716 | 3 |
| H3BO3 | 27 | 290,000 | 0.001 |
| KH2PO4 | 940,000 | 34,019 | 2,800 |
| Na2HPO4 | 2,200,000 | 12,000 | 19,000 |

## 13.5 Equipment

To assess US capacity to produce the needed amount of equipment, I initially used data from Unibio and several other sources on the equipment needed for SCP production to estimate the cost of different pieces of equipment. I then took this price, scaled to the level needed to make enough SCP to meet US calorie needs, and compared these prices to the annual revenue of industries that currently produce these pieces of equipment. The degree of specificity vs. generality for the reference industries varies somewhat (e.g., for fermenters, I only found information on the annual sales of manufacturers of high pressure tanks, not of fermenters specifically. But for spray dryers, I found an estimate of the annual sales of spray dryer manufacturers in the US). I have now updated these findings with the equipment list and pricings from the Synonym TEA. The TEA includes 59 pieces of equipment, but only 15 of these contribute >1% of the total equipment cost. Together these 15 pieces of equipment account for over 82% of the total cost of equipment, so I've focused on these for now. The table below shows the cost of these pieces of equipment, added up across the over 700 facilities needed to meet all US calorie requirements from only natural gas-fed microbial food. The table also shows the reference industry used to compare the production cost of each piece of equipment to current production levels. For some pieces of equipment, I have combined them since they are very similar and so are compared to the same reference industry.

| Equipment | Total cost of equipment (million $) | Reference Industry | Annual sales of reference industry (million $) | % of reference industry |
|---|---|---|---|---|
| Production Ferm UHT Package | 2,500 | Ultra high temperature processing packages | 4,600 | 54 |
| Tanks* | 28,000 | Heavy gauge metal tanks | 13,061 | 215 |
| Thermal Oxidizer | 1,300 | Thermal Oxidizers | 9,600 | 13 |
| Inactivation HTST | 1,600 | Beverage | 7,700 | 21 |

| | | pasteurization machines | | |
|---|---|---|---|---|
| Centrifuge | 4,700 | Industrial centrifuges | 3,300 | 141 |
| Evaporator | 7,900 | Industrial evaporators | 18,700 | 42 |
| Spray Dryer | 30,000 | Spray dryers | 1,880 | 1,600 |
| Ferm Chillers | 20,000 | Industrial chillers | 312 | 6,300 |
| Cooling Tower | 2,400 | Cooling towers | 645 | 380 |
| Boiler Package | 2,500 | Boilers | 7,458 | 33 |

\* includes Pre-Seed Skid, Production Fermentor, Product Packaging, Reclaimed Water Tank, DSP Packaged CIP System

The table above shows that the current production capacity of US manufacturers is sufficient for some but not all pieces of equipment within a short time period, going with the economic proxies of prices and sales. Of course, not all manufacturers of a given type of equipment are currently making equipment that would meet the specifications needed for an SCP production plant. On the other hand, history suggests that manufacturers can quickly begin producing substantial amounts of products that are quite different from their typical outputs. As discussed in the main text section on methane-fed microbial scale-up, US car manufacturers quickly began producing tens of thousands of ventilators at the start of the COVID-19 pandemic, and during WWII automakers were able to pivot to making novel types of planes within a year of orders being placed. This gives me some confidence that manufacturers without any experience making the types of equipment listed above could, perhaps quickly, begin making these inputs to microbial food production. This could mean that the industrial base drawn on to make the needed equipment may be much larger than indicated in the table above. Still, the need to quickly make ~16x the value of spray dryers as currently made each year and or over 60x the annual value of industrial chillers sales may be beyond the capacity of US manufacturers in time to start up food production before US food reserves run out, even if they pivoted away from non-essential goods and focused on microbial fermentation equipment.

To get a sense of how the demands for equipment given above stack up against overall US manufacturing capacity, one can look at more general references for manufacturing output than the narrow industry comparisons in the previous table. The production of most, but not all, pieces of equipment listed above would be classified as machine manufacturing under the North American Industry Classification System (NAICS). This sector has an annual revenue of almost $337 billion, more than 3x the ~$100 billion cost of all the equipment shown above. Another reference is the amount that industries similar to microbial fermentation spend each year buying new equipment. US manufacturers overall spend an average of ~$140 billion on new equipment used to manufacture different products, but more narrow reference classes show that much less is spent by industries more similar to the microbial production process of interest here. One fairly narrow reference is the chemical manufacturing subsector classified by NAICS as "other basic organic chemical manufacturing", which includes biofuels not made at petroleum refineries. Here, the industry only spends an average of ~$4 billion annually on new equipment. Another narrow reference class is the brewing industry, which only spends ~$1 billion

annually on new equipment. Other reference industries that may buy similar equipment are general chemical manufacturing, which spends ~$23 billion on new equipment each year, food manufacturers (~$16 billion), and gas and oil (~$12 billion).

The references above leave me uncertain about whether the US could manufacture the needed equipment within the time that food reserves would last (22 months assuming worst-case scenario timing but good management and rationing). I think that a closer look at some historical precedent would be informative, and I see this as a promising direction for future work.

One uncertainty that a closer look at historical precedents could help resolve is whether the method that I have used — comparing the dollar value of pre-disaster outputs from various industries to the value of goods that they would need to produce post-disaster — has much validity.[49] From a very preliminary look, the pivoting of automakers to make ventilators during the COVID-19 pandemic and planes during WWII suggests that they may be able to make a value of new goods similar to or greater than the value of goods that they were producing prior to pivoting. However, these comparisons are complicated by the change in prices during disasters, so I would want to see more research before becoming too optimistic about the ability of manufacturers to efficiently and quickly pivot to making new equipment in a disaster.

Looking back at the use of automakers to produce ventilators during the COVID-19 pandemic, a Ford car parts plant in Rawsonville, MI, went from no ventilator production to making 50,000 within just 5 months. This production fulfilled a $336 million federal contract. Since it was fulfilled in just 5 months, this suggests that the manufacturing plant would be able to make ~$806 million worth of ventilators in a year, This is ~73% of the ~$1.1 billion pre pandemic annual output of the Rawsonville plant, based on the contribution to local GDP from Automobile and Light Duty Motor Vehicle Manufacturing (NAICS 33611). Ford estimated that they could have continued production at a rate of 30,000 ventilators per month after the initial 5-month ramp-up, which would put their total annual output in the first year at over $1.7 billion, assuming the same ventilator price of $6,720 implied by their initial contract. This suggests that the car parts manufacturer could pivot to making an amount of product worth more than it was previously able to make, although I suspect that prices were inflated due to the high demand for ventilators at the height of the pandemic, so this may not be a fair comparison between pre- and post-pandemic production value. Another reason that the comparison may not be quite fair is that Ford partnered with GE HealthCare, and the Rawsonville plant may have received resources from other parts of the Ford corporation or others, so the lost output of these other players should also be tallied for a proper comparison.

---

[49] One example that may point at price being a poor proxy, and which likely argues in favor of the US having enough production capacity, is the production of stainless steel fermenters. I used the reference industry of heavy gauge tank manufacturers and compared annual sales from this industry to the cost of making enough reactors and holding tanks. However, stainless steel is a much more expensive material than many other metals that are commonly used in making heavy gauge tanks, so this comparison may overstate how large of a fraction of the reference industry's manufacturing capacity is needed to make SCP equipment. Stainless steel is 5–10x more expensive per kg than other types of steel. Also, the Bureau of Economic Analysis's (BEA) Input-Output data shows that heavy gauge tank manufacturers purchase ~$1.7 billion from US iron and steel producers. This is ~1.9% of the revenue of this industry, and since the US makes ~82 Mt of steel each year, this suggests that the industry buys ~1.6 Mt of steel annual, which is ~3x the ~550 kt of steel used for stainless and carbon steel tanks and fermenters in the Synonym TEA. The table above showed that these cost ~2.5x the annual sales of heavy gauge tank manufacturers, but this suggests that these manufacturers already use far more material for making equipment each year than they'd have to in this case.

Looking briefly at the output of automakers producing planes during WWII, their prewar annual production was valued at ~$3 billion, and yet within the span of a year these manufacturers fulfilled an order of $9.2 billion (all in 1940s dollar value). Again, this may largely reflect an inflated price of planes during WWII, and from a brief review of historical sources, I'm not sure if other industries also contributed to the fulfillment of the federal orders. So this comparison may not be quite fair.

## 13.6 Construction

To evaluate whether the needed facility construction could be completed in a short period of time (under two years) by the US construction industry, I use a similar method as in the previous section on equipment manufacturing, comparing the estimated cost of construction to current construction revenue. The total construction costs for making all of the needed facilities, according to the Synonym TEA, is ~$179 billion. As a first pass I had used data from Unibio on their projected facility costs and other sources, but here I'll stick to the figures from the TEA model since they broadly agreed with my initial findings.

As quick points of comparison to the overall construction cost of ~$179 billion, annual revenue of the US construction industry is ~$1.8 trillion. As subsets of the construction sector, the building construction sector has annual sales of ~$804 billion, and industrial building construction is worth ~$30 billion annually. The TEA model shows the following breakdown of construction costs, if scaled to the 704 100-ktpa plants needed to meet all US calories. The table also shows the annual revenue of reference industries, and the percent of this revenue represented by the total estimated expense for each type of expense. Note that I'm comparing total costs to the annual revenue of reference industries, but the construction would not have to be completed within one year.

| Expense | Total cost (billion $) | Reference industry | Annual revenue (billion $) | % of annual revenue needed |
|---|---|---|---|---|
| Piping | 46.2 | Plumbing, heating, and AC contractors | 207.9 | 22.2 |
| Buildings | 36.7 | Industrial buildings | 30.3 | 121.1 |
| Civil/Foundations | 16.8 | Heavy civil engineering construction* | 22.7 | 74 |
| Steel | 8.4 | Steel production | 18.2 | 46.2 |
| Instrumentation | 33.6 | Building equipment installation contractors | 32.8 | 102.4 |
| Electrical | 29.4 | Electrical contractors | 173.8 | 16.9 |

| | | Drywall and | | |
|---|---|---|---|---|
| Insulation | 4.2 | Drywall and insulation | [44.1](#) | 9.5 |
| Paint | 4.2 | Painting and wall cover contractors | [26.3](#) | 16 |

\* excluding roads, bridges, etc.

The table shows that for most of the subcomponents of the construction costs, the total cost is less than the current annual revenue of the reference industries. The exceptions are building construction and instrumentation, but even here the total cost does not significantly exceed annual revenue, and a better comparison would be the output of the reference industries over a period slightly longer than a year. However, construction and equipment installation at a large-scale fermentation plant is not completely analogous to other construction activities that are more common in the US economy, so the comparisons above may implicitly assume a greater level of fungibility of construction capacity than is realistic. Here again more research would be beneficial, including references to historical analogies and input from more experts. As a quick note, the number of new factories built each year in the US **doubled** from 1942 to 1945. This suggests that construction rates of potentially complex facilities can be increased quickly, but it is possible that the number of factories may not be a good indication of total output, as the newer factories may have been smaller or less complex. In absolute terms, the factories of the 1940s may also have been less complex than today's, even adjusted for contemporary construction capacity. Therefore, I do not take this as especially strong evidence that the US could quickly begin making hundreds of new fermentation facilities. However, combined with the economic proxy data from above, I am tentatively optimistic that construction capacity would not be a major bottleneck to making the needed facilities in a disaster.

## 14.0 Natural gas-fed microbial food plant startup timelines

As noted above, there are several ways that the time needed to engineer, build, and bring to full operation a natural gas-fed microbial food production plant could be shortened. The most obvious step would be to waive permitting requirements. Another fairly dependable way that start-up times could be reduced is to expedite site purchase agreements and contract negotiations, although these are unlikely to be able to be completely eliminated even in an urgent disaster. Other steps could include shortening the delivery time of major pieces of equipment, expediting plant assembly, less comprehensive equipment and systems checks, and more aggressive ramp-up schedules.

Based on a schedule created by the Synonym Bio team, the tables below summarize the time savings that may be possible given some assumptions about speeding up different steps, for both a first of its kind plant and a repeat plant. In this first table, I show the time savings possible from each assumption independently, without making other changes (e.g., construction times decrease, but there are no changes to permitting or equipment sourcing times). In the second table, I show how much time can be saved by adding several assumptions together. I first show how much time can be saved by making fairly weak assumptions, such as that permitting and negotiation times can be eliminated or reduced. I consider these relatively weak assumptions because I am not counting on any physical or designing process to be accelerated. I then combine all of my assumptions, which I consider to be an aggressive set of assumptions, since I assume that several physical and design processes can be sped up substantially relative to typical time requirements. Note that these total time savings are greater than the time savings implied by just adding the time savings from individual assumptions, shown in the first table. This is because the time schedule model developed by Synonym has many processes carried out in parallel, and changing the time of one still often leads to the overall progress of the plant being bottlenecked by another process. Below the tables, I discuss some of the assumptions about time-saving possibilities that I make, although note that these are fairly tentative, and this is an area of uncertainty and ongoing research for me.

Time savings for different assumptions, holding all else constant

| Assumption | Time saved for novel plant (months) | Time saved for repeat plant (months) |
|---|---:|---:|
| No permitting | 0 | 0.7 |
| Reduce site selection and negotiations by ⅔ | 0.3 | 0.7 |
| Reduce engineering time by ¼ | 0.5 | 0 |
| Reduce long lead time equipment by ½ | 0 | 0 |
| Reduce construction time by ⅔ | 2.7 | 6.9 |
| Reduce commissioning by ½ | 0 | 1.1 |
| Reduce ramp-up time | 5.5 | 4.5 |

Time savings for different assumptions, combined

| Assumption | Time saved for novel plant (months) | Time saved for repeat plant (months) |
|---|---:|---:|
| Weak assumptions (no permitting, reduce site selection and negotiations time by ⅔, reduce ramp-up time) | 5 | 5.3 |
| Strong assumption (all assumptions combined) | 20.7 | 19.6 |

I want to emphasize that I do not have strong justification for making the assumptions above. For speed-ups in construction time, I looked to ALLFED, who used research on the impact of shift work on construction productivity to estimate that going from a typical 40-hour work schedule to a rotating 24/7 construction schedule reduces construction times to ~32% of the original estimate.[50] For reduced ramp-up times, I discussed with experts how a plant might adopt a more aggressive but still realistic schedule for going from mechanical completeness to full production by bringing more parts of the plant on at once and starting the seed train (the part of the plant used to culture methanotrophs to inoculate the main production tanks) before other parts of the plant have been completed and tested. I suspect that equipment lead times, at least for some plants, could be significantly reduced, since current quotes assume that providers have to fulfill other orders, and other manufacturers may be able to produce the needed equipment (appendix 13.5 Equipment), but the estimate that lead times could be reduced by ½ is basically a placeholder until I have better information on how much time could actually be saved by prioritizing orders and taking other steps to speed up delivery. I think it is reasonable to assume that site identification, purchase negotiations, and other steps could be expedited significantly, but I don't know by how much, and the same goes for engineering.

Overall, the findings presented here should be taken as at best indicating that there may be significant potential for reducing plant start-up times, but should not be read as literal predictions of post-disaster schedules. I hope to conduct, or see others carry out, more research into how much start-up times could be realistically reduced in disaster settings, accounting for both possible speed-ups in construction schedules and various bottlenecks that might emerge from trying to build hundreds of new plants simultaneously.

# 15.0 Sugar-fed fungal and bacterial food

One way to use sugar from lignocellulose to provide more nutrition is to use it as a substrate for SCP production. This has been done commercially for decades by the company Quorn, using fungi. I have not looked into this technology in the same depth that I looked into methanotrophic SCP. However, below I walk through several rough estimates of the costs of making microbial foods through sugar fermentation, and I provide some additional data and calculations in tab 11 of this Google Sheet. I find that, when accounting for the costs of making enough sugar, the CAPEX required to feed the US with this process is probably higher than that needed to make enough food through natural gas fermentation. However, as covered in the next section on fungal SCP nutrition and as mentioned above, some food products made through sugar fermentation may be valuable supplements to a diet otherwise consisting of methanotrophic microbial food. This means that in scenarios where dead plant biomass can be harvested to make sugar, some would likely be used as a substrate for fermentation.

While the processes do overlap, there are significant differences between gas and sugar fermentation for biomass production. To estimate the equipment requirements and costs of microbial food production from lignocellulosic sugar, I relied mostly on this paper, which describes SCP production using an agricultural feedstock.

The paper models a facility with a single 50 $m^3$ fermenter. This is likely far smaller than the optimal facility size that would be deployed in a scenario where a significant amount of food is produced through lignocellulosic sugar fermentation. This is because economies of scale make larger facilities more efficient, at least up to a point. Since there exist fermenters around 10x larger than the one studied in the paper, and facilities can have multiple fermenters, I suspect that using the facility size suggested by the paper will overestimate the cost of food production using this process. I attempt to correct for this by estimating what a larger facility that might produce 100 ktpa (same approximate size as the MOB SCP facilities described earlier) would cost. While it is possible to use a scaling formula to estimate larger-sized equipment based on the cost of smaller equipment (described below), this approach is limited in its applicability to equipment that is significantly smaller or larger than a given starting data point, so these estimates should be taken as very tentative.

---

[50] During shift work, per-hour productivity declines as responsibility for tasks shifts from one worker to the next and personnel move in and out of the construction area. In the case of a 24/7 rotating schedule, it is estimated that per-hour productivity falls by ~25% compared to a standard 40 hour work week.

First, to establish how much SCP a given sized fermenter would likely produce, I reviewed several studies on sugar-fed microbial food production. The table below shows what these studies found for the yield of microbial biomass per kg of substrate (sugar) and the production rate per unit volume. The table also shows how much a 50 m³ fermenter would produce in a year and how much fermenter volume would be needed to make 100 ktpa. The fermenter volume calculation uses the growth rates from different studies and also uses the same assumptions used to calculate the fermenter volume needed for methane-fed microbial food production, i.e., that total fermenter working volume is 70% and there is 10% facility downtime.

| Source | g SCP/g substrate | Production in g/L h | Production in a 50 m3 fermenter (tonnes/y) | Volume needed for 100 ktpa (m3) |
|---|---|---|---|---|
| Wu et al., 2018 (fungi) | 0.40 | 0.86 | 237.3 | 21,070 |
| Voutilainen et al., 2021 (fungi) | 0.50 | 1.02 | 281.5 | 17,765 |
| Voutilainen et al., 2021 (fungi) | 0.31 | 0.74 | 204.2 | 24,486 |
| Voutilainen et al., 2021 (fungi) | 0.37 | 0.82 | 226.3 | 22,097 |
| Risner et al., 2023 (fungi) | 0.31 | 2.33 | 642.9 | 7,777 |
| Ugwuanyi, 2008 (bacteria) | 0.40 | 1.15 | 317.3 | 15,756 |

The data in the above table show that a 100 ktpa facility using sugar fermentation would likely require significantly more fermentation volume than a natural gas fermentation facility producing the same amount of microbial food, which I estimated above would require about 5,000 m³ of fermenter volume.

To see how feasible it would be to provide the amount of sugar needed to ferment enough biomass to feed the US on lignocellulosic sugar fermentation as its only food source, I calculated in the table below how much wood would be needed to provide enough sugar, based on the studies above and the 0.36 mass conversion efficiency of wood to sugar from earlier. The table also shows how much total fermenter volume would be needed to feed the US on sugar-fed microbial food. For estimating these quantities, I assumed that the US would have to make ~86 Mt of fungal biomass (accounting for 13% waste), based on the calorie density of fungal SCP.

| Source | Wood needed to feed the US (Mt/y) | Total fermenter volume (million m3) |
|---|---|---|
| [Wu et al., 2018](#) (fungi) | 597 | 18.1 |
| [Voutilainen et al., 2021](#) (fungi) | 478 | 15.3 |
| [Voutilainen et al., 2021](#) (fungi) | 771 | 21.1 |
| [Voutilainen et al., 2021](#) (fungi) | 646 | 19 |
| [Risner et al., 2023](#) (fungi) | 771 | 6.7 |
| [Ugwuanyi, 2008](#) (bacteria) | 597 | 13.6 |

The table shows that total wood harvests (in most scenarios considered, harvests would be of dead plant biomass) would have to be increased from the current ~310 MT/year. Total fermenter volumes are also quite high, much higher than, say, the 3.1 million $m^3$ of fermenter volume that I estimate is used in US breweries or the ~0.6 million $m^3$ used for ethanol production. This means that a lot of fermenter capacity would have to be made specifically for fungal SCP production rather than just being repurposed from other industries.

Looking at the costs and equipment requirements for sugar fermentation, I look at both how much it would cost to produce 86 Mt going with the small facility studied in the [paper](#) mentioned above, and how much it would cost to make this much with large (100 ktpa facilities). When using a different sized piece of equipment, the cost can be estimated with the formula

$$Cost_2 = Cost_1 * (Size_2/Size_1)^n$$

where $cost_1$ is the cost of the original equipment, $cost_2$ is the cost of the new piece of equipment I'm trying to estimate, $size_1$ and $size_2$ are the old and new equipment respectively, and n is a factor that is usually ~0.6, but varies. The same paper on SCP production from agriculture feedstocks provides different values of n for different pieces of equipment, varying from 0.34 to 0.82, sourced from the literature. I used these to scale a facility up 363x to have ~18,000 $m^3$ of fermenter volume, since this is the average of the values found above. A two order of magnitude scaling is likely well outside the bounds of the applicability of the scaling equation from above, and likely gives a fairly significant underestimate of the costs of such a large facility. When I later multiply these costs to the scale needed to feed the whole US, I take the cost implied by a small, 50 $m^3$ fermenter facility to be an upper estimate and the costs implied by a scaled up 100 ktpa (18,000 $m^3$ fermenter) facility to be a lower estimate.

| Equipment Type | Equipment Size in Paper | Scaling Exponent | Cost of Equipment for a 50m3 fermenter plant ($) | Cost of equipment for 100 ktpa facility ($) |
|---|---|---|---|---|
| Centrifugal pump | 16,000 kg/h | 0.58 | 2,500 | 77,000 |
| Flow distributor | 8,000 kg/h | 0.48 | 2,100 | 36,000 |
| Valve solids and liquids | 8,000 kg/h | 0.6 | 2,600 | 88,000 |
| Solid packed | 1,200 und/h | 0.6 | 76,000 | 2,600,000 |
| Fermenter | 50,000 L | 0.56 | 35,000 | 4,600,000 |
| Fermenter | 10,000 L | 0.56 | 19,000 | 2,500,000 |
| Digester | 50,000 L | 0.56 | 43,000 | 1,200,000 |
| Balance tank | 1,000 L | 0.57 | 2,000 | 58,000 |
| Storage tank | 10,000 L | 0.52 | 20,000 | 430,000 |
| Plate and frame exchanger | 8,000 kg/h | 0.6 | 12,000 | 420,000 |
| Centrifugal equipment | 16,000 kg/h | 0.67 | 83,000 | 4,300,000 |
| Filter equipment (80 mesh) | 8,000 L | 0.59 | 6,400 | 210,000 |
| Decorticator | 150 kg/h | 0.6 | 5,500 | 190,000 |
| Gear pump | 2,500 L/h | 0.34 | 5,100 | 38,000 |
| Mixing valve — liquids | 16,000 kg/h | 0.82 | 2,000 | 250,000 |
| Sterilize | 5,000 L/h | 0.6 | 12,000 | 4,100,000 |
| Ion exchange equipment | 8,000 L/h | 0.62 | 17,000 | 640,000 |
| Dryer | 86,000 L | 0.53 | 83,000 | 1,900,000 |
| Dryer | 8,000 L/h | 0.6 | 10,000 | 340,000 |
| Air compressor | 4,000 kW | 0.69 | 7,800 | 450,000 |
| Solid conveyor | D × L: 30 cm × 5 m | 0.58 | 3,300 | 100,000 |
| Solid conveyor | D × L: 30 cm × 5 m | 0.58 | 2,000 | 60,000 |
| Condenser | 16,000 L/h | 0.79 | 13,000 | 130,0000 |
| Absorption column | 8,000 L/h | 0.78 | 28,000 | 2,700,000 |
| Degassing column | 8,000 L/h | 0.78 | 26,000 | 260,0000 |
| Mechanical skinning | 150 kg/h | 0.6 | 5,200 | 1,800,000 |
| Microfiltration | 80 m2 | 0.6 | 34,000 | 120,0000 |

| Rotary drum filter | 80 m2 | 0.39 | 30,000 | 300,000 |

The table above shows a total equipment cost of ~$0.7 million for a 50 m$^3$ fermenter facility and ~$32.9 million for a 100 ktpa (18,000 m$^3$ fermenter) plant. Installed equipment costs are often used to estimate total plant costs through a Lang Factor method. This approach uses a multiplier (the Lang factor) for a typical plant to see how much the whole facility would cost given the equipment costs. The paper that I have been using in this section finds a Lang factor of ~4.3. This is fairly consistent with typical values assumed in the literature, where in general chemical processing, Lang factors of about 3.1 for processed solids, 4.7 for liquids, and 3.6 for mixed products is usually assumed. Going with ~4.3, a 50 m$^3$ fermenter plant should cost ~$3.0 million, and a 100 ktpa (18,000 m$^3$ fermenter) plant should cost ~$140.8 million. Feeding the whole US should therefore have a CAPEX cost between $121 and $802 billion. This is broadly comparable to the $530 billion CAPEX cost estimated for feeding the whole US with natural gas fermentation. However, this does not account for the cost of making enough sugar. Averaging across the studies from above, producing the needed ~86 Mt of fungal biomass requires about 232 Mt of sugar. Earlier, I discussed that a DOE review of lignocellulosic sugar production found that the CAPEX cost for a new sugar production facility may be ~$2,502/tonne. This would suggest that making 232 Mt of lignocellulosic sugar would cost almost $580 billion. This would put the total cost of making SCP from sugar-fed SCP at ~$701 to $1,382 billion, or about 1.3 to 2.6x the cost of producing enough SCP through natural gas fermentation to feed the US population (around $530 billion, see appendix 13.1 Facility construction and operation costs).

Just going on CAPEX, these findings suggest that methane-fed microbial food is probably a better option even in cases where large amounts of dead plant matter can be harvested. However, my approach here was rough, and I'm quite uncertain. I also conservatively assumed that all sugar-fed microbial food production and all lignocellulosic sugar production required new facilities, rather than repurposing existing infrastructure. These foods could generally use infrastructure from similar existing industries, like brewing or ethanol production. As a very rough demonstration of how much costs might be reduced if some existing infrastructure was used, I'll assume that all ethanol and pulp and paper capacity in the US is repurposed to make lignocellulosic sugar, and all brewing capacity is used to make sugar-fed SCP. I assume that breweries are not used for sugar because I estimated above that breweries may contribute only about 12% of total sugar production capacity available if existing industries are repurposed to make lignocellulosic sugar, so using them to make SCP would not significantly detract from sugar production potential. I assumed that US ethanol is used for sugar production and not fermentation because they may have only ~18% of the fermenter volume as breweries, so would not be as valuable for fermentation. These are very rough assumptions and are not meant to be taken as literal prescriptions.

Going with the tentative assumptions above, the US would have the capacity to make about 29 Mt of sugar from retrofitted facilities, and perhaps ~17 Mt of sugar-fed SCP in breweries, if all could be repurposed to make SCP. This represents about 12% and 20% of the needed sugar production and fermentation capacity, respectively. From appendix 6.0 Lignocellulosic sugar production, retrofitted sugar production has an average cost of ~$710/tonne, so this would bring the cost of producing 232 Mt of sugar from ~$580 billion to ~$530 billion. A GFI report finds that retrofitted fermentation facilities can cost up to 70% less than new ones. If this holds in this example, then retrofitting facilities to make ~20% of the needed SCP would bring the total cost down to ~$104 to $690 billion. These rough estimates give a combined sugar and fermentation CAPEX cost of ~$634 to ~$1220 billion, which is still higher than the CAPEX estimated for methane-fed SCP production.

## 15.1 Fungal SCP nutrition

The table below shows the nutritional value of a 2,100 kcal/day diet of mycoprotein based on data from a presentation about Quorn. As above, the recommended intakes come from ALLFED, and I've colored nutrients that are provided in a safe amount as green, those that are too low as yellow, ones that are too high red, ones that are above the recommended values but where those values are not dangerous as blue, and left one nutrient for which there was no data as blank.

| Nutrient | Unit | Amount per 2,100 kcal | Min daily intake | Max daily intake |
|---|---|---:|---:|---:|
| Protein | g | 271.8 | 46 | 400 |
| Fat | g | 74.1 | 23 | 200 |
| Omega-3 | g | 9.9 | 1.35 | 5 |
| Omega-6 | g | 0 | 10 | 23 |
| Carbohydrates | g | 222.4 | 50 | 500 |
| Ca | mg | 1,050 | 50 | 16,500 |
| Fe | mg | 12.4 | 10 | 120 |
| Mg | mg | 1,111.8 | 400 | |
| P | mg | | 550 | 4,000 |
| K | mg | 2,470.6 | 3,500 | 7,000 |
| Na | mg | 308.8 | 200 | 2,300 |
| Zn | mg | 222.4 | 9.5 | 40 |
| Cu | mg | 12.4 | 0.09 | 10 |
| Mn | mg | 148.2 | 2.05 | 11 |
| Se | µg | 494.1 | 55 | 400 |
| Vitamin C | mg | 0 | 0.9 | 4,000 |
| Thiamin | mg | 0.2 | 0.88 | |
| Riboflavin | mg | 5.7 | 1.2 | |
| Niacin | mg | 8.6 | 12 | 500 |
| Pantothenic acid | mg | 6.2 | 5 | 20 |
| Vitamin B6 | mg | 2.5 | 1.3 | 500 |
| Folate | µg | 800 | 166 | 1,000 |
| Vitamin B12 | µg | 0 | 0.9 | 25 |
| Vitamin A | µg | 0 | 500 | 8,000 |

| Vitamin E | mg | 6 | 10 | 1,000 |
| Vitamin D | μg | 0 | 3.8 | 15,000 |
| Vitamin K | μg | 52 | 70 | |

From this data, it seems that fungal SCP may have worse overall nutrition than methane-fed microbial food, but it does have a substantial amount of omega-3, an essential fatty acid that appears missing in methanotrophs, so it could be cultivated as a dietary supplement for that purpose.

# 16.0 Note on storing wood

There may be scenarios where, after a disaster, societies can access non-crop plant biomass, but not for long (e.g., because the threat in question destroys or makes the biomass somehow unuseable). Non-crop plants, if accessible, could greatly increase the time that societies would have to scale up the production of industrial foods by acting as a stopgap food source. Wood would be the most important source of plant biomass because of its abundance (making up ~70% of all terrestrial plant matter), general resistance to degradation, and ability to be used to make food, such as sugar or mushrooms. To assess how much wood could be harvested and stored if accessible for just a short period of time, I looked at harvest speeds and available storage space.

As covered in appendix 6.0, providing all US calories with lignocellulosic sugar requires ~214 Mt of wood per year. Providing all US calories with mushrooms (appendix 18.5), if grown on logs with low-tech cultivation techniques, requires ~2,224 Mt of wood per year if using shiitakes, and 3,021 to 5,677 Mt per year if growing oyster mushrooms. This means that storing enough wood to make the sugar needed to provide all US calories for one year would, given the average[51] wood density of ~0.58 tonnes/m3, take up ~369 million m³ of storage space. For growing a year's worth of shiitake mushrooms, the volume would be ~3,835 million m³, and 5,209 to 9,788 million m³ for oyster mushrooms. I estimate that the US has 9.9 to 14.9 billion m³ of warehouse space. This is based on the floor space of US warehouses, currently a little over 1.6 billion m², and the typical height of warehouses being about 6 to 9 m. This means that if all of the storage space in warehouses was suitable for maintaining wood in usable condition and wasn't needed for other goods, all of the wood needed for a year's worth of calories from lignocellulosic sugar would take 2.5% to 3.7% of US warehouse space. Storing the wood needed to make a year's worth of calories from shiitake mushrooms would take 25.9% to 38.8% of the current US supply of warehouse space, and doing this for oyster mushrooms would take 35.0 to 98.9% of US warehouse space. Warehouses actually only represent a small fraction, ~5%, of all US building floor space, but probably a higher percent of building volume given higher ceilings. From this quick check, it seems that storing all of the wood needed to provide a year's worth of calories from sugar should very likely be doable, while doing so for the wood needed to provide a yearlong supply of calories from mushrooms may be possible but quite challenging, displacing many other goods.

The US currently harvests 535 million m³ of wood per year. Going with the wood density of ~0.58 tonnes/m³ from above, this gives a harvest mass of ~310.3 Mt. The average monthly harvest is

---

therefore ~25.9 Mt. If harvest rates did not increase, then each month's harvest would gather enough wood to provide ~12.1% of US calorie needs from sugar, but only ~1.1% from shiitake mushrooms. After a year's worth of harvesting at normal rates, the US could store enough wood to provide almost 1.5 years of calories for 332 million Americans from lignocellulosic sugar, but only ~1.7 month's worth of food from shiitake mushrooms. I do not currently have a method for estimating how much more quickly the US could harvest wood if determined to do so. My best guess is that the speed could increase at least 25% based on the US logging industry only operating at ~80% capacity in recent years. Going with this figure, a month's worth of harvest could get the wood needed to provide just ~1.5% of US calories if used to grow shiitake mushrooms, but ~15.1% of calories if used to make sugar. In a year's time, if using full harvesting capacity, the US could get enough food to make 2 months' worth of calories from mushrooms and almost 22 months of calories from sugar.

While shiitake mushrooms are fairly nutritionally complete (appendix 8.3), sugar is just a source of energy and has no other nutritional value. See appendix 6.1 for a discussion of how much sugar could be included in different post-disaster diets while still allowing the population to meet their minimum macronutritional needs from other foods.

## 16.1 Note on wood decay

In a loss of agriculture disaster that also kills trees, the amount of wood that could be accessed will depend on harvest rates, covered above, and how quickly wood decays. The expected rate of wood decay depends significantly on species type and climate conditions. In a proper treatment of the question of how much wood would be available after a disaster, future research could use established relationships between wood type and climate variable, combined with data on the distribution of wood species and climate regimes in the US and elsewhere, to arrive at fairly accurate estimates. For a very rough first pass intended to illustrate the approximate rate of wood decay, the ~24 Gt of wood in the US can be modeled as decaying according to an exponential function

$$Wood\ in\ year\ (t)\ =\ 24Gt\ \cdot\ e^{-\frac{ln(2)t}{T_{1/2}}}$$

where $T_{1/2}$ is the half life of wood. This value averages 18 years for softwood and about 10 years for hardwood. Since US forests are approximately half hardwood and half softwood by wood biomass, I'll use an average value of 14 years for this exercise.

Currently, the US harvests wood at a rate of ~310 Mt per year. At this rate, and given the decay rate above, the US would be able to harvest wood for almost 24 years. In that time, it would harvest ~7.4 Gt of wood – ~31% of the wood currently in the US – before running out. If the US could double its rate of wood harvesting, then it could harvest for almost 17 years, collecting ~10.4 Gt of wood in total. Current US harvest rates would provide enough wood to make 1.5 years worth of calories from sugar or 0.6 years worth calories from sugar-fed fungal food (15.0) or 0.14 years worth of mushrooms.

# 17.0 Industry availability for nonagricultural food production by small groups if society collapses

The main body of this report is concerned with whole-of-society survival in a loss of agriculture scenario. However, a disaster that makes agriculture impossible could lead to a breakdown in society and significant loss of life, perhaps even if it is technically feasible to produce enough food for large populations. If this happens, maintaining the infrastructure, material supply, and manufacturing capacity necessary for substantial food production without agriculture would be quite challenging. This would make it much less likely that large populations — or even small groups of survivors, which I focus on below — endure a loss of agriculture scenario than in cases where society maintains its organizational and industrial capabilities. I believe that societal collapse is a possible outcome of a loss of agriculture disaster, and that this would make it difficult to produce enough power, acquire enough materials, and make the needed equipment to produce the highly scalable microbial foods discussed in the main text of this report. In appendix 18.0, I discuss what food options may be available without access to these inputs. Here, I review the challenges of maintaining infrastructure, supply chains, and manufacturing capacity in societal collapse scenarios, with further detail provided in subappendices.

Above, I focused on methane-fed microbes as a nonagricultural food option that is especially scalable in the US given the country's large natural gas supply. For smaller groups who survive societal collapse, however, the most suitable food options will depend on what resources they can acquire. US natural gas production is distributed across more than 500,000 wells, although most do not individually provide enough gas to feed many people via microbes. There are over 31,000 wells big enough to feed more than 10,000 people, about 4,000 that could feed over 100,000, and just 31 that could feed over a million people. While less easily scaled up given current carbon capture and hydrogen gas production levels in the US, microbial food could also be grown on $CO_2$ and $H_2$. In appendix 17.1, I discuss the possibility of using different $CO_2$ capture and $H_2$ production sites to grow microbial food. I find that there are sites that could support many people, but only a few of them. Other food options could be available to smaller groups after societal collapse, but they all face similar challenges of relying on equipment that would be hard for small groups to produce and infrastructure that they may not be able to maintain.

Another key input to producing food without plants is electricity. For now I'll continue to use natural gas-fed microbial food production as a model food source that would be available in the scenarios considered in this report. Making enough natural gas-fed microbial food to feed one person requires about 6% of the US's current per capita electricity production. Given this, many US power plants could support tens of thousands to several million people (see appendix 17.2 for more detail). However, many types of power plants could be difficult to maintain. This difficulty comes from the need to both maintain fuel supplies for some types of plants and regularly fix or replace key pieces of equipment that — like the equipment directly needed for food production — may be hard to manufacture without modern supply chains. Important pieces of equipment for gas turbine plants need to be replaced every few years, so this would be a fairly imminent problem for people in post-collapse scenarios. A loss of power could also make manufacturing and repair much more challenging, putting survivors in a precarious state. See appendix 17.2 for a further discussion of the maintenance needs for power plants.

A final challenge to making microbial foods on nonagricultural substrates like natural gas is the need for various nutrients, some of which are only mined or processed at a few facilities in the US. For example, magnesium is only mined at one site in Utah, and cobalt is only mined at two. (Furthermore, cobalt minerals are generally produced overseas.) In both cases there is secondary production (recycling) at other sites, but overall, given how critical these minerals are for microbial food production, this concentration may severely limit small groups of survivors ability to make microbial food. See appendix 18.4 for more detail.

The challenges of maintaining industry in a world where society has largely collapsed suggest that many of the production options most likely available for small groups of survivors will be fairly low tech, meaning that they require little to no newly manufactured equipment or power. Even fairly rudimentary pieces of equipment, like standby generators, may be hard to maintain for long and will likely not provide much help in the conditions considered here (see 17.4). One large scale food production option that may be available, even with limited capacity for manufacturing or power generation, could be lignocellulosic sugar and organisms, like yeast, that can grow on it. The advantage of this technology is that there already exist facilities with most of the needed equipment, and some of them generate the power that they need onsite through burning waste biomass. Their gas turbines would likely still face the maintenance issues as other power plants, and a bigger concern would probably be accessing the various chemicals needed for delignifying wood and saccharifying cellulose and hemicellulose into glucose and xylose. These sugars would also have to be fermented into various other foods, adding to the material and facility complexity needs. Still, I discuss in appendix 18.1 how a small number of people may be able to overcome some of these challenges, but this likely doesn't change the overall picture that low tech food production options are needed in a loss of industry and agriculture scenario. Wood also may not be available in all cases considered in this report.

## 17.1 Methane and $CO_2$ infrastructure

If a group of survivors was able to maintain a power source and were looking to use gas fermentation to make single-cell protein, they would also have to access the carbon and hydrogen feedstocks required for SCP production. When looking at feeding all Americans, using natural gas to feed methane oxidizing bacteria (MOB) appears more scalable than using $CO_2$ and $H_2$ to feed hydrogen oxidizing bacteria (HOB). However, for an isolated group of survivors, scaling up isn't the concern, and they are limited only by what would be locally available, so here I investigate the size of individual sources of both natural gas for MOB SCP and $CO_2$ and $H_2$ for HOB SCP. Some relevant data and information is provided in tab 12 of this Google Sheet.

Natural gas is extracted from wells. The Energy Information Agency shows that the US has ~381,000 wells, with substantial variation in size. In tab 12 of this Google Sheet, I use the same EIA data to calculate how many people's worth of food (2,100 kcal/day, plus 13% waste, for 1 year) could be produced with the annual amount of natural gas from wells of different sizes. This calculation uses the same natural gas -> SCP mass conversion (~1.52 kg methane/kg SCP) from Unibio that I used earlier. I find that ~67% of wells do not provide enough natural gas to make enough food for 1000 people (every year). About 28% of wells produce enough natural gas to feed 1,000–10,000 people. The largest 31 wells produce, on average, enough to feed a little over 1 million people each. Based just on this, there are several locations in the US where, if an isolated group of survivors could build and maintain the needed infrastructure, they could access enough natural gas to sustain a fairly large population. I should note that, like power plants, natural gas wells require maintenance, recommended at a frequency of roughly once per year. It may therefore be difficult for an isolated group of survivors to keep a gas well running, although I suspect that building and maintaining an SCP production facility would be much harder.

For making MOB SCP, one study finds that you need 11.02 kWh/kg SCP. Budgeting for 13% waste, I estimate that each person needs ~221.3 kg of SCP/year, so about 2,438 kWh are needed per year per person's worth of food made at an SCP plant (of course this would vary with the scale of the plant, but for now I'm holding this figure constant). If a group of survivors was using natural gas to produce this electricity, then this would require ~512 m³ of natural gas, or ~389 kg. This is slightly more than the amount of natural gas, ~337 kg, needed to directly produce a person's annual requirement of MOB SCP. Therefore, if a group of survivors used a natural gas well to both provide natural gas as a feedstock for SCP and run a natural gas power plant to power SCP production, then the number of people that a well could support would be ~½ of the number who could be supported by the natural gas well if it was only used as a source of methane feedstock. Survivors would also need electricity for many other purposes beyond operating an SCP plant, and I don't mean to propose reliance on one well as an optimal solution. I present these figures mostly to illustrate a general sense of population scale possible given existing infrastructure.

For HOB SCP, there are several potential sources of $CO_2$, but production appears more likely to be limited by $H_2$ availability, unless survivors could make new $H_2$ plants. Based on this study of SolarFood's production process, it looks like producing 1 kg of HOB SCP requires ~1.76 kg of $CO_2$. Assuming that HOB SCP has similar nutrition to MOB SCP, then a person would also need to eat ~221.3 kg of SCP per year, so feeding one person for a year would require ~389.5 kg of $CO_2$. HOB SCP production needs 0.56 kg $H_2$/kg SCP, so a year's worth of SCP for one person would require ~123.9 kg of $H_2$. The same study of SolarFood's production process shows the company using ~17.83 kWh/kg SCP, so about 3,549.8 kWh are needed to produce enough SCP for one person-year of calories.

For $CO_2$ supplies, ~57% of $CO_2$ captured globally is used for urea production, and 30–35% is for enhanced oil recovery. Looking at US urea production, since about 130 Mt of $CO_2$ is used at urea plants worldwide, and the US accounts for ~6.9% of global urea production, US urea plants should capture ~9 Mt of $CO_2$. Since the US has 22 urea plants, each should capture about 407,000 tonnes $CO_2$. From the figures above, this single-plant total should be enough to feed a little over 1 million people, if $CO_2$ supply was the only limiting factor. This Clean Air Task Force map shows 8 operational industrial sector $CO_2$ capture sites, with a combined annual capture capacity of 17.6 Mt of $CO_2$. The largest is a gas processing plant in Wyoming that captures 7 Mt of $CO_2$/year, or enough to feed almost 18 million people, again if $CO_2$ supply was the only limiting factor. The smallest location shown is a gas processing plant in Michigan that could capture 400,000 tonnes of $CO_2$/year, enough to feed just over 1 million people. The US is yet to bring on any direct air capture sites beyond a pilot scale, but two that are listed in the map could feed a substantial population: one in Texas would capture 1 Mt of $CO_2$/year, enough to feed over 2.5 million, and one in Wyoming would be five times as large.

Hydrogen supplies are more limiting, given existing infrastructure. Total US annual $H_2$ production is ~10 Mt, or about a quarter of what would be needed to feed all Americans. About 40% of $H_2$ is generated by industrial gas companies, 25% at refineries, 20% at ammonia plants, and 15% at chemical/methanol plants. Looking at ammonia production plants, since there are 35 ammonia plants in the US (across 16 states), and these produce roughly 2 Mt, the average production per plant should be ~35,100 tonnes/y, or enough to feed ~442,000 people. A potentially interesting industrial site is an industrial gas production plant in Port Arthur, TX. This plant produces ~173,200 tonnes of $H_2$ gas per year, and can also capture about 1 Mt of $CO_2$ per year. Combining both a $CO_2$ and an $H_2$ supply would be valuable for a group of survivors trying to produce HOB SCP. The $H_2$ supply from the Texas plant is sufficient to feed ~1.4 million people, and the $CO_2$ supply could feed over 2.5 million (if not limited by $H_2$ supply). The plant also produces ~2 Mt of steam and 0.35 tWh of electricity per year. Since HOB SCP production requires 7.6 kg of steam per kg of SCP, this amount of steam could be used to make enough SCP to feed ~1.2 million people. Making this much SCP (265.5 Mt per year) would require ~4.3 tWh, so a group of survivors this large would need to maintain a sizable power plant, and of course maintaining a large industrial gas production plant would probably be quite challenging. But the availability of feedstocks suggests that there are some locations that could sustain a large population.

A small group of survivors could also potentially build facilities to produce $H_2$ and to capture $CO_2$, although this seems likely to be quite difficult, especially for a small group cutoff from modern supply chains who are in a hurry to get SCP production online before food supplies run out. I therefore haven't looked into this possibility much, but it may be worth exploring in later work. I'll note that capturing $CO_2$ from fossil fuel power plants is an expanding technology, and the Clean Air Task Force [map](#) from above shows many planned sites in the US scheduled to begin $CO_2$ capture in the next 10–20 years. Natural gas and coal plants [emit](#) an average of 407.3 and 988.8 kg of $CO_2$ per MWh of electricity produced, respectively. The $CO_2$ requirements for one person's year long supply of HOB SCP are therefore equal to 965 and 394 tWh of electricity from a natural gas or coal plant, respectively. Since both are lower than the electricity required to produce a person-year's worth of HOB SCP (see above), this shows that a $CO_2$ capture system that reclaims most of the $CO_2$ emitted by a fossil fuel power plant could provide more than enough $CO_2$ given how much SCP could be produced from the electricity available. Of course, if other sources of electricity are available that don't emit $CO_2$ or that aren't outfitted with a $CO_2$ capture system, then electricity may not be limiting. This note was mostly designed to illustrate that $CO_2$ capture at power plants could be quite useful, but of course HOB SCP production also requires a $H_2$ supply.

## 17.2 Power plants

Generating electricity seems broadly useful for survival, and would open up several food production options, such as running bioreactors for SCP production, growing algae or crops indoors, or turning fossil fuels into edible fats. Unfortunately, a brief review suggests that the maintenance of most power plants requires regular replacement of key pieces of equipment. A very shallow review found little historical precedent for efforts to maintain a power source without access to modern supply chains. One example which is illustrative of the challenges that small groups of survivors may face in maintaining power supplies can be seen in the US military's [difficulty](#) maintaining power at military bases when cut off from the electrical grid, in some cases needing over a week to get their backup generators running during drills or natural disasters. These failures are spurring the US to outfit more of their bases with [independent microgrids](#) and develop small, [mobile nuclear reactors](#) that are meant to function for up to three years in hostile environments. Similarly, an effort to operate a nuclear reactor at an Arctic base experienced nearly [continuous failures](#), prompting its premature retirement.

To get a sense for the size of US power plants, I show in the table below the typical sizes of different types of municipal power plants. I also list the number of workers at each plant to give a sense of the level of direct labor input needed to maintain plant operations. I show these calculations in [tab 13](#) of this spreadsheet.

| Power plant type | tWh of electricity produced in the US/y | Number of utility-scale plants in the US | Average tWh/plant | Total # of workers | Workers/plant | Workers/tWh |
|---|---|---|---|---|---|---|
| Natural gas | 1,687 | 835 | 2.0 | 420,83 | 50 | 25 |
| Coal | 832 | 157 | 5.3 | 70,831 | 451 | 85 |
| Petroleum | 23 | 726 | 0.03 | | | |
| Other gasses | 12 | 1 | 12.0 | | | |
| Nuclear | 772 | 31 | 24.9 | 55,562 | 1,792 | 72 |
| Wind | 435 | 700 | 0.6 | 120,164 | 172 | 276* |
| Hydro | 262 | 909 | 0.3 | 53,029 | 58 | 202 |
| Solar | 146 | 2,500 | 0.06 | 333,887 | 134 | 2,287* |
| Geothermal | 16 | 93 | 0.2 | 8,222 | 88 | 514* |

\* note that in these cases, unlike with other power sources, the numbers very likely include workers involved in construction and not just maintenance, so this is almost certainly an overstatement of the workers required to maintain power output

In the next table, I show how many people's worth of annual food supplies (providing 2,100 kcal/day for 365 days, not accounting for waste) could be met by these different types of municipal power plants, if all of the electricity generated from a plant was used for food production. HOB SCP refers to hydrogen oxidizing bacteria single-cell protein (where the carbon source is $CO_2$ and the hydrogen source is $H_2$ gas), and MOB SCP refers to methane oxidizing bacteria SCP (where carbon and hydrogen come from natural gas). Based on modeling done by Synonym Bio, MOB SCP production requires ~3.47 kWh/kg, and HOB SCP requires 3.45 kWh/kg (of course these exact figures would vary depending on several characteristics of a production plant, but for now I use these estimates). I also show how many people could be provided current US per capita electricity production (~12,940 kWh/person) by different municipal power plants. This may be a useful conservative benchmark for how much electricity per capita is needed in an industrial society capable of manufacturing the equipment needed for a gas fermentation plant and for maintaining all of the needed supporting infrastructure.

| Power plant type | Population that could be fed with one typical plant, HOB SCP (thousands) | Population that could be fed with one typical plant, using MOB SCP (thousands) | Number of people who could be serviced current per US capita electricity supplies by a typical plant (thousands) |
|---|---|---|---|
| Natural gas | 3,000 | 3,000 | 200 |
| Coal | 7,900 | 8,000 | 400 |
| Petroleum | 45 | 45 | 3 |
| Nuclear | 37,000 | 37,000 | 2,000 |
| Wind | 930 | 930 | 50 |
| Hydro | 430 | 440 | 20 |
| Solar | 90 | 90 | 5 |
| Geothermal | 260 | 260 | 10 |

If survivors can maintain only one plant, a nuclear plant appears likely to be able to maintain the largest population, but coal fired power plants also look promising, followed by natural gas.

The table below shows the average age and typical lifespan on nuclear, natural gas, and coal power plants.

| Power plant type | Average age (years) | Typical life span (years) |
|---|---:|---:|
| Coal | 39 | 50–60 |
| Natural gas | 22 | 40–50 |
| Nuclear | 36 | 30–40 |

At a high level, the table suggests that many existing US plants are near their expected retirement, and it may be a risky bet for a group of survivors to try and maintain these for a long period of time. But survivors may be able to select a plant which is in newer condition, giving themselves a while to figure out alternatives. However, an OSTI report finds that the typical service life for key pieces of equipment common to most power plants ranges from 2–5 years: on average, a main generator stator lasts 2 years, a unit transformer 3 years, a circuit breaker 4 years, and a motor 5 years. Given that the procurement time for these parts ranges from 8–18 months, I suspect that they are fairly difficult to make, and may need to be customized to a given power plant. This would make it difficult to scrap and repurpose parts from other plants, and require hard-to-source materials. It is possible that the recommended replacement schedules are designed to maintain high plant efficiencies and are not strictly necessary. I would like to talk to an expert to get a better sense of this, and have not concluded my research here. However, there does not seem to be massive variation in power plant efficiencies within the US or between countries, which is not what I would expect if part replacements were simply to optimize efficiency but weren't necessary. For example, the top 10% of coal power plants in the US are 37.4% efficient, compared to the 32% average. Page four of this report on cross-country comparisons of fossil fuel power plant efficiencies also does not appear to show massive differences in efficiency.

Even getting the fuel needed for power plants might prove difficult, although for coal and natural gas plants I imagine that it would be possible even without significant new production. A typical coal plant has ~40 days of fuel in stock, so survivors would have to resupply regularly, but the amount of coal that the US has in storage should be sufficient for a while. US coal storage ranged from 44.4 to 78.9 Mt in 2022. Since it takes ~0.52 kg of coal to generate 1 kWh of electricity, an average coal plant producing 4.9 tWh/year requires ~2.5 Mt of coal, so the amount of coal in stock is enough to run a typical plant for around 17–31 years on existing stocks if all of them could be accessed by survivors. Natural gas plants require ~0.21 m³ of natural gas to produce one kWh, so a typical plant producing 2.1 tWh per year needs about 10 billion m³ of natural gas. The US stock of natural gas ranges from ~45.3 to ~101.9 billion m3, or enough to power one typical plant for about 45 to 102 years if all of it could be accessed, though I suspect that not all of it could be accessed by an isolated group of survivors. I have not investigated this, but intuitively the infrastructure for maintaining coal mining and transportation seems easier than natural gas, since natural gas requires substantial processing before being pipeline-ready, but I don't actually know the details of what is involved in either case. Nuclear power plants generally need to be refueled every 18–24 months, and the US only produces ~3% of its uranium supplies, so survivors would probably have a difficult time sourcing, let alone enriching, an ongoing supply of uranium themselves. Wind and electric energy is probably too intermittent to serve as the main power source for a small group of survivors, but hydroelectric power could be useful.

Overall, it seems that the infrastructure for most scalable food production options is difficult for a small group of survivors to maintain. This argues for the importance of maintaining the population and coordination needed to keep an industrial society running, but also points to some preparation options for a potential scenario where a group of survivors has the benefit of advanced planning and can store equipment and fuel.

## 17.3 Note on microgrids

As noted above, access to a power source is critical for most food production options in a disaster where crops cannot be grown outdoors, and it is necessary for all options if no non-crop plants or fish can be accessed. If a group of survivors was able to maintain a power plant, it may be useful to control a plant that is set up to operate independently of the larger grid, as much of the US grid would likely collapse in a scenario involving significant population loss. The US has 769 such installations, called microgrids. The most common application of these microgrids is commercial use (~42%), followed by community or city use (14%), college or university (11%), hospital or healthcare (9%), and military (7%). These installations have a total capacity of just a little over 1,000 megawatts, or just 0.08% of total US electricity generation capacity. While the total capacity is low, the individual capacity generation of even the largest microgrids is large enough to support a midsized population (hundreds of thousands) if all of the electricity was used to run a single-cell protein plant. However, no single microgrid appears large enough to run a 100 ktpa plant, which is the size I modeled elsewhere. There is nothing dictating that SCP plants must be this size, but to take advantage of scale, an effort to feed the whole US would likely build large plants. This means that these plants probably couldn't be maintained by isolated groups operating a single power source.

The largest microgrid is used by an Air Force base in Georgia. If all of its power generation capacity was used toward MOB SCP, it could power a plant making ~66 ktpa, enough to feed ~290,000 people. The site uses a mix of natural gas (providing ~56% of its capacity) and solar (~44% of capacity). The next four largest microgrids could power an SCP plant making ~28 to 45 ktpa.

I've discussed elsewhere that maintaining a utility-scale power plant, at least one that uses a turbine generator, would likely be difficult for a group of survivors cut off from modern supply chains and manufacturing capabilities. This is because various critical parts need to be replaced fairly regularly (at least every 2–5 years). Accessing fuel would be a more immediate concern for most types of power plants. Given this, solar farms are an attractive alternative since solar panels typically last for 25–30 years, and the solar inverters needed to convert DC to AC current to use solar-generated power in the wider grid last 10–15 years. There is also no need to maintain a fuel supply. There are 130 microgrid solar installations in the US, 128 of which also have some electricity storage capacity. The largest solar microgrid, located in Nevada, is one of the largest microgrids in the country, able to generate enough power to run a MOB SCP plant producing ~44 ktpa. The average solar microgrid size, however, could only make ~630 tonnes of MOB SCP, which could only feed ~2,840 people. This is also assuming that the energy efficiency of such small-scale production would match that of a large SCP plant, but that assumption is unlikely to hold for operations orders of magnitude smaller than the one I am considering for US-scale production.

## 17.4 Note on electric generators

Access to power increases the number of food production options available to survivors. Even outdoor algae cultivation requires some power to stir a pond, although even if microalgae makes up all of someone's diet, its production requires only ~478 kWh per person per year, or just about 3.7% of the current US per capita electricity consumption, and some power may be helpful in preparing and storing food, although is probably not critical. In a case where isolated individuals can access a significant amount of power, they may be able to use more high-tech food production options, like growing algae or even conventional crops indoors, although these options — like gas fermentation — require further infrastructure and equipment that I suspect may be out of reach for people in a post-societal collapse scenario. Also, electricity access seems helpful for general survival needs and the ability to coordinate with others by powering radio etc., but I haven't investigated these other aspects of post-disaster resilience and recovery. From a shallow review, it seems that the US has on hand many standby generators that could likely be operated by individuals or small groups, but that these are unlikely to be able to sustain continuous power output over the long term, and that access to fuel could be limiting.

If population declines enough or the US does not maintain high levels of coordination, it may not be able to sustain its current power grid. Electricity provision currently employs ~378,000 people, not including workers in adjacent sectors like construction and manufacturing who are needed to maintain power plants and power lines, so most of this infrastructure should be expected to become inoperable in a collapse scenario. In appendix 17.2, I explore the possibility of larger groups operating power plants, but here I assume that standby generators powered by natural gas or diesel are the most likely power source that survivors would be able to operate.

Two estimation methods similarly suggested that the number of standby generators in the US is probably ~4 million. An industry report found that about 4.75% of owner occupied homes have a standby generator, and there are ~85 million owner-occupied homes in the US In 2021, around 143,000 new generators were sold in the US, with a compound annual growth rate of only 0.5% since 2015. Since generators typically last 25–40 years under normal operating conditions, if the annual sales figure has held roughly constant for the last few decades, then there should be between 3.6 and 5.7 million generators in the US.

A 2007 DOE report found that standby generators in the US had a combined annual power production capacity of 1,752 tWh, if they were run at full capacity 24/7 for a year. This would be 42.3% of total electricity at the time, and the report projected a 5% annual growth in electricity generation capacity, which would bring it to 2,409 tWh today, or 56.1% of total US electricity production (again, if the generators were actually run at full capacity year-round). Running generators at full capacity doesn't appear sustainable. Despite normally lasting 25–40 years, this is only assuming 10,000–30,000 hours of actual run time. If run continuously, this is only 1.1–3.4 years, so even a generator in new condition wouldn't run very long if heavily relied on. Fuel supplies could also be an issue. While efficiency of gas-powered generators varies, they typically need ~0.3 kg (0.37 m$^3$) of natural gas per kWh produced. The amount of natural gas stored in the US ranges from ~45.3 to ~101.9 billion m$^3$ over a year (low in February, high in November). If all 4 million generators were run at full capacity and were all powered by natural gas, then they'd only be able to run for 23–51 days on existing natural gas supplies. US diesel stocks are significantly less promising. A diesel-powered generator generally needs ~13.5 gallons of diesel per kWh produced, so the US stock of ~4.5 billion gallons of diesel could only power all 4 million generators for less than a day. This all suggests that distributed food production by isolated individuals powered by a fleet of generators does not seem feasible, and even if a small group of people had access to large amounts of fuel, existing generators would probably break down quite quickly and so wouldn't be a long term power source.

# 18.0 Low-technology food production options without agriculture

If limited to foods that can be produced without newly made equipment or electric power, options are quite limited. As covered in the discussion of other microbial food substrates, producing food on some nonagricultural feedstocks is less technologically complex than using gas fermentation. However, there is limited information on the foods produced from the fermentation of these feedstock, or the equipment needed. Even if these biomass fermentation processes are theoretically possible in low tech settings, obtaining the right starting fungal or bacterial culture may be difficult, production may be bottlenecked by limited access to important micronutrients, and there may be a very limited number of people with the expertise needed to pull this off. There are several other options available, at least in some cases. These include fishing, growing mushrooms or invertebrates on dead plant biomass, and growing algae or aquatic plants in open ponds. All of these options have significant potential for calorie production, but present serious nutritional challenges and are subject to failures due to contamination that would be hard to manage. In some cases, however, survivors may have a significant amount of time to establish new food production due to the size of various stored food depots in the US.

There are about 8,000 off-farm storage facilities for grains and oilseeds in the US. As detailed in appendix 19.0, these facilities hold around 65% of the country's major crops (in September). In September, the average facility should have enough food to meet one person's calorie needs for about 22,500 years. Therefore, a group of 1,000 people could meet their calorie needs for over 20 years, without spoilage. On-farm storage is also significant, although spread across many more farms. As I note in my discussion of food spoilage, major grains can be stored for over ten years in good conditions. However, without temperature and moisture control, the shelf life can go down to under a year. Acting with foresight, some groups of survivors may be able to provide good storage conditions by moving some grains underground. This technique has been used historically, and allowed for storage of grains for over ten years without electrical environmental controls. Oilseed crops — most importantly, soy — are higher in fat than grains and tend to preserve less well. Soybeans can be dried, extending their shelf life up to three years in room temperature conditions, or can be kept for over ten years if frozen. Frozen conditions may not be consistently attainable for survivors in a societal collapse scenario, but it seems that many years-worth of calories, especially from grains, could be accessed by small populations if food reserves aren't largely eaten through prior to major population losses. This means that I might expect survivors to have a long time to develop alternative food production options.

While a look at the nutritional profile of fish shows several missing essential vitamins and low levels of needed minerals, there is precedent for populations subsisting almost entirely on fish, and limited cases of people living exclusively on seafood for extended periods of time (appendix 7.1 Fish nutritional value). Achieving a complete diet on seafood requires eating organs and other products that are not currently consumed in large amounts in most cultures, but these were staples of traditional diets in some arctic areas. This suggests that in cases where the threat doesn't affect marine ecosystems and initial attempts to survive don't totally deplete near shore environments, then some groups may subsist on fish, possibly indefinitely. However, the knowledge of how to catch and prepare enough seafood to make a complete diet may be fairly rare in modern society, and the base rate of hunter-gatherer groups dying off appears high from the archaeological record, so this isn't a surefire bulwark against starvation.

As noted above, the most widely cultivated mushrooms in the US, agaricus mushrooms, cannot grow directly on wood and require compost. Other common varieties like shiitake and oyster mushrooms can be grown on wood logs. As long as there is dead organic matter, people should be able to make compost, although woody biomass takes a while to decompose even in favorable conditions. For now I focus on wood-grown mushrooms, but agaricus mushrooms are actually more nutritious so may be favored. Producing enough oyster mushrooms to meet one person's calories for a year requires roughly 5–10 tonnes of wood. From available studies, shiitake mushrooms take several times this amount of wood, but these studies are often not optimizing for high yields since organic matter is not generally a limiting factor to production in most contexts. Either way, this is probably a feasible amount of wood for someone to harvest even working manually, with 10 tonnes of wood being equal to about 95 average trees, although one large oak tree, for example, could be more than twice this weight. It also seems that someone could produce enough spawn to inoculate logs from a fairly small amount of starting mushrooms (about 1.6 to 3.2 kg, see appendix 8.1 for discussion and calculations). Mushrooms can be harvested from a given log usually about twice per year, so if survivors could subsist on stored food or some other source for about six months, they should be able to hold out for a mushroom harvest. It's important to note that yields can vary significantly, and harvest failures are common. In one of the studies used to calculate yields, a full 30% of trials failed to produce any mushrooms (these no-yields were averaged in for calculating the overall yield). People without experience growing mushrooms would likely struggle, and mushrooms are an imperfect food source, so this is again a tenuous survival option (see appendix 18.5 for a discussion of mushroom yields on wood, and appendix 8.3 for a discussion of mushroom nutrition).

Another food production option that I looked at is growing algae or aquatic plants in open ponds. In a scenario where the threat doesn't preclude outdoor algae cultivation, there is still the challenge of yields being extremely limited by nutrient supply and subject to variation from pest infestations that are hard to control even in modern settings. Microalgae would also be hard to harvest given their small size, and current techniques like using chemical flocculants may not be accessible to those surviving without industry. Aquatic plants like lemnaceae, colloquially "duckweed", could be a better option, but pose similar nutrient challenges and may not be viable in the scenarios considered if the threat affects all plants. Lemnaceae can have very high yields, potentially requiring just about 100 m² of pond surface area to produce enough food for one person for a whole year. However, most unfertilized freshwater in the US does not have enough nitrogen to support this rate of growth. Given the approximate normal range of total nitrogen levels in US freshwater, even if there were no pests or herbivores and lemnaceae could be grown in a monoculture without competition, growing enough to feed one person for a year might instead require roughly 1,000–10,000 m² of shallow open pond space. Maintaining production over subsequent years may require moving to new bodies of water, potentially putting this food production option further out of reach as a primary food source. Growing small amounts of aquatic plants may still be valuable, however, to create a more balanced diet than what is possible with e.g., mushrooms. As noted in my discussion of mushroom nutrition, common varieties of mushrooms do not provide enough omega-3 fatty acid, an essential nutrient. Lemnaceae are extremely high in omega-3, higher than common algae like chlorella or spirulina, and one only needs to eat about 43 to 72 g of Lemnaceae per day, dry weight (79–105 kcal/day), to get the minimum daily amount of omega-3. This is less than 5% of the amount of consumption I assumed was needed in the previous calculations to be the sole food source, so this level of cultivation may be feasible. See tab 15 of this Google Sheet for more detail.

The last low-tech food option that I considered is feeding wood to invertebrates. There are many animals that can digest wood and so could be cultivated if dead wood biomass was available. I find that termites might be able to produce human-edible calories more efficiently than mushrooms; the number of termites needed to feed one person for a year potentially needs to eat just about 2–3 tonnes of wood per year (appendix 18.3 Xylophagic animals). A central issue with termites as a major food source, however, is their lifecycle. One person would need to eat the termites coming from ~100 different mature colonies to get their daily calories, and a new colony might take five years to reach this maturity. Termite colonies also "die" after about 15 years, so managing dozens of colonies to maintain a stable food source seems like a risky bet. There are other options, such as marine invertebrates known as shipworms, that are already cultivated for food — mostly in the Philippines — and beetles. I provide a brief discussion of these options in appendix 18.3, but have not investigated thoroughly.

If there is no access to modern industry, supply chains, and infrastructure, food production without crops is still possible. However, it likely requires that aquatic ecosystems remain roughly intact and/or that dead plant biomass is available. Even so, the options are fairly unreliable, and survivors would be in a precarious state, similar to that of hunter-gatherers of the past, and long term survival is far from guaranteed.

## 18.1 Lignocellulosic sugar production without industry

As noted in appendix 6.0, the US does not currently produce a significant amount of lignocellulosic sugar for human consumption. (Only one company that responded to a DOE request for information reported making food-grade sugar from inedible plant fibers.) Yet the US does have several industries whose plants could be repurposed to make sugar from wood. The two main options are pulp mills and ethanol plants. Pulp mills separate out fibers from wood such that cellulose and hemicellulose are free of lignin and can be made into edible simple sugars. Ethanol plants currently turn agricultural products into sugar and then ferment this into ethanol. Some ethanol facilities in the US start with cellulosic feedstocks and turn these into sugars for fermentation, so they may be more easily repurposed for lignocellulosic sugar production. In fact, one facility in the US already rates its cellulosic sugar products as food grade.

The US has 176 pulp mills. These mills produce a total of ~49.7 Mt of pulp each year. This means that the average mill produces ~282 kt of pulp. ALLFED's work found that a pulp mill that produces ~234 kt of pulp each year could make ~11 kt of sugar, suggesting that a 282 ktpa pulp mill could make a little over 13 kt of sugar. This is enough to provide the calories needed by ~69,000 people each year. Of course, sugar does not provide any nutrients, so this wouldn't actually support such a large population, since much of the sugar would have to be used as a feedstock for fungal or bacterial food production through fermentation. Very roughly, 1 kcal of sugar could provide ~0.32 kcal of fungal SCP, so the above figure should be cut in third if trying to estimate how many people would actually be fed, bringing it to ~22,000.[52] This suggests that repurposed pulp mills could support at least several thousand people, although repurposing and maintaining the supporting infrastructure would still likely be quite challenging.

Cellulosic ethanol plants may be even more promising, since they should require fairly little repurposing effort. The largest cellulosic ethanol plant in the US — located in Iowa — produces enough sugar to meet the calorie needs of roughly 1.2 million people. Even the smallest produces enough glucose to meet the calorie needs of ~38,000 people. Again, these figures should be cut by ~one third if concerned about more than just calorie provision. Evaluating their promise would require a more detailed assessment of the engineering and manufacturing required to retrofit a facility, which I have not yet looked into.

---

[52] Since sugar could make up some fraction of a diet, not all sugar calories would need to be turned into SCP. So this may be a bit low, but I'm also not accounting for waste.

There are a little over 16,000 people working in ethanol production in the US, collectively making around 15 billion gallons of ethanol. Since the largest cellulosic ethanol plant makes only 30 million gallons, if a plant's workforce scaled linearly with production, there should be ~22 employees operating this production plant. Obviously this linear scaling assumption isn't quite right, but I think it suggests that not many people are needed to operate a large ethanol plant.

In appendix 6.0, I cover two methods of sugar production. One required cellulase, but appeared more scalable for the whole US, while another did not need any enzymes, but did require more chemical inputs. I am unsure which process would be easier for a small group of survivors to use, but I suspect that enzyme production is more complicated than managing several additional chemical inputs. Also, since this operation would be at a much smaller scale than one trying to feed the entire US, it may not run into the same material supply constraints that I noted above. However, the ease of implementing one production process over another in different repurposed facilities may be a larger driver of the decision of which to implement. For now, I'll assume that the enzyme-free process is used. Referencing the same life cycle assessment of a pilot scale plant that I cited earlier, the table below shows the inputs required per tonne of glucose produced, which I've scaled to the size of the largest cellulosic ethanol plant in the US.

| Input | tonnes/tonne ethanol | Amount/tonne glucose | Tonnes/year for a 30m gallon/y plant |
|---|---|---|---|
| Sulfuric acid | 2.2 | 1.0 | 236,270 |
| Lime | 1.34 | 0.6 | 141,762 |
| Anthraquinone | 0.1 | 0.05 | 11,813 |
| Sodium hydroxide | 0.03 | 0.01 | 2,363 |
| Sodium sulfide | 0.06 | 0.03 | 7,088 |
| Sodium carbonate | 0.2 | 0.1 | 23,627 |
| Sodium sulfate | 0.01 | 0.005 | 1,181 |

Note that the last four ingredients came from this description of the chemical constituents of green liquor, since the study only said that green liquor was used.

I've gone through what is needed to produce each material listed above and the amount of material that is already in stock in the US. In short, the last four materials are probably pretty easy to recover or make, anthraquinone doesn't actually appear necessary for production, and sulfuric acid looks somewhat difficult to make, but the US has fairly large stores of the needed materials.

The US currently produces ~22.8 MT/year of sulfuric acid, or ~96x what is needed per year in the scenario above. I couldn't find how much sulfuric acid is currently in stock. There are several routes to making sulfuric acid that are used industrially, but a common one in the US is the wet sulfuric acid process, and another is the contact process. Both involve burning sulfur and reacting it with water in the presence of a vanadium pentoxide catalyst. The amount of sulfur in stock in the US is about 120,000 tonnes (only ~1.5% of current annual production), which is enough to make ~1.5 years of sulfuric acid for the plant above, assuming that the survivors could access all of it. Apparently, the $Va_2O_5$ catalyst is hard to substitute, and the US stopped domestic primary production of vanadium in 2020, although secondary production continues in Arkansas and Ohio, recovering it from petroleum, spent catalysts, and ash. However, the US has 260 tonnes of vanadium-containing minerals in stock. I couldn't find how much of this is $Va_2O_5$, but $Va_2O_5$ makes up ~47% of all vanadium imports, so if it is stored at the same rate that it is imported, then there should be ~120 tonnes of it in stock. I also couldn't find how much is needed for sulfuric acid production, but if all 2,000 tonnes imported into the US each year are used only for sulfuric acid production (which should be an overestimate of how much is needed), then you'd need just ~88 grams per tonne of sulfuric acid, or ~20.8 tonnes/year for the amount needed above. Thus, current stocks should be enough for at least 5.7 years if survivors could access them.

The US currently makes ~25,000 tonnes of anthraquinone per year, which is only ~2x what is needed for the one plant above. I've tracked a bit of the production process involved, but a USDA report finds that adding anthraquinone to pulping probably only increases cellulose and hemicellulose (the constituents that need to be extracted from lignocellulose for sugar production) yields by 1–4%, so this is probably not a vital input.

The US produces ~17 MT/year of lime, or ~120x what is needed for the plant above. I wasn't able to find a figure for how much is in stock, but lime kilns operate in 28 states, so the materials should be fairly easily accessible for survivors.

Sodium hydroxide is currently produced at a rate of ~11.6 MT/year, which is over 5,000x the amount needed for the plant above. Given this, I strongly suspect that a large amount should be in stock, but I haven't found a figure confirming this. Even if stocks were to run out, the production process seems relatively basic. Sodium hydroxide is made by reacting sodium carbonate with calcium hydroxide, and I haven't seen any indication that it requires other difficult-to-procure materials like a catalyst. The US produces 13.9 Mt of sodium carbonate at 4 facilities in Wyoming and 1 in California, and it has 290,000 tonnes in stock. Producing enough sodium hydroxide for the plant above requires ~6.2 kt of sodium carbonate per year, so the US has plenty in stock. Calcium hydroxide comes from mixing water and lime, and as discussed above, I suspect that lime is fairly easy to come by, and the lignocellulosic sugar plant would only need 4.4 kt of calcium hydroxide per year. Some sodium carbonate (8,170 tonnes per year) is needed on its own for sugar production, but this still seems likely to be easily obtainable given how widely lime is made in the US and that 17 Mt are currently made each year.

The US probably makes ~35x more sodium sulfide (~250 thousand tonnes per year) than what is needed for the plant above, based on its market share of the global market for sodium sulfide. Sodium sulfide is made by reacting sulfur and sodium. The amount of sodium hydroxide needed is ~7.2 kt and the amount of sulfur needed is ~2.9 kt per year, so survivors should be able to access plenty from existing reserves per the discussion above.

The US produces ~300,000 tonnes/year of sodium sulfate, with two primary producers (1 in Texas and 1 in California), and 11 plants in 9 states recover it from other processes. Given that only 1,108 tonnes are needed per year, I suspect that enough is in stock for many years, but I haven't been able to confirm this. In addition to being mined, the sodium sulfate can also be made by reacting sodium chloride with sulfuric acid, which, per the discussion above, should be accessible for survivors. So I would be surprised if it proves too difficult to get enough sodium sulfate.

Overall, I haven't found an obvious material supply reason that a small group of survivors shouldn't be able to maintain a lignocellulosic sugar production plant, but I suspect that the larger hurdle would come from having to make repairs to the facility, since manufacturing new parts may be beyond the abilities of a small group cut off from modern supply chains. It also seems that if transportation options are limited in a post-collapse scenario, gathering the needed materials may well prove impossible.

## 18.2 Outdoor-grown algae

Microalgae, such as cyanobacteria or eukaryotic green algae, can be grown in photobioreactors (PBRs) in scenarios where outdoor cultivation is not possible due to some threat. However, such cultivation is fairly resource- and energy-intensive, and so it's probably outside of the capabilities of a small group of survivors post–societal collapse. In some threat scenarios, outdoor cultivation may be possible, and this is likely a more accessible option for such survivors.

The National Renewable Energy Laboratory (NREL) report that I referenced above (9.3.2) in my evaluation of cyanobacteria and green algae grown in PBRs also describes microalgae production in outdoor ponds. NREL reports that the aerial yield of the species studied in its report, *Scenedesmus*, ranged from 9.7 to 15.8 g/m²/day. Given the calorie density of *Scenedesmus* (5,070 kcal/kg), and that ponds can generally operate for only about 230 days per year due to the need to regularly clean them and restart cultivation after contamination, feeding one person for a year would require ~48 m² to 78 m² of pond space. This seems like a feasible amount of area for one person to attend to, as traditional small-scale farmers generally farm several ha of land. However, the ponds reported on by NREL are raceway ponds, which are artificially circulated to increase algae exposure to light and nutrients. The ponds also receive substantial fertilization from nutrient fertilizers like ammonia and phosphate, as well as captured $CO_2$. The yields reported by NREL may therefore not be achievable by survivors who cannot access power sources or fertilizers, and much larger areas may be necessary.

One reference point that may be more appropriate for comparison to what could be achieved in low-tech survival situations may be microalgae growth on wastewater. When grown on wastewater, edible cyanobacteria like *Spirulina* and *Chlorella* produce ~3.3 g/m²/day. Given the calorie density of *Chlorella*, ~4,100 kcal/kg, each person would need ~155 m² of pond area to grow enough chlorella to meet their calorie needs. Wastewater is still likely more nutrient rich than what could be maintained if survivors of a societal collapse did not have access to fertilizers.[53] As an idealized illustration that microalgae cultivation by survivors of a societal collapse in a low-tech setting is likely quite challenging purely on the grounds of how much area would be needed, the table below shows how much area would be needed to support one person from microalgae grown in typical freshwater, based just on bioavailable nitrogen abundance. In the US, bioavailable nitrogen levels in freshwater typically range from ~2.8 to ~29.3 g N/m³, based on the abundance of ammonia and nitrate and nitrites. As I did in my earlier assessment of microalgae grown in PBRs, I present yields under pessimistic nutrient requirement assumptions (those given in an LCA off *Spirulina* production by Ye et al. 2018) and optimistic nutrient requirement assumptions (those provided in the same NREL study of *Scenedesmus* referenced above). For a review of the nutrient requirement assumptions of each, see appendix 9.3.2. I also assume that survivors are working with a pond that is uniformly 0.3 m deep, since algae growth is limited to the upper layer of water. For now I only focus on nitrogen availability in freshwater, but other nutrients may be even more limiting of microalgae growth.

---

[53] Using the nitrogen uptake rates found for algae grown in wastewater (max ~75%), the N that needs to be absorbed by the ~187 kg of chlorella that one person would need to eat if algae was their only food source (~10.5 kg, based on the composition of algal biomass), and the amount of N that a typical human adult produces in waste each year (4.55 kg), one person's waste is only sufficient to grow enough algae to provide ~43% of their annual calorie needs.

| Assumption | Area needed to feed one person (m2) |
|---|---|
| Pessimistic (*Spirulina*, low N uptake) | 8,890 to 94,180 |
| Optimistic (*Scenedesmus*, high N uptake) | 370 to 3,930 |

It seems likely that this much area could not be reliably cultivated by a single person.

Beyond likely nutrient limitations, relying on outdoor-grown algae as a major food source also seems challenging because of the difficulty of harvesting microalgae and of controlling infections by zooplankton and other pests. There don't appear to be especially effective pest control strategies, although adding ammonia can help. Harvesting microalgae is difficult because they are small (<20 μm in diameter) and have a similar density to water. Current harvesting methods generally rely on very fine filters and flocculation (adding chemicals that help bind algae together to make larger particles). These methods aren't necessarily inaccessible to an isolated individual or small group of survivors, but certainly raise the bar and make algae a less attractive option. I looked at a larger edible aquatic plant, *Lemna gibba*, grown on wastewater to see if it would be a suitable food source while avoiding the challenges that come with harvesting small algae, but it appears less viable on a few metrics. It has a much lower N uptake in wastewater (6%), grows at only 1.2 g/m2/day, and is less calorie dense than chlorella (~2,450 kcal/kg versus ~4,100), so it would require much more area and a higher nutrient input.

## 18.3 Xylophagic animals

While in a loss of agriculture disaster I'm generally assuming that no new plants can grow, survivors would likely have access to wood from both dead trees and buildings and furniture. I'm focused on wood rather than other dead plant matter for now because wood represents ~70% of terrestrial plant biomass, and will probably last longer before decaying. Turning wood into sugar requires a substantial amount of industrial inputs to separate hemicellulose and cellulose away from lignin and then break these fibers into simple carbohydrates (glucose and xylose), so I do not think that an isolated individual or small group of survivors could use this food source. However, they may be able to cultivate animals that eat wood. Termites probably have a fairly high feed to biomass production rate, are already eaten in several cultures, and are fairly abundant. From a shallow investigation, I find that the amount of wood needed to feed enough termites to meet one person's annual calorie needs isn't that high, but that getting a new colony to be nearly large enough to be a significant source of food would take many years and that one person would have to rely on well over 100 colonies. Mostly out of interest and not as a practical matter, I used literature values of the typical density of termites in different biomes and the coverage of different biomes in the US (from FAO, using MODIS satellite data) to estimate that there are perhaps ~34 Mt of termites in the US, or enough for 47% of all US calorie needs for 1 year given their (fresh biomass) calorie density of 3,525 kcal/kg.

I couldn't find a study looking at the conversion rate of wood eaten to termite body mass, but other insects seem to have a conversion efficiency of ~13%, going from dry matter eaten to fresh body mass. Since one person would need to eat ~0.6 kg of termites/day to get 2,100 calories, the termites would have to eat ~4.6 kg of dry wood, or a little under 1.7 tonnes in a year. Wood is usually [75–92%](#) dry weight, so this brings the total amount to around 1.8 to 2.3 tonnes for a year. This is only ~3.1 to 3.8 $m^3$ of wood, so could easily be stored by an individual, and just working with a chainsaw allows one to harvest up to ~[8 $m^3$/hour](#), so it would take very little time to harvest years' worth of wood supplies. If termite production was scaled to serve as the sole food source for all Americans, the US would have to harvest wood at 1.9–2.4x its current rate, so this would actually be a potentially feasible large scale food source, if it weren't for the limitations imposed by termite lifecycles.

Termites have a [typical](#) body mass of 2–7 mg. Assuming that the average is ~5 mg, one person needs to eat ~119,000 termites per day. Even once a colony is mature, a queen only produces ~1,000 eggs per day. Even if someone could eat all of the termites at the rate that they were produced, they'd need more than 100 colonies to sustain them. Since only the queens reproduce, most of the termites in a colony can be eaten without directly reducing their reproductive capacity, but I assume that a colony can't be sustained if it is losing members at the same rate that they are being produced, so the actual number of colonies would have to be much higher. One of the reasons that this poses a challenge is that a colony [usually takes](#) ~5 years to reach maturity. A colony starts off with only ~50 individuals, and the queen lays [only](#) a few eggs per day at the beginning of a colony's "lifecycle". Thus, survivors would have to rely on other food sources for many years or start with a mature colony. I don't yet have a great understanding of the termite colony lifecycle, but they [appear](#) to "die" (a point at which most of the individuals, most importantly the queen, have died, after having spawned other colonies) after maybe ~15 years, depending on many factors. This further complicates the task of managing a sufficient population of termites for survivors.

There are quite a few other organisms that are able to eat wood, and one that is already cultivated for food are [shipworms](#).[54] I haven't looked into what is needed to maintain these gastropods, and there doesn't seem to be a lot of data on their cultivation. If they had the same feed conversion rate as sea snails ([18%](#)), and the same nutrition as [oysters](#), then to feed enough of them to make one year's worth of food for one person would take 9 tonnes of wood (15 $m^3$). Given that shipworms are marine, I assume that cultivating them requires seawater, which potentially adds other complications. Moreover, if someone is on the coast or at sea, they're probably better off fishing (assuming fish populations are little affected in this total loss of agriculture scenario). More research is needed.

## 18.4 Microbial food substrate options

In appendix [9.0](#), I found that SCP fed on gaseous substrates appeared among the most promising options for food production at a large scale. However, this type of SCP production is a very industrially intensive process that would be too complicated for isolated small groups to carry out. Even for larger groups of survivors, manufacturing the bespoke equipment needed to make a gas-fed SCP plant and maintaining the needed infrastructure appears quite challenging. However, there are other SCP technologies that may not be suitable for large scale production but could be more easily undertaken by small groups of survivors. Generally, these have received much less attention in the literature than has natural gas or $CO_2$ fed SCP, at least in the last few decades. I therefore have less detailed information on equipment requirements and the general production process for these options than I do for gas-fed SCP. This means that my assessment of the viability at small scales is less confident, and this section should be read more as a guide to which options warrant further investigation, possibly with small grants, rather than as complete assessments of production options.

---

[54] Credit to ALLFED for inspiration

This initial review has four main findings. First, there are at least five other substrates for SCP cultivation that could be used for food production: peat, paraffin wax, diesel, methanol, and various plastics. Second, one of these substrates, diesel, is currently produced in large enough quantities to potentially feed the whole US. Third, peat, gas oil, diesel, and methanol are stored in concentrated and large quantities such that even in scenarios where survivors fail to maintain the infrastructure needed to extract or produce the substrates, enough could be accessed at one facility to sustain a sizable population for decades. And fourth, existing stocks of the nutrient inputs needed for production are generally concentrated in only one or a few areas of the US, such that they may well limit production.

Overall, I think that SCP cultivation on peat hydrolysate and possibly diesel are the most plausible options for small-scale survival scenarios if there is no access to dead plant biomass and fishing is impossible. However, cultivation may be limited by the availability of nutrient inputs like salts and fertilizers. Survivors might also have difficulty extracting and processing the SCP even if they can produce it. If basic supporting infrastructure could be maintained, like utility plants and manufacturing, then other substrates discussed are possibly competitive with natural gas or $CO_2$ and $H_2$ as feedstocks for SCP production.

### 18.4.1 From peat

Peat is acidic, partially decomposed organic matter, with up to 5% of the mass consisting of simple carbohydrates like glucose and xylose. A peat hydrolysate can be made through a combination of acid and heat treatment, and the resulting solution is suitable for SCP production using a batch process to cultivate acidophilic fungi. For now, I'll refer to this study that used an acidophilic fungus (which at the time hadn't been classified, but which was tested for toxicity and nutrition content). Other fungi are also usable if the hydrolysate is mixed with something alkaline to raise the pH. One advantage of keeping the growth media acidic, though — especially in the low-tech context of a small group of survivors — is that it reduces the risk of contamination, since the media is less hospitable to other organisms.

The study found that after 14 days, the SCP concentration was around 15 g/L. The authors note that concentrations had roughly doubled from 7 days of fermentation, so the fungi may still have been in a linear growth phase. This means that yields could have been higher with a longer batch time, but this was not investigated. For now, I'll assume that each batch only lasts 14 days, and will further conservatively assume that there is no recycling of the media after the fungal SCP is harvested. This probably overstates the amount of nutrients needed for SCP production, since recycling is very likely possible. Given the macronutrient content of the fungi cultivated in the study, it should have ~3,573 kcal/kg, meaning that a day's worth of calories would require ~590 grams of SCP. Given the production of 15 g/L over 14 days, one batch would have to be ~973 L (a little under 1 m³) to provide enough food for 14 days for one person, and a year's worth of SCP would require a total of ~25.3 m³ of substrate media per person.

Based on the concentrations of peat in the hydrolysate provided in the study (4.7% to 7% of the media by weight), each person would need to use ~1.2 to 1.8 tonnes of peat per year. This is likely feasible at a small scale. However, if scaled up to feed the whole US, this would require a total of ~400 to 600 Mt of peat per year. Currently, the US only mines ~420 kt of peat/year, so a per capita consumption of 1.2 to 1.8 tonnes would represent a several thousand fold increase in annual production. There are also only about 150 Mt of peat that are considered economically recoverable in the US, although total reserves are much larger, at ~120 Gt.

The table below shows the annual requirements per person for other chemical or nutrient inputs to SCP production on peat hydrolysate, estimated from figures in the aforementioned study.

| Input | Amount/person-year (kg) |
|---|---|
| $K_2HPO_4$ | 300 |
| $(NH_4)_2SO_4$ | 220 |
| $MgSO_4$ | 26 |
| $H_2SO_4$ | 280 |

I have estimated how much of each of the above inputs is stored in the US, how many sites it might be stored at, and how large individual stocks are. All of these are extremely tentative estimates. For reserves, I looked at the ratio of inventory to revenue for industries involved in producing each input. For example, US phosphatic fertilizer manufacturers have inventories of slightly over $1 billion and an annual revenue of around $5.9 billion, for an inventory equaling ~17% of revenues. I then multiplied this by the current production of phosphate in the US to estimate how much might be in stocks. I also looked at how many production plants or mines make the input. In the case of phosphate, there appear to be 10 operational US mines, so I assume that each has 1/10 of the total US inventories of phosphate. This is a very simplified estimation method, and doesn't account for stocks that might exist in stores from other industries that purchase the input. (In the case of fertilizer, I expect farms to have some amount on hand, and retail distributors and merchants would also likely have some amount in stock.) Nonetheless, I think this provides a rough estimate of the feasibility of a small group of survivors finding stocks of the needed input for small-scale SCP production from peat. I show the results in the table below.

| Input | NAICS | Inventory % | Production | Inventory | # plants | Inventory / plant (or mine) |
|---|---|---|---|---|---|---|
| $K_2HPO_4$* | 325312 | 17.3 | 349.0 kt potash<br>506.2 kt phosphate | 60.4 kt potash<br>875.7 kt phosphate | 5 potash<br>10 phosphate | 12.1 kt potash<br>87.6 kt phosphate |
| $(NH_4)_2SO_4$ | 325311 | 8.4 | 3 Mt | 250.5 kt | 35 | 7,157 tonnes |
| $MgSO_4$ | 325180 | 11.2 | 86.3 kt | 9.7 kt | 4 | 2,416 tonnes |
| $H_2SO_4$ | 325180 | 11.2 | 22.8 Mt | 2.6 Mt | 77 | 33.2 kt |

* I couldn't find information on dipotassium phosphate production specifically, so I looked at potassium fertilizer (potash) and phosphate.

The next table shows how many person-years worth of inventory is at individual plants, based on the estimation method from above.

| Input | Inventory/plant (or mine) | Amount/person-year (kg) | Thousand person-years in inventory |
|---|---|---|---|
| K2PO4 | 12.1 kt for potash<br>87.5 kt for phosphate | 300 | 41 or 300 |
| (NH4)2SO4 | 7,157 tonnes | 220 | 32 |
| MgSO4 | 2,416 tonnes | 26 | 93 |
| H2SO4 | 33.2 kt | 280 | 120 |

The results from above suggest that some plants likely have ample supplies of different inputs needed for SCP production from peat hydrolysate. However, the figures on the number of production facilities or mines in the US also suggest that some inputs may not be accessible except in a few places. Any one group of survivors may therefore have trouble accessing the needed inputs. In addition, equipment requirements may prove to be a bottleneck to efficient production for survivors without access to modern supply chains and manufacturing facilities.

### 18.4.2 From paraffin wax

Paraffin wax occurs naturally in all oil, but the concentration varies considerably. Extracting paraffin wax from oil involves cooling oil to the point where wax crystals form and then separating these by either filtration or centrifugation. The wax is then melted and reformed several times to help remove whatever oil remains. The product can then be used as a carbon source for both fungal and bacterial SCP cultivation. The biomass to feedstock yield is 0.95. Given the same assumption about calorie density from above (3,400 kcal/kg, based on Quorn SCP), this means that a person-year's worth of wax is ~237.3 kg. US production of paraffin wax is ~63.5 kt, or enough for a little under 268,000 people. This production estimate is from 2002, and I haven't found a more up-to-date source that provides a production estimate, but I would be fairly surprised if production had increased to anywhere near high enough to be scalable for a large population. For small-scale production, supplies may be quite concentrated, since in 2002 there were only two US producers. Paraffin wax producers are classified under the NAICS code for petroleum refineries, and this industry keeps inventories equal to ~5.5% of its revenues, so I estimate that the US has ~3.5 kt of paraffin wax in stock. If this is split across two manufacturing plants, then individual stores should be ~1,750 tonnes, or enough to feed ~7,370 people for 1 year. Thus without the ability to separate paraffin wax from oil, the number of survivors that could eat paraffin wax–fed SCP based on existing stocks would be quite limited.

The production process for SCP grown on paraffin wax is also fairly energy intensive, apparently requiring over 49 MWh/tonne of SCP. If scaled to the whole US, this energy intensity would mean the US would have to devote almost 86% of its current electricity production to SCP cultivation. Despite this, the Soviets reportedly had the capacity to make 860,000 tonnes per year of paraffin-fed SCP, enough to meet the calorie needs of over 3.8 million people (if it was suitable for human consumption- as with current SCP production, the Soviet's mainly used their process to make livestock feed). There may be lower tech ways of using this feedstock that would be more suitable for survivors cut off from modern infrastructure, but I so far have not found any description of such a process. Even if there was, the lack of readily available paraffin wax makes this an unappealing option.

### 18.4.3 From diesel

Diesel is made either from biomass or petroleum. The latter process is carried out at a refinery through fractional distillation, where oil is heated and then cooled, allowing fractions with different evaporation temperatures to be separated out. Diesel is typically fractionated out at around 250–300˚C. In a [paper](#) describing an SCP cultivation using diesel, the authors used batch fermentation, with one batch lasting seven days. The optimal diesel concentration in the growth media was found to be 33.50 to 50.25 g/L, producing ~15.86 g of dry cell weight/L. At this production rate, assuming a kcal/kg content of 3,400 (same as other fungal SCP) and going with the same 7-day batch setup described in the paper, one person would need a 232.8 L fermenter, and 52 batches per year means 12.1 m³ total fermentation media. From this, I made the following table showing per capita input requirements.

| Input | g/L | kg/person-year |
|---|---|---|
| Diesel | 33.50 to 50.25 | 474.9 to 712.3 |
| NH$_4$Cl | 2 | 28.3 |
| KH$_2$PO$_4$ | 1 | 14.2 |
| KCl | 0.5 | 7.1 |
| MgSO$_4$ • 7 H$_2$O | 0.5 | 3.5 |
| MnSO$_4$ • 4 H$_2$O | 0.05 | 0.5 |
| FeSO$_4$ • 7 H$_2$O | 0.005 | 0.04 |

The amount of diesel in stocks in the US at any given point is [~4.1](#) billion gallons, or about ~14.3 Mt tonnes given diesel [density](#) of 3.22 kg/gal. This is enough to feed between 23.6 and 34.5 million people for a year, assuming there was no other limiting factor. With current US [production](#) of diesel at ~236.5 Mt, more than enough SCP could be made to meet US calorie requirements. Although this is not the focus of this report, the table below makes a quick comparison between annual input requirements for scaling up diesel-fed SCP and current US production of the inputs.

| Input | Amount to feed US (kt) | Current US annual production (tonnes) | % of current annual production needed (%) |
|---|---|---|---|
| Diesel | 160,000 to 240,000 | [240,000](#) | 67 to 100 |
| NH4Cl | 9,400 | [50](#) | 19,000 |
| KH2PO4 | 4,700 | [4,300](#) | 110 |
| KCl | 2,400 | [260](#) | 910 |
| MgSO4 • 7 H2O | 1,200 | [86](#) | 1,300 |
| MnSO4 • 4 H2O | 170 | [14](#) | 1,200 |

| FeSO4 • 7 H2O | 13 | 18 | 74 |
|---|---|---|---|

The table above shows that, given the requirements laid out in the paper on diesel-fed SCP, current US production levels for several inputs are not adequate, in two cases by more than an order of magnitude. However, I suspect that other compounds would be suitable for supporting SCP growth, and that the US likely has adequate production capacity for these. For example, while the US only produces about 50,000 tonnes of ammonium chloride, it produces over 4 million tonnes of ammonium nitrate and ammonium sulfate (combined) and many different chlorine compounds; many of these may be able to substitute for $NH_4Cl$. The US could also likely make much more ammonium chloride, since it can be made from ammonia and hydrochloric acid, of which the US makes 5.2 Mt and 2.5 Mt, respectively. The same likely goes for the other compounds listed above, but I'll save an investigation for later, since scaling up to meet total US demand is not the focus of this section.

I suspect that an analysis of the inventories of the industries that produce these inputs, and the number of production plants, would produce a conclusion similar to that for peat hydrolysate-fed SCP: that there may be a small number of plants that have the needed inputs stored in large enough quantities to sustain a small population for a long time, but that these are probably not colocated.

An even bigger barrier may be operating an SCP fermenter. Little detail was given about the fermentation vessel or other technical parameters, so I would need to do more research to better understand the process involved. The paper does specify that the fungal SCP were separated from the fermentation media by centrifugation and then oven dried. Per my discussion on standby generators, small groups of isolated survivors may have limited access to electricity to power devices like a centrifuge. I am unsure if lower tech options, like filtering through readily available materials, might be suitable. SCP grown on diesel is fungal, so the organisms are larger than bacterial SCP and thus better suited for being captured through filtration, but I have not found information on the use of this separation technique.

### 18.4.4 From methanol

In appendix 9.2.1, I discuss using methanol as a substrate on scalable food production and found that current US methanol production is not sufficient to meet more than ~5% of US calories. This implies small groups of survivors may be able to use methanol to grow enough SCP. However, examples of SCP production on methanol generally use advanced techniques with bespoke equipment, like special U-Loop or airlift reactors, similar to gas-fed SCP cultivation. This probably limits the utility of methylotrophic SCP for small groups of survivors, but there may be less advanced production options that I haven't encountered so far.

Because the technological complexity makes methanol-fed SCP less attractive as an option for a small group of survivors, and current methanol availability limits its use at a large scale, I haven't researched other input requirements very thoroughly. I'll briefly note the availability of methanol and ammonia inventories at an average production plant, following the same approach as above. Inventory percentages again came from the economic census: current annual methanol production was found to be ~5.7 Mt, and annual ammonia production is ~5.2 Mt. Requirements per person-year were calculated based on yield data from this paper, going with the assumption that the methylotrophic SCP has the same kcal/kg content as methanotrophic SCP, 3,982.

| Input | Amount in | Number of US | Inventory/plant | kg/person-year | Person-years |
|---|---|---|---|---|---|

| | inventory (kt) | plants | (kt) | | from an average plant (thousand) |
|---|---|---|---|---|---|
| Methanol | 526.1 | [18](#) | 29.2 | 342.6 | 85 |
| Ammonia | 426.4 | [32](#) | 13.3 | 98.2 | 136 |

The table above suggests that the stocks in different production facilities may be large enough to support a small number of survivors for an extended period of time. I haven't included a list of micro nutritional inputs, but suspect that doing so would paint a similar picture as the peat hydrolysate-fed SCP, i.e., that the inputs could technically be sourced from existing inventories at a few plants, but that doing so is probably impractical for small and isolated groups.

### 18.4.5 From plastic

The Defense Advanced Research Project Agency (DARPA) has two projects — [Cornucopia](#) and [ReSource](#) — working on tech that may be relevant, especially to a small group of survivors scenario. Neither have produced published results at the time of writing this (11/14/2023). One of the teams on the ReSource project, however, has [published](#) some findings on biomass production using different plastics as a carbon source. Given their ubiquity, the ability to use plastics as a substrate for food production would be really advantageous for isolated groups of survivors. The data available in the paper isn't really intended to demonstrate the parameters that I am interested in, like yield or nutrition, so I had to make several fairly bold assumptions.

The paper reported $OD_{600}$ measures for microbes cultivated in a medium where plastics were the only carbon source, but did not report the yield in mass or mass per unit volume. I assume for now that the $OD_{600}$ measures translate to g/L as E Coli ([0.36 g/L](#) per 1 $OD_{600}$), but this may well be quite off. If this assumption does hold, then the paper implies the following biomass densities for the four types of plastic investigated as substrates, after cultivating the plastics in a media for 10 days. The table also shows approximate yields, defined as g biomass/100 g plastic.

| Plastic | Biomass concentration (g/L) | Yield (g biomass/100 g plastic) |
|---|---|---|
| Alkene mixture* | 0.11 | 2.08 |
| Bisphenol A | 0.63 | 0.95 |
| HDPE | 0.14 | 2.6 |
| Disodium terephthalate | 0.04 | 0.72 |

* hexene, decene, hexadecene, and eicosene in equal parts

The researchers cultivated a mix of microbes, and their focus was not on producing edible biomass, so I do not know if any of the organisms they grew were safe to eat or what their nutritional content might be. Mostly out of curiosity, I used the above yield figures to see what the feedstock needs would be if trying to feed one person or the US population on biomass grown on these plastics, and compared this to the current annual production of the plastics. To calculate this, I assumed that the biomass had the same kcal/kg, 3,982, as methanotrophic SCP, but of course this is another rough assumption.

| Plastic | Tonnes to feed 1 person for a year | Mt to feed the US for a year | Current US annual production (Mt) |
|---|---|---|---|
| Alkene mixture | 9.3 | 3,100 | 0.6 |
| Bisphenol A | 20.4 | 6,800 | 1 |
| HDPE | 7.4 | 2,500 | 10 |
| Disodium terephthalate[55] | 26.8 | 8,900 | |

I strongly suspect that a substantial amount of the materials listed above are available in fairly large quantities in existing products, so access to the substrates would not be limited to only what is produced in a given year. However, the 2–3 orders of magnitude difference between the amount required and the amount produced in a given year makes it seem extremely unlikely that there is enough of any of the feedstocks in the US for biomass production to scale enough to meet US calorie needs (given my various assumptions laid out above). Since the paper was not trying to cultivate biomass for food production, I also suspect that yields could somewhat easily be made much higher, such as by cultivating only one instead of many species, but this is only speculation.

Other micronutrient inputs were also used, shown in the table below. Broadly, these seem similar to those used in SCP cultivation, but the amount needed per person is much higher because the yields (g/L) were much lower. I only looked at the case of HDPE, since that had the highest yield of the substrates studied, and given the yield of 0.14 g/L, one person-year of food would take almost 1.4 million L of culture media to produce.

| Input | Concentration (g/L) | Amount needed/person-year if growing biomass on HDPE (kg) |
|---|---|---|
| $MgSO_4$ | 0.2 | 275 |
| $CaCl_2$ | 0.02 | 27.5 |
| $KH_2PO_4$ | 1 | 1,374.9 |
| $(NH_4)_2HPO_4$ | 1 | 1,374.9 |
| $KNO_3$ | 1 | 1,374.9 |
| $FeCl_2$ | 0.05 | 68.7 |

---

[55] I was not able to find a figure for current US annual production of disodium terephthalate.

This paper wasn't written to provide parameters for assessing plastic as a feedstock for food production, but the takeaways do not look promising at either a large or a small scale. However, I have a great deal of uncertainty about these rough findings, and I plan to monitor other research outputs from the two DARPA projects as they come along.

## 18.5 Cellulolytic mushrooms

Mushrooms can grow on either dead or alive wood as their only substrate, and they are a fairly nutritionally complete food source. This makes them an attractive food option for isolated groups of survivors who do not have access to modern manufacturing, supply chains, or infrastructure. While mushrooms can grow only on wood, modern cultivation typically involves mixed substrates. The substrate mix used in commercial cultivation often contains a higher nitrogen content source, such as soy powder or grain bran.[56] Neither substrate would be available in the scenarios I am considering. Current mushroom farming practices are also fairly energy and resource intensive, as covered in appendix 8.0. To get a better sense of how efficient low-tech cultivation could be in a small group of survivors scenario, I looked for studies that did not include these inputs.

One study on oyster mushroom cultivation on logs found wet weight mushrooms yields ranging from 131 to 200 g wet mushrooms/kg of dry log weight in the first 6 months, depending on the log species. Given that king oyster mushrooms have 410 kcal/kg when fresh, a person would have to eat ~1,870 kg/year to get 2,100 kcal/day. The yields reported in the study therefore suggest that growing enough king oyster mushrooms to meet someone's calorie requirements requires ~4.7 to 7.2 tonnes of dry wood per year. Since wood is typically 75–92% dry, the amount of fresh wood needed is probably between ~5.1 to 9.6 tonnes. This is likely a feasible amount of wood for one person to harvest, since this equates to ~12 to 23 m3, and a person with a chainsaw can harvest up to ~8 m³ of wood/hour. However, survivors may not have long-term access to the fuel or electricity needed to run a chainsaw, but this is still probably a feasible amount of wood for someone to collect, although I haven't investigated manual harvesting and processing labor requirements. For reference, one large oak tree may weigh more than 20 tonnes.

The yields of oyster mushrooms suggest that even if production could otherwise be scaled up, the US would have to substantially increase its annual wood harvests to rely on wood-grown mushrooms as a major calorie source. A per capita annual consumption of 5.1 to 9.6 tonnes of wood represents a roughly 5–10x increase in current US per capita wood harvest rates. If this per capita wood consumption was scaled to 332 million Americans, this would equal ~7 to 13% of total US wood biomass per year.

A study on shiitake cultivation on logs left outdoors reported a high level of variation in yields, with average yields substantially lower than those found for oyster mushrooms. The maximum yield in this study was about 1.9 g of dry shiitakes/kg of oak logs. Dry shiitakes have ~2,960 kcal/kg, so for someone to get 2,100 kcal/day for a year from just dry shiitakes, they'd have to eat ~259 kg. Given the maximum yield in the study, this means that feeding someone for a year would require ~13.6 tonnes of wood. This was gathered for a single harvest, but shiitakes can be grown on the same log for 2–3 harvests per year, cutting the wood requirements to ~4.5 to 6.8 tonnes per person per year. However, this study found that in 6 out of 20 trials, no shiitake mushrooms grew on the logs despite the logs being inoculated with shiitake spawn. Low or no shiitake yields were associated with the presence of other fungal species. Including the 30% of trials where no shiitakes grew, the average dry weight yield was just ~0.54 g of dry shiitake/kg of oak logs. Going with this average yield brings the wood requirement, accounting for 2–3 harvests/year, to about 16 to 23 tonnes per person per year. Another [study](#) found lower yields, where a production of ~2.7 g of wet shiitake mushrooms/kg of oak log would imply a per capita wood consumption (assuming 2–3 harvests per year from the same log) of 281 to 422 tonnes. If shiitake yields are in fact this low, the wood required for getting enough food may make them out of reach for small groups of people post–societal collapse, and also make them a poor choice for large-scale production.

---

[56] For reference, I estimate that bran is ~2.56% N using the protein content and standard protein to N conversion factor of 6.25, while hardwood is reported as ranging from 0.36 to 0.65% N by mass, so bran has ~4–7x higher N concentrations.

I have also assumed that all of the wood would be suitable for mushroom cultivation, but it looks like shiitake mushrooms, the most nutritionally complete of the three species I looked at, [can't grow](#) on softwood. Softwood makes up about [48%](#) of US forest biomass, with the rest being hardwood, which should generally be suitable for shiitake growth. Oyster mushrooms, however, [do grow](#) on pines and other softwoods, so with a mix of mushroom species, the US should be able to use most of its tree species to cultivate mushrooms. If using just one fungal species, isolated groups may be restricted by their geography, since [93%](#) of hardwood tree mass is in the Eastern US.

## 19.0 Amount of food stored on farms and off farm facilities

The USDA provides [data](#) on how much of major grains and oilseeds[57] — wheat, corn, barley, oats, and soy — are stored on- or off-farm. Combined with [data](#) on the total number of off-farm storage facilities,[58] the number of farms producing these major crops (which I only found for [corn](#), [oats](#), [soy](#), and [wheat](#)), and nutrition data from the USDA [database](#), I made the following table showing the average number of person-years of calories are available on- and off-farm for major crops, and how many different facilities there are in the US. The table uses crop storage figures for September, which, as I covered above, is roughly when food stores are lowest in the US due to the crop calendar for major grains and oilseeds. I provide the relevant data and calculations in tab 14 of this [Google Sheet](#).

| Crop | Number of off-farm facilities | Number of farms | Person-years of food on average off-farm facility | Person-years of food on average farm |
|---|---|---|---|---|
| Barley[59] | 198 | | 55,245 | |
| Corn | 3,987 | [304,801](#) | 20,145 | 271 |
| Oats | 70 | [73,684](#) | 43,640 | 32 |
| Soy | 889 | [303,191](#) | 32,006 | 44 |
| Wheat | 2,924 | [240,000](#) | 55,376 | 199 |

[57] These crops make up ~28% of the US diet currently, not counting the use of crops as livestock feed, and these crops make up ~70% of the crop calories stored in the US in September.

[58] The data from USDA just shows the total number of off-farm storage facilities. I assumed that the number of facilities used for different crops is proportional to the amount of each crop in storage (during September).

[59] I have not found a source for the number of farms that grow barley in the US.

Sizes of food storage sites of course vary substantially, and the table only shows an average value. Also, a single food source would not provide adequate nutrition, and the table only expresses the size of food stores in terms of how long the food would meet an average adult's calorie needs. When the table reports, for example, that there are ~20,000 person-years of calories at the average off-farm corn storage facility, that means there are enough calories, though not enough essential nutrients, to feed 1,000 people for 20 years if the food doesn't spoil. This table provides some sense of how much food survivors of a societal collapse might be able to find at just a few locations, if reserves haven't been drawn down prior to collapse.

# 20.0 Large-scale food production with other technologies and in other countries

To better understand food production options in a loss of agriculture scenario, future research should evaluate the feasibility of scaling up nonagricultural food production in countries besides the US, including research into other foods not thoroughly explored in this report. Here, I provide some high-level analysis of the potential to scale up several nonagricultural foods outside of the US, but leave an in-depth investigation of such a scale up to future work.

As noted above, the work reported here focused on the US as a test case, because finding that nonagricultural food production is not possible in the US would have provided strong reason to be skeptical of its viability in other countries. This is because the US is especially well positioned to respond to a loss of agriculture: its comparatively large per capita food reserves ([3.0](#)) give it a longer runway to begin nonagricultural food production, and its low levels of import dependence, high natural resource availability, and significant domestic industrial strength ([20.1](#)) make it more capable of the massive amounts of construction and manufacturing needed to scale up nonagricultural foods, especially in disasters where international trade falls. Since I find that the US may in fact be able to replace agriculture with other food sources before existing food reserves run out, there is some reason to hope that other countries may also be able to respond to a loss of agriculture disaster in a similar way. In appendix [20.2](#), I explore several high-level indicators of the feasibility of large-scale nonagricultural food production in other countries, and in [20.3](#) I provide a first-pass qualitative assessment of overall feasibility in several selected countries.

Much as future research should look beyond the US, it should also look to other potentially scalable nonagricultural foods beyond methane-oxidizing bacteria single-cell protein (MOB SCP), which was the focus of this report. MOB SCP was chosen as an illustrative case study of a food that does not require agricultural inputs because the major inputs to its production, like natural gas and nutrients for the growth media and equipment manufacturing and materials for facilities, are already available in sufficient quantities in the US to make enough calories to feed the country. In [9.0](#), I found that there were no other foods for which the major material inputs were already produced at a sufficient level to allow them to scale to meet all US calorie needs. However, as discussed in [10.0](#), this analysis misses the possibility of increasing the production of these inputs to make more foods viable options. For example, the US does not currently capture enough $CO_2$ or produce enough $H_2$ to make hydrogen-oxidizing bacteria single-cell protein (HOB SCP). However, Synonym's analysis finds that building the facilities to produce these gasses would cost ~$65 billion, or just ~11% of the ~$589 billion needed to build all of the other facility components needed for HOB production, so this may well be feasible given my earlier assessment of SCP facility construction in the context of US industries ([13.0](#)).[60]

To provide a high level picture of the feasibility of scaling up other foods in various countries, I make several simplifying assumptions and look only at some gross measures of feasibility, rather than replicating the detail that I provided on US MOB SCP production in this report. Future work should extend this analysis by relaxing these assumptions and going into greater detail on food production requirements. For now, I assume that:

- All countries only access the natural resources in their borders, i.e. operate under autarky
- For each food evaluated, it provides all of a country's calorie needs, even if it wouldn't on its own be nutritionally complete and would require supplementation
- All countries' populations are initially unaffected by the disaster, but they are then vulnerable to starvation if new food cannot be produced before existing stores run out
- Wood is available, but as noted elsewhere, I am also interested in cases where wood or other plant biomass is not available.

I look at total CAPEX compared to the size of a country's industry, current production and size of reserves of major raw material inputs like fossil fuel feedstocks or mineral nutrients, current production of some major materials needed for facility construction like steel, import reliance measured as total imports divided by national GDP, and electricity production. This contrasts with the more granular assessment of specific equipment needs compared to various industries that I provided for MOB SCP production in the US (with Synonym Bio's help). The work presented here should generally be seen as narrowing the search space for future research rather than pointing to specific conclusions.

## 20.1 Comparing the US to other countries

To frame this preliminary discussion of the potential to scale nonagricultural food options in various countries, I provide an overview here of how the US compares globally on metrics important to consider for foods produced via chemical synthesis, growing photosynthesizers in protected environments, and cultivating microbes on non-plant substrates ([9.0](#)). All of these options require significant amounts of energy, fossil fuel feedstocks, and/or nutrient fertilizers to make. As explored for MOB SCP, the capital needed to manufacture the equipment and build the facilities required for food production is also considerable, with upfront costs equal to almost 2% of US GDP ([13.1](#)). Even if non-crop plant biomass is available, this doesn't [significantly change the picture](#), with the production of sugar ([6.0](#)) and sugar-fed fungal food ([15.0](#)) still requiring substantial capital investment. These material and economic requirements motivated my selection of gross indicators discussed in the subsections that follow, where I compare the US to other countries on these indicators.

An important takeaway from comparing the US to other countries is that while other countries may be similarly well positioned to scale up alternative foods, there are probably not countries that are able to exploit a *different* set of foods from those available to the US, since there do not appear to be countries that have clear advantages over the US in their industry, energy, or natural resource sectors.

---

[60] The $H_2$ production process assumed in Synonym's analysis was steam methane reforming using natural gas, which is the process used to make [~47%](#) of $H_2$ globally, followed by using coal, oil, and electrosynthesis of water. Steam methane reforming requires ~3.6 tonnes of natural gas/tonne of $H_2$ produced. Since the US produces [~860 Mt](#) of natural gas per year, and would need ~34 Mt of $H_2$ per year for HOB SCP ([9.2.2](#)), this would only require ~14% of US annual natural gas supplies, slightly less than what is needed for MOB SCP ([9.2.3](#)).

Something missed in comparing the US to other countries on these indicators one by one is that while the US is not unique on any single indicator, there are few countries that have a combination of resources and industrial capacity comparable to the US.

### 20.1.1 Electricity

For several food production options — such as indoor agriculture or electrosynthesis options for carbohydrate production — the main disqualifier for further consideration was the high electricity inputs needed. According to the US Energy Information Agency (EIA), the US ranks 9th in per capita electricity production, putting it at ~3.5x the global average. Only one country, Iceland, produces >2x the amount of electricity per capita as the US, so with that possible exception, other countries would likely also face difficulties scaling up foods that require high electricity input. I exclude these foods from further consideration for now, but investigating the possibility of quickly scaling up electricity generation in an emergency, so as to make these foods more feasible, is an important possible extension of this work. The table below shows the top 10 countries by electricity production per capita. In 20.2 I put these figures in context for the per capita electricity requirements for scaling certain food production technologies.

| Country | kWh/capita |
|---|---|
| Iceland | 51,306 |
| Norway | 22,950 |
| Bahrain | 21,185 |
| Qatar | 16,368 |
| Finland | 15,173 |
| Kuwait | 15,058 |
| Canada | 14,336 |
| Sweden | 12,324 |
| United States | 12,015 |
| Luxembourg | 10,167 |

### 20.1.2 Industry

The need for large amounts of manufacturing and construction to make nonagricultural food means that the value of a country's industry per capita is a useful high level indicator of its ability to scale up food production in an agricultural disaster. Using the World Bank's data on the value (in USD) of the manufacturing and construction sectors in different countries, here called industry, the US has the 14th-highest value of its industry per capita, putting it at ~3.6x the global average. The table below shows the top 20 countries by the value of their industry per capita. In 20.2 I put these figures in context for the per capita cost of scaling different food production technologies.

| Country | Value of manufacturing and construction sectors divided by population (USD) |
|---|---|
| Ireland | 45,691 |
| Qatar | 35,013 |
| Switzerland | 25,453 |
| Norway | 24,773 |
| United Arab Emirates | 19,376 |
| Denmark | 13,390 |
| Kuwait | 13,034 |
| Australia | 12,900 |
| Austria | 12,559 |
| Sweden | 12,323 |
| Iceland | 11,652 |
| Germany | 11,521 |
| Republic of Korea | 11,277 |
| United States of America | 10,952 |
| Japan | 10,880 |
| Canada | 10,826 |
| Finland | 10,573 |
| Luxembourg | 10,428 |
| Oman | 9,977 |
| Netherlands | 9,548 |

### 20.1.3 Import dependence

In a global disaster, international trade may be significantly reduced. This may require countries to operate autarkically. For some countries, this would make it extremely difficult to maintain their current industry, let alone to rapidly scale up novel food production technologies. The World Bank provides data on countries' imports as a percent of national GDP. This is a very high-level indicator of potential import dependence for critical inputs to scaling food production, as it does not capture specific goods and services relied upon by different countries. However, it provides some indication of which countries would least likely be handicapped by a drop in global trade. Here again the US stands out, ranking 4th, tied with Russia and only behind Sudan, Turkmenistan, and Argentina, of countries for which there is recent data. The table below shows the ten countries with the lowest import levels as a percent of national GDP.

| Country | Imports as a % of GDP |
|---|---|
| Sudan | 1.1 |
| Turkmenistan | 12.5 |
| Argentina | 15.4 |
| Russia | 15.6 |
| United States | 15.6 |
| Gabon | 16.6 |
| Nigeria | 16.9 |
| China | 17.5 |
| Ethiopia | 18.3 |
| Brazil | 19.3 |

### 20.1.4 Fertilizers

For food production using algae or microbial fermentation, various nutrients are needed. The most important, called macronutrients, are nitrogen, phosphorus, and potassium (N, P, and K, respectively). Here it is helpful to look at both current production and reserves.[61] For phosphorus and potassium, the main sources used for fertilizer production are mineral reserves of phosphate and potash, while nitrogen is generally made through the Haber-Bosch process using natural gas. The table below shows the US ranking in per capita production of N, P, and K fertilizers as well as the US rank in per capita reserves of phosphate, potash, and natural gas. I provide more information on the production and reserves of nutrients in other countries in 20.2.2, and I put these values into context for different food production options. Data on production quantities comes from the UN Food and Agriculture Organization (FAO), data on reserves for phosphate and potash come from the US Geological Survey, and reserves data for natural gas comes from the Central Intelligence Agency's World Factbook.

| Category | US global ranking |
|---|---|
| N fertilizer production per capita | 21 |
| P fertilizer production per capita | 13 |
| K fertilizer production per capita | 20 |
| Natural gas reserves per capita | 11 |
| Phosphate reserves per capita | 21 |
| Potash reserves per capita | 20 |

---

[61] For resources like minerals or fossil fuels, reserves are generally recorded for "economically available reserves", meaning reserves that would be worthwhile to exploit, given current economic conditions (i.e., the price of a commodity and the cost of extracting it). This means that as the price goes up or down for a commodity, the reserves increase or decrease. In the scenarios considered in this report, the value of inputs to food production would generally be expected to increase significantly, so it would be worthwhile for the countries to exploit resources that are not currently economical. This means that economical reserves undercount the availability of reserves for these purposes. For phosphate and potash, I used data on total estimated geological deposits, but for natural gas I only used economically available reserves, given the data that I could find.

While the US only ranks modestly in the above categories, it is more notable for being one of only nine countries in the world that has current domestic production and reserves for all three macronutrients. The others are Argentina, Belarus, Brazil, China, Israel, Jordan, Kazakhstan, and Russia.

### 20.1.5 Fossil fuels

Fossil fuels are important inputs to several nonagricultural food technologies, not just as energy sources but also as feedstocks. For example, natural gas can be fed to methane-oxidizing bacteria to make SCP, and natural gas, coal, or oil can be sources for both the H₂ gas needed for HOB SCP and the Fischer-Tropsch or paraffin wax that can be used to make synthetic fat (9.1.1). As with nutrient fertilizers, it is important to consider both current production and reserves in different countries. The table below shows the US rank in per capita production and per capita reserves of natural gas, coal, and oil. Production data comes from the EIA for natural gas, coal, and oil; reserves come from the CIA's World Factbook for natural gas and oil and the EIA for coal.

| Category | US global ranking |
|----------|------------------:|
| Natural gas production per capita | 15 |
| Coal production per capita | 52 |
| Oil production per capita | 19 |
| Natural gas reserves per capita | 11 |
| Coal reserves per capita | 9 |
| Oil reserves per capita | 25 |

As with fertilizer production and reserves, the US ranks only modestly in fossil fuel indicators, and unlike with fertilizers, it is not in as exclusive a club for having current production and reserves for all three fossil fuels, since 44 other countries also have this attribute.

### 20.1.6 Wood resources

In scenarios where woody biomass is available, it could be a valuable resource for food production, allowing countries to make both lignocellulosic sugar (6.0) and sugar-fed fungal food (15.0). Here, three main indicators are relevant for assessing lignocellulosic food production potential in different countries: current per capita wood harvests, per capita wood stocks, and per capita availability of repurposable infrastructure. Repurposable infrastructure refers to facilities used for making beer, ethanol, or pulp and paper that have been proposed as candidates for retrofitting to make sugar from lignocellulosic biomass. As discussed in appendix 6.0, tab 7 of this Google Sheet shows calculations for how much sugar can be produced from existing US facilities if all were repurposed to make sugar. I use this and country-level production data on beer (FAO), pulp and paper (FAO), and ethanol (DOE) to estimate the per capita availability of repurposable infrastructure. Having facilities that can be retrofitted to make sugar is not necessary for scaling up food production from woody biomass, but it does shorten construction time and reduces costs.

The table below shows the US global ranking for current per capita wood harvests, wood reserves, and repurposable infrastructure. In 20.3.4 I put these figures into context for scaling up wood-based food production in the US and in other countries.

| Category | US global ranking |
|---|---:|
| Per capita wood harvests | 30 |
| Per capita wood reserves | 42 |
| Per capita repurposable infrastructure | 4 |

## 20.1.7 Food reserves

The amount of time that a country has to begin producing nonagricultural food in a disaster is determined by how much food it has stored, assuming there is no international trade. As discussed in appendix 2.0, food stocks in the US vary substantially throughout the year, but data from the FAO — the organization with the most comprehensive international data on food stocks that I am aware of — only shows stock levels for crop commodities right before harvest. For the US, I supplemented the FAO's annual data on stored food with USDA quarterly crop stores for six major US crops to estimate food stores across the year (2.1). For other countries, both FAO and USDA food stock data is only annual, so I do not have estimates for within-year variation in stock levels worldwide. This means that I have not analyzed how the timing of a disaster or the possibility of advance warning (2.6) would impact how long food stores would last outside of the US.

While FAO global food stock data appears fairly comprehensive — more so than USDA data — contradictions between the two, and gaps in both, suggest caution in taking the information at face value. The FAO records food stock measurements for all UN-recognized countries, with data on 399 food commodities worldwide. USDA international data covers 50 stored food products and is missing data for many countries.[62] For this reason, it is only possible to provide a globally comprehensive estimate of how long food stores would be able to feed the populations of different countries by using FAO data. However, the two data sources sometimes provide conflicting estimates for how much of the same product is stored in a given country. For the US, the two estimates are generally quite close (within about ±10%). For other countries, the estimates can be more than an order of magnitude apart. Generally, it seems that FAO provides higher estimates for crop stock levels, but not ubiquitously. For example, FAO shows Kazakhstan's 2021 wheat reserves to be ~10x what is shown by USDA, but USDA estimated Indonesia's 2021 palm oil reserves to be ~10x higher than FAO's estimate. The FAO generally provides more data for most countries, but it shows zero soybean reserves for Brazil in 2021, despite Brazil being the world's leading soybean producer according to FAO's own production data. The USDA shows Brazil as having had almost 30 Mt of stored soybean that year (equal to ~23% of Brazil's 2021 soybean production) though it does not provide stored food data even for several OECD countries, such as Hungary, Ireland, Denmark, and others.

---

[62] The difference in the number of covered products is not as stark as this comparison suggests, because the FAO data breaks out many commodities that are treated as one product by USDA into different products, such as differentiating between various types of sugar. Also, globally, stored calories are fairly concentrated in a few products: the top 3 stored products account for ~48% of all stored calories, the top 10 account for ~70%, and the top 50 account for ~91%. The USDA data generally focuses on major food products, so for countries for which it has data, it captures most of the stored calories represented in the FAO's data, barring disagreements between the two data sources.

Given these data limitations, I am skeptical of many of the estimates of how long food supplies would last different populations based on FAO stock levels. A full list of these estimates, where I multiplied food stock levels by the calorie content of foods and divided by the daily calorie needs of a country, can be found on tab 16 of this Google Sheet (I discuss my methodology more in appendix 2.0). To show the discrepancies between FAO and USDA data, the table below shows how many days worth of stored food — assuming low waste (2.3) and a 2,100 kcal/day diet (4.0) — are suggested by FAO and USDA data for the world's ten most populous countries (covered in appendix 3.0). The table also shows the percent difference between the estimate based on FAO data and the one based on USDA data.

| | Months worth of stored food | | |
|---|---|---|---|
| Country | FAO | USDA | % difference FAO vs USDA |
| India | 4.3 | 6.5 | -33.8 |
| China | 20.1 | 19.4 | 3.9 |
| United States | 18.5 | 16.3 | 13.5 |
| Indonesia | 20.1 | 4.3 | 370.8 |
| Pakistan | 3.9 | 2.4 | 60.8 |
| Nigeria | 3.8 | 1.2 | 225.0 |
| Brazil | 5.8 | 13.4 | -56.5 |
| Bangladesh | 2.7 | 1.8 | 47.3 |
| Russia | 9 | 7.7 | 16.9 |
| Mexico | 7.7 | 5.3 | 44.8 |

Note that the data in the table above does not include a calculation of food supplies from living animals nor does it account for stock variations across the year. As reviewed for the US, this actually provides an unrealistically low estimate of stock levels, because crop stocks are at their lowest at different times of the year for different crops, so reporting only annual minimums for all crops undercounts the total supply at any given time of year.

Going with FAO's food stock data, the US has the 45th largest food reserves per capita of 163 countries for which FAO had complete data. Without another comprehensive data source, it is hard to corroborate these estimates. I suspect that the reserve levels for many countries are overstated, but I am unsure. On a calorie basis, the US produces ~10% of the world's food — while having ~4% of the world's population — so it should be expected that the US would have larger per capita food reserves than other many other countries, but small, food exporting countries may be expected to have larger reserves.

Taking the FAO food stock data at face value, it is clear that most countries would be significantly better able to respond to a loss of agriculture disaster if they had larger food reserves. In 20.2, I point to various countries that plausibly could scale up nonagricultural food technologies. Of these, only Malaysia, Kazakhstan, Romania, Canada, Argentina, and Finland have food reserves that would last longer than two years if a disaster hit at a bad time with respect to crop harvests.[63] Given that two years is close to the shortest start up time that I identified for natural gas-fed microbial food plants (14.0), this means that even among countries with advanced industry and abundant natural resources, there may not be enough time to scale up food production given current food reserves.

## 20.2 Nonagricultural food production options across the world

In this subsection, I review the prospects for nonagricultural food production options across the world. I provide data on a few select countries for different technologies in this text, but information on all countries for which I could find relevant data can be seen on tabs 17–20 in this Google Sheet. The countries selected to be shown in the tables that follow were chosen based on several indicators for different technologies. However, these indicators do not capture all of the variables that impact the feasibility of scaling up nonagricultural food production. A country's ultimate prospects for doing this will depend on institutional stability, food stores, and many other factors that go beyond the simplistic analysis below.

My review of nonagricultural food production technologies in the US finds that while MOB SCP may be one of the more scalable options, HOB SCP, synthetic fat, and microalgae (grown outdoors in photobioreactors to avoid contamination from a possible biological threat) are probably also viable candidates. In cases where wood is available, sugar and fungal foods are also options. Given that other countries likely face similar constraints as the US when it comes to scaling other food options explored in appendix 9.0, I do not expect there to be many other technologies that could scale in other countries, but research to identify other possible food sources in a loss of agriculture scenario would be valuable.[64]

There are two main reasons to be interested in considering other food options besides MOB SCP. The first is that other countries have different economic and natural resource conditions than the US, so different foods may be more favorable in those cases. The second reason to be interested in other foods is that diversified food production would provide better post-disaster nutrition, hedge against the risk that a given technology runs into an unforeseen failure, and allow countries to optimize resource allocation across different technologies that may require different inputs, helping to avoid bottlenecks that could arise from focusing heavily on a single food source.

### 20.2.1 Gas fermentation for microbial foods

HOB SCP production appears slightly less favorable than MOB SCP in the US, given that natural gas is fairly abundant in the US and HOB SCP production would require scaling up $H_2$ production and $CO_2$ capture. However, once these inputs are available, the technologies to make both products are extremely similar. Since there are multiple routes to making H$_2$ besides using natural gas, such as using coal, HOB SCP may be preferable for countries that do not have large amounts of natural gas but do produce a significant amount of other fossil fuels. Therefore I review both HOB and MOB SCP production potential across the world.

---

[63] Note that this estimate does not include slaughtering livestock or drawing on stopgap food sources like extra fishing or lignocellulose sugar.

[64] In addition to the technologies discussed in 9.0, lignocellulosic foods explored in 6.0 and 15.0, and low-tech options described in 18.0, I also looked briefly into a few other fermentation-based options that do not require plant-derived substrates, such as turning natural gas into ethanol or glycerol and using these as carbon sources for microbial foods. One reason to potentially favor such options over HOB or MOB SCP is that they do not require gas fermentation and instead use liquid substrates. Liquid fermentation facilities are generally less capital intensive and less difficult to construct, and could also take advantage of retrofitting existing fermentation facilities and equipment more easily. However, for the non-plant-derived substrates that I explored, production rates — measured in biomass produced per unit volume per unit time — appear much lower than those for gas fermentation. Since the products have similar caloric values per unit mass, this means that using these liquid substrates would require significantly more fermentation volume, probably making them impractical.

I use Synonym Bio's techno-economic assessments (TEAs) of HOB and MOB SCP, combined with nutrition information on MOB from Unibio and on HOB from this paper to calculate per capita CAPEX and input requirements for both technologies. Using the same country-level data described in 20.1, I found the fraction of a country's industry — defined as the value of their manufacturing and construction sectors — and current resource production that would be required to make either HOB or MOB SCP. I also calculated how long a country's fossil fuel reserves[65] would last, assuming that they were used exclusively for making SCP. For HOB SCP, I assumed that a country used either natural gas or coal to make H$_2$, going with whichever input could produce more H$_2$ gas in that country, given its current level of production and the mass conversion rate from fossil fuels to H$_2$. I did not assess using electrosynthesis to make H$_2$, because making enough H$_2$ for the US this way would take ~41% of the country's current electricity generation. Even for the eight countries with a higher per capita electricity production than the US (20.1.1), only Iceland would have to use less than 20% of its current electricity generation just to make enough H$_2$ to feed its population with HOB SCP. Given the current global electricity landscape, this is probably not a very scalable option.

The table below shows data on the production requirements for MOB SCP in selected countries, if MOB SCP production was scaled up to the level needed to feed the whole population. These ten countries were chosen based on how promising the technology appears given their economic and natural resource conditions. A full list of countries for which I could find complete data can be found on tab 17 of this Google Sheet. Note that the gross indicators of feasibility I've chosen do not represent a complete assessment of SCP production in different countries, but instead provide a high-level picture of which countries I expect are most likely to succeed in scaling up MOB SCP if they had enough time.

| Country | % Natural gas | % Electricity | % Industry | % Current N fertilizer production needed | Imports as a % GDP | Years natural gas reserves would last | Month food reserves would last |
|---|---|---|---|---|---|---|---|
| Qatar | 0.8 | 4.7 | 4.6 | 3.7 | 31.5 | 21,566 | 12.5 |
| Norway | 1.6 | 3.3 | 6.5 | 33.5 | 27.1 | 891 | 14 |
| Canada | 9.6 | 5.3 | 15 | 36.3 | 33.7 | 125 | 54.5 |
| United States of America | 14 | 6 | 14.6 | 94.8 | 15.6 | 587 | 18.5 |

[65] As noted earlier, the term "reserves" for natural resources generally refers to the proven amount of a resource that would be economic to exploit given current market conditions. In the scenarios considered here, the value of the resources needed to make food would likely increase substantially, so countries would try to exploit many sources of materials that are not currently economic to extract. This means that the reserves data I am citing conservatively underestimates the true size of fossil fuel resources available to countries. However, in calculating how long reserves would "last", I also made the unrealistic assumption that fossil fuel resources were used only as feedstocks for microbial food production.

| | | | | | | |
|---|---|---|---|---|---|---|
| Saudi Arabia | 14.5 | 8.3 | 18 | 42.8 | 23.3 | 544 | 20 |
| Oman | 5.8 | 10.8 | 16 | 10.1 | 41.4 | 367 | 11 |
| Russia | 10.9 | 11.7 | 51.6 | 45.4 | 15.6 | 774 | 9 |
| Australia | 18.5 | 8.3 | 12.4 | 238.2 | 19.7 | 78 | 13.5 |
| Kazakhstan | 21.6 | 15.9 | 39.7 | 422.1 | 26.3 | 295 | 61 |
| Venezuela (Bolivarian Republic of) | 16.9 | 37.8 | 23.3 | 516.2 | 31.4 | 465 | 6 |

The table below shows data on the production requirements for HOB SCP in selected countries. As above, these countries were chosen based on how promising large-scale HOB SCP production looks based on my chosen indicators. A full list of countries can be found in tab 18 of this [Google Sheet](). The columns referring to fossil fuel use refer to either natural gas or coal, depending on which is more advantageous for a given country to use.

| Country | % Coal or natural gas needed | % Electricity needed | % Industry | Years coal or natural gas would last | % Current N fertilizer production needed | Imports as % GDP | Months food reserves would last |
|---|---|---|---|---|---|---|---|
| United States of America | 14.2 | 6.2 | 15.6 | 532.8 | 118.6 | 15.6 | 18.5 |
| Qatar | 0.8 | 4.8 | 4.9 | 19,580 | 4.6 | 31.5 | 12.5 |
| Canada | 0.04 | 5.5 | 16 | 243.7 | 45.5 | 33.7 | 54.5 |
| Saudi Arabia | 14.6 | 8.6 | 19.2 | 494.2 | 53.5 | 23.3 | 20.0 |
| Australia | 0.004 | 8.6 | 13.2 | 7,894.1 | 298 | 19.7 | 13.5 |
| Oman | 5.8 | 11.2 | 17.1 | 332.8 | 12.6 | 41.4 | 11.0 |
| New Zealand | 0.1 | 9.9 | 20.4 | 2,086.5 | 237 | 29.6 | 16.0 |
| China | 0.03 | 16.6 | 37.4 | 134.7 | 226.8 | 17.5 | 20.0 |
| Russia | 11 | 12.1 | 55 | 702.6 | 56.8 | 15.6 | 9.0 |
| Germany | 0.03 | 11.8 | 14.8 | 611.7 | 294.3 | 49 | 21.0 |

### 20.2.2 Microalgae

In the US context, I found that scaling up the production of microalgae — such as cyanobacteria or green algae — to the level needed to meet all US calorie needs would require an increase in the annual production of several nutrient fertilizers, especially potassium. In appendix 9.3.2, I lay out two sets of assumptions, one optimistic and one pessimistic, for how much of each nutrient would be needed. The optimistic assumptions are based on a report by the National Renewable Energy Laboratory (NREL), and the pessimistic assumptions are based on a life cycle assessment (LCA) that used data from a commercial Chinese spirulina production facility. As reviewed above, the differences come from different nutrient requirements for the two microalgae species that the studies looked at, and different assumptions about nutrient recycling and uptake efficiency.

Looking at nutrient fertilizer availability, only 17 countries worldwide have known reserves of natural gas, phosphate, and potassium, though many more countries have reserves of at least one of these resources.[66] As noted in appendix 20.1.4, of these 17 countries, only 9 currently produce N, P, and K fertilizers according to FAO data. If countries are restricted to true autarky, and the data that I found on mineral nutrient and natural gas reserves is exhaustive, then at most 17 countries could scale up microalgae production. This number could expand significantly if the assumption of autarky was relaxed, other mineral nutrient reserves were found, or other methods for acquiring nutrients were deployed. These other technologies could include extracting nutrients from seawater or from P- or K-containing rocks. A very cursory review suggests that these approaches are too nascent and resource intensive to be deployed at a large scale, but technological advances could change that. For now, I'll stick with the conservative list of countries that could scale up microalgae production without international trade, based on their known natural resources.

I used data from NREL's study to construct the table below (divided into two parts for clarity) for production requirements for eight countries whose industry and natural resources suggest that they may be able to scale up microalgae cultivation. NREL described several photobioreactor (PBR) designs: tubular, helical, flat panel, and hanging bags. Based on their data, when scaled up to the level needed to produce enough microalgae to meet US calorie needs, the flat panel appears to be the most feasible PBR design. I've based the following calculations on the assumption that a flat panel PBR design is used, but other configurations may prove better. An important trade-off I made in selecting the flat panel design is that it is considerably more capital intensive per unit of annual algae production than a tubular design. However, it requires ~2.6x less low-density polyethylene (LDPE) to make enough flat panel PBRs compared to using a tubular design. Since both designs would require the US — and other countries — to use more than their current annual production of LDPE to make enough PBRs, I expect that the trade-off is worthwhile, especially since scaling up LDPE production could add extra time in a scenario where countries would be racing to produce new food before existing reserves ran out. However, future research should examine this assumption.

In addition to indicators used for other food above, I show in the tables below what fraction of a country's domestic steel production would be needed to build all of the facilities for scaling algae production to meet its calorie needs, and do the same for the LDPE. I also calculate the country's potential production of polyethylene (PE) based on its current production of either natural gas or naphtha, the two main feedstocks used to make PE. For countries that do not currently produce one or more of the macronutrients needed for algae cultivation but have reserves for these nutrients, I show "NA" in the column displaying the percent of a country's current fertilizer production that would be needed, meaning that the country would have to start production from scratch (assuming complete autarky). Data for all countries that have domestic reserves of natural gas, phosphate, and potash can be found in tab 19 of this [Google Sheet](#).

---

[66] From my data sources (see 20.1.4), there are 102 countries with known natural gas reserves, 51 with known phosphate reserves, and 28 with known potash reserves.

| Country | % Electricity needed | % Industry | % LDPE | % Steel | Months food reserves would last |
|---|---|---|---|---|---|
| Argentina | 25.6 | 27.2 | 328.3 | 17 | 43.5 |
| Canada | 4.8 | 7.7 | 131.2 | 5.4 | 54.5 |
| China | 14.6 | 17.9 | 362.4 | 2.5 | 20.0 |
| Iran | 21.5 | 46.4 | 76.6 | 5.2 | 12.0 |
| Israel | 9.8 | 10.1 | 673.2 | 54.7 | 16.5 |
| Russia | 10.7 | 26.3 | 279.1 | 3.5 | 9.0 |
| United Kingdom | 15.1 | 10.3 | 254.5 | 22.1 | 8.0 |
| United States of America | 5.4 | 7.4 | 221 | 7.2 | 18.5 |

| Country | % Current N | % Current P | % Current K | % PE potential | Imports % GDP |
|---|---|---|---|---|---|
| Argentina | 65.1 | 162.7 | 19,587.9 | 83.5 | 15.4 |
| Canada | 5.9 | NA | 0.6 | 15.8 | 33.7 |
| China | 29.2 | 26.3 | 66 | 133.4 | 17.5 |
| Iran | 27.8 | NA | NA | 27.1 | 24.8 |
| Israel | 161.8 | 6.6 | 0.9 | 42.9 | 28.8 |
| Russia | 7.3 | 9.9 | 3.1 | 17.9 | 15.6 |
| United Kingdom | 125.8 | NA | 156.4 | 112.5 | 36.1 |
| United States of America | 15.3 | 19.6 | 222.8 | 23 | 15.6 |

### 20.2.3 Synthetic fat

Of the foods that I reviewed made through chemical synthesis (9.1), synthetic fat produced from Fischer-Tropsch or paraffin wax is probably the most promising for producing a lot of calories (9.1.1). As discussed above, Fischer-Tropsch wax can come from natural gas or liquified coal, and paraffin wax occurs naturally in crude oil. Synthetic fat production in a loss of agriculture scenario also requires making synthetic glycerol, which can be done using propylene, a propane derivative. Using figures provided by a synthetic fat company, Savor-it, a paper authored by some of Savor-it's scientific staff, and a paper by ALLFED, I calculated the percent of a country's industry, propylene production potential based on current natural gas production, and Fischer-Tropsch or paraffin wax production potential based on current oil, natural gas, or coal production. For wax production potential, I assumed that only one fossil fuel source was used, and choose whichever one would yield a higher amount of wax for each country. To calculate wax production potential, I used the conversion figures provided by Savor-it's paper for coal and natural gas, and I used the global average paraffin wax content in crude oil to calculate wax production potential based on current oil production. Using only one fossil fuel resource underestimates the wax production potential for some countries, but I chose this simplifying assumption to be conservative, and also to reflect that quickly scaling up additional production pathways poses additional challenges.

The table below shows the production requirements for ten countries whose economy and natural resources appear to position them to scale up synthetic fat production. I also show how long fossil fuel reserves would last if used only for synthetic fat production. It should be noted that while the figures below assume that synthetic fat meets all of a country's calorie needs, it would not be possible to eat only fat. The need for dietary diversity holds for all of the food options reviewed there, but it is most important when considering a food that contains only one macronutrient and no micronutrients. Based just on macronutrient requirements, fat could provide only ~80% of a 2,100 kcal/day diet, and intake should probably be lower for better health. Therefore these figures should be seen as illustrative of the requirements for including synthetic fat as a major component of post-disaster diets, but should not be taken literally. A full list of countries can be found on tab 20 of this Google Sheet.

| Country | % Propylene production potential | Years oil, natural gas, or coal reserves would last | % Industry | % Current coal, oil, or natural gas needed to make wax | Month food reserves would last |
|---|---|---|---|---|---|
| United States of America | 9.9 | 76,249 | 6.8 | 7.8 | 18.5 |
| Saudi Arabia | 10.2 | 70,732 | 8.3 | 1.2 | 20 |
| Norway | 1.1 | 115,749 | 3 | 0.9 | 14 |
| Australia | 13.1 | 10,137 | 5.8 | 10.4 | 13.5 |
| Qatar | 0.5 | 2,802,265 | 2.1 | 0.4 | 12.5 |
| Turkmenistan | 2.6 | 728,440 | 24.2 | 2.1 | 9.5 |
| Russia | 7.7 | 100,560 | 24 | 5.1 | 9 |
| Canada | 6.8 | 1,626 | 7 | 3 | 54.5 |
| Venezuela | 12 | 60,406 | 10.8 | 9.5 | 6 |
| Oman | 4.1 | 47,628 | 7.4 | 1.5 | 11 |

## 20.2.4 Lignocellulosic foods

In cases where wood is available, countries could produce sugar, sugar-fed fungal food, or mushrooms grown on wood. I found that mushrooms grown on wood are fairly inefficient ([18.5](#)), so the two other options are generally preferable, at least in the context of a functioning industrial society. As with synthetic fat, sugar provides only calories and no other nutrients. In appendix [6.1](#) I discuss how much sugar could be included in a post-disaster diet. The amount depends on the nutritional value of other food sources. If combined with fungal SCP ([15.1](#)), then sugar could potentially comprise up to almost 70% of a diet, but this would exceed the amount of sugar in any recorded diet that I have seen ([6.1](#)), and such a diet would be deficient in several important micronutrients. As a simplifying assumption, however, here I'll assume that countries use wood to make a diet that by calories is 50% lignocellulosic sugar and 50% fungal SCP.

I use the capital and resource inputs for the production of lignocellulosic sugar and fungal SCP described in appendices 6.0 and 15.0, respectively, to provide the following table. I show several select countries that may be well positioned to scale up lignocellulosic food production based on their economy and forest resources. A full list of countries for which I could find data can be found [here](#). In the table below, I show how long a country's total stock of wood would last if no new wood grew, no wood decayed, and all wood was used to make lignocellulosic food. These simplifying assumptions are meant to provide a rough sense of the current forest resources in countries, but are not meant as literal predictions.

| Country | % Current wood harvest needed | Years wood last | % Industry | % Current N fertilizer production needed | Imports % GDP | Month food reserves would last |
|---|---|---|---|---|---|---|
| United States of America | 139.9 | 64 | 11.2 | 35.2 | 15.6 | 18.5 |
| Norway | 77.9 | 120 | 5 | 12.4 | 27.1 | 14 |
| Finland | 16.1 | 232 | 11.6 | 37.4 | 47.7 | 29.5 |
| Canada | 50 | 621 | 11.4 | 13.5 | 33.7 | 54.5 |
| New Zealand | 26.3 | 424 | 14.7 | 70.2 | 29.6 | 16 |
| Russia | 140.2 | 293 | 39.7 | 16.8 | 15.6 | 15.6 |
| Austria | 87.8 | 69 | 9.8 | 30.4 | 61.6 | 13.5 |
| Argentina | 451.9 | 35 | 41.1 | 149.4 | 15.4 | 43.5 |

| Poland | 161.9 | 37 | 26.9 | 24.1 | 61.2 | 23.5 |
| Romania | 210.7 | 64 | 38.4 | 42.6 | 49.8 | 54.5 |

## 20.3 Initial guesses for viability of nonagricultural food production potential in select countries

Below, I provide a very rough qualitative assessment of the feasibility of different food production options in several selected countries. This assessment is based on the indicators of feasibility for different nonagricultural food production options reviewed in 20.2, and available in tabs 17–21 of this Google Sheet. My assessment also takes into account the amount of time that countries would have to scale up new food production options based on their existing food stores (tab 16, and also provided below for select countries). However, as covered in 20.1.7, there may be reason to doubt some of these food store estimates based on data quality issues.

Here, I am evaluating the ability of countries to scale up a given food production technique to meet the calorie needs of its current population before existing food reserves are used up. To simplify, I assume in each case that only one food is used to meet all calorie needs. However, relying on a single food carries many nutritional risks, and is certainly not possible for some of the foods considered here, like synthetic fat, which provides calories but no carbohydrates, protein, or micronutrients. I also assume that countries operate in a state of autarky. The feasibility of this for different countries is roughly captured in my assessment by including imports as a percent of GDP (from the World Bank) as part of my analysis. In the case of lignocellulosic-based foods, namely sugar (6.0) and fungal SCP (15.0), I provide my assessment of feasibility conditional on scenarios where woody biomass is available.

**I want to reemphasize the roughness of this qualitative assessment.** It is based on guesswork, and thus substantially less robust than other analysis described in this report. Follow-up research will likely modify this preliminary assessment considerably. Research into other countries' prospects for food scale-up is an important future research direction.

In the table below, I provide my qualitative assessment of the prospects for fifteen countries that appear to have the best chance of scaling up at least one type of food production option to meet all of their country's calorie needs, based on the characteristics discussed in 20.2. I show very rough probability estimates of the likelihood that each of these fifteen countries has the capacity to scale up individual food options — gas fermentation, synthetic fat, microalgae grown in photobioreactors, and lignocellulosic-based foods — before their existing food supplies are exhausted. A country could feed its population if it succeeds in scaling up at least one food option, so these subjective probabilities allow for a calculated guess of overall survival likelihood.

Beyond the individual probabilities that I assign to country and technology pairs, there are two other factors that I take into account in my estimate of survival likelihood for each country. First, I apply an across-the-board haircut of 35% on the probability of any food tech succeeding, to account for technical challenges I have not modeled into my initial tech analysis in this document. Second, I apply a 25% haircut to each country's odds of surviving, to account for the possibility of society-wide institutional failure that inhibits an effective disaster response. Combining these haircuts with the probabilities that I assign to country and technology pairs, I show in the table my overall assessment of the likelihood that each of the fifteen countries scales up at least one nonagricultural food production method before its food supplies run out.

| Country | Months worth of food stores | Probability a country could scale up each food production option before running out of food[67] | | | | Overall assessment of survival likelihood |
|---|---|---|---|---|---|---|
| | | Gas fermentation for microbial foods | Synthetic fat from fossil fuels | Microalgae | Lignocellulosic-based foods | |
| United States of America | 18.5 | 30% | 30% | 30% | 35% | 45% |
| Canada | 52.9 | 30% | 30% | 15% | 40% | 45% |
| Saudi Arabia | 18.9 | 20% | 25% | <5% | <5% | 20% |
| Australia | 19.4 | 20% | 25% | <5% | <5% | 20% |
| Finland | 29.6 | <5% | <5% | <5% | 40% | 20% |
| Norway | 14.0 | 15% | 15% | <5% | 15% | 20% |
| New Zealand | 16.1 | 10% | 10% | <5% | 20% | 20% |
| Kazakhstan | 60.9 | 5% | 25% | <5% | <5% | 15% |
| Argentina | 43.7 | <5% | 20% | 5% | <5% | 10% |
| China | 20.1 | 10% | <5% | 10% | <5% | 10% |
| Qatar | 12.5 | 10% | 10% | <5% | <5% | 10% |
| Libya | 19.9 | 0% | 15% | <5% | <5% | 10% |
| Israel | 16.5 | 0% | 10% | <5% | <5% | 5% |
| Russia | 9.0 | <5% | 5% | <5% | 5% | 5% |
| Kuwait | 15.3 | <5% | 10% | <5% | <5% | 5% |

The table above assumes that countries operate in a state of autarky. However, there are many cases where different countries have complementary assets and attributes. Because of this, trade may make nonagricultural food scale-up significantly more feasible in different parts of the world. I have not evaluated this possibility here, but research into how trade could change the prospects of responding to a loss of agriculture disaster is an important extension of this work.

Overall, my first pass analysis suggests that there are few countries that are currently as well positioned as the US to scale up nonagricultural food production before existing reserves run out, and under the assumptions of this report, no country appears to have a surefire path to success. However, I consider the results of this rough analysis surprising, in a positive way: there are many countries with plausible paths to feeding their populations using a diverse set of existing technologies, even if all plants died and could not be replanted. The fact that multiple countries have plausible "shots on goal" at survival suggests that society as a whole may be less vulnerable to a sudden and total loss of agriculture than one might naively expect, though successes across countries are likely highly correlated.

---

[67] Not accounting for the technological and institutional failure haircuts described in the text and incorporated in the rightmost column

Moreover, ending this vulnerability with heightened confidence may now be within reach. The work presented in this report further suggests several ways that individual countries, and humanity as a whole, can improve their resilience to a loss of agriculture (see above). These include additional investigation, development of relevant technical and industrial capacities, and investment in larger food stores.